\documentclass[12pt]{article}
\makeatletter \@addtoreset{equation}{section} \makeatother
\renewcommand{\theequation}{\thesection.\arabic{equation}}
\newcommand{\ba}{\begin{array}}
\newcommand{\ea}{\end{array}}
\newcommand{\beq}{\begin{equation}}
\newcommand{\eeq}{\end{equation}}
\newcommand{\bea}{\begin{eqnarray}}
\newcommand{\eea}{\end{eqnarray}}

\def\bce{\begin{center}}
\def\ece{\end{center}}

\def\nonu{\nonumber}

\def\pa{\partial}
\def\al{\alpha}
\def\be{\beta}

\def\de{\delta}

\def\ep{\epsilon}

\def\th{\theta}

\def\si{\sigma}

\def\eps6{{\displaystyle \mathop{\epsilon}^{6}}{}}
\def\g6{{\displaystyle \mathop{g}^{6}}{}}

\def\nab6{{\displaystyle \mathop{\nabla}^{6}}{}}

\def\0{{\sst{(0)}}}
\def\1{{\sst{(1)}}}
\def\2{{\sst{(2)}}}
\def\3{{\sst{(3)}}}
\def\4{{\sst{(4)}}}
\def\5{{\sst{(5)}}}
\def\6{{\sst{(6)}}}
\def\7{{\sst{(7)}}}
\def\8{{\sst{(8)}}}

\def\ba{\begin{array}}
\def\ea{\end{array}}
\def\beq{\begin{equation}}
\def\eeq{\end{equation}}
\def\be{\begin{equation}}
\def\ee{\end{equation}}

\def\Tr{\mathop{\rm Tr}}

\def\eps{\epsilon}

\def\th{{\theta}}

\def\ba{\begin{array}}
\def\ea{\end{array}}
\def\beq{\begin{equation}}
\def\eeq{\end{equation}}
\def\be{\begin{equation}}
\def\ee{\end{equation}}

\def\Tr{\mathop{\rm Tr}}

\def\eps{\epsilon}

\def\th{{\theta}}

\def\eps6{{\displaystyle \mathop{\epsilon}^{6}}{}}

\def\nab6{{\displaystyle \mathop{\nabla}^{6}}{}}

\newcommand{\bean}{\begin{eqnarray*}}
\newcommand{\eean}{\end{eqnarray*}}

\begin{document}
\thispagestyle{empty} \addtocounter{page}{-1}
   \begin{flushright}
KIAS-P10009 \\
\end{flushright}

\vspace*{1.3cm}
  
\centerline{ \Large \bf   Holographic ${\cal N}=1$ Supersymmetric
  Membrane Flows }
\vspace*{1.5cm}
\centerline{{\bf Changhyun Ahn 
 {\rm and} Kyungsung Woo }
} 
\vspace*{1.0cm} 
\centerline{\it  
Department of Physics, Kyungpook National University, Taegu
702-701, Korea} 
\vspace*{0.8cm} 
\centerline{\tt ahn@knu.ac.kr 
\qquad wooks@knu.ac.kr
} 
\vskip2cm

\centerline{\bf Abstract}
\vspace*{0.5cm}

The M-theory lift of ${\cal N}=2$ 
$SU(3) \times U(1)_R$-invariant RG flow via a combinatorical use of  
the 4-dimensional  flow and 11-dimensional Einstein-Maxwell
equations
was found previously. 
By taking the three
internal coordinates differently
and preserving only $SU(3)$ symmetry from the ${\bf CP}^2$ space, 
we find a new 11-dimensional solution 
of ${\cal N}=1$ $SU(3)$-invariant RG flow interpolating from ${\cal N}=8$
$SO(8)$-invariant UV fixed point to ${\cal N}=2$ $SU(3) \times U(1)_R$-invariant IR
fixed point in 4-dimensions. 
We describe how the corresponding 3-dimensional  
${\cal N}=1$ superconformal Chern-Simons matter theory deforms.
By replacing the above ${\bf CP}^2$ space with 
the Einstein-Kahler 2-fold,  
we also find out new 11-dimensional solution 
of ${\cal N}=1$ $SU(2) \times U(1)$-invariant RG flow 
connecting above two fixed points in 4-dimensions. 

\baselineskip=18pt
\newpage
\renewcommand{\theequation}
{\arabic{section}\mbox{.}\arabic{equation}}

\section{Introduction}

The 3-dimensional 
${\cal N}=6$ $U(N) \times U(N)$ 
Chern-Simons matter theory \cite{ABJM}
with level $k$  describes 
the low energy limit of $N$ membranes at 
${\bf C}^4/{\bf Z}_k$ singularity.
The ${\cal N}=8$ supersymmetry
is preserved for $k=1, 2$. 
The matter contents and the superpotential  of this theory
are exactly same as the ones in the theory for 
D3-branes at the conifold in 4-dimensions \cite{KW}. 
The 3-dimensional membrane 
theory is related to the 4-dimensional ${\cal N}=8$ gauged
supergravity  theory \cite{dN} via AdS/CFT 
correspondence \cite{Maldacena}. 
The holographic ${\cal N}=2$ 
$SU(3) \times U(1)_R$-invariant renormalization group(RG) flow 
connecting the ${\cal N}=8$ $SO(8)$
ultraviolet (UV) point 
to the ${\cal N}=2$ $SU(3) \times U(1)_R$ infrared (IR) point 
has been studied in 
\cite{AP,AW,AR99} long ago while
the ${\cal N}=1$ $G_2$-invariant RG flow  
from the ${\cal N}=8$ $SO(8)$  UV point 
to the
${\cal N}=1$ $G_2$ IR point has been described in 
\cite{AW,AI}.
The former has $SU(3) \times U(1)_R$-symmetry
and the latter has $G_2$-symmetry, around IR region.
The 11-dimensional M-theory lifts of these two RG flows 
have been found in \cite{CPW,AI}  by solving the
Einstein-Maxwell equations explicitly in 11-dimensions.

The mass deformed $U(2) \times U(2)$
Chern-Simons matter theory with $k=1, 2$ 
preserving the ${\cal N}=2$ 
$SU(3) \times U(1)_R$ symmetry has been found 
in \cite{Ahn0806n2,BKKS} while
the mass deformation for this theory preserving the ${\cal N}=1$ $G_2$
symmetry  has been found in \cite{Ahn0806n1}.  
The non-supersymmetric 
RG flow equations preserving two $SO(7)^{\pm}$ symmetries 
have been studied in \cite{Ahn0812}.  
The holographic ${\cal N}=1$ $SU(3)$-invariant
RG flow equations connecting  ${\cal N}=1$ $G_2$ point 
to  ${\cal N}=2$ $SU(3) \times U(1)_R$ point in 4-dimensions have been 
studied in \cite{BHPW}.  
Moreover, the other holographic supersymmetric
RG flows have been 
found and  
further developments on  
the 4-dimensional gauged supergravity (see also \cite{GW02,PW03}) 
have been made
in \cite{AW09,Ahn0905}. 
The spin-2 Kaluza-Klein modes around 
a warped product of $AdS_4$ and a seven-ellipsoid having the
${\cal N}=1$ $G_2$ symmetry are discussed in \cite{AW0907}. 
The gauge dual with 
${\cal N}=2$ $SU(2) \times SU(2) \times U(1)_R$ symmetry 
for the 11-dimensional lift of 
$SU(3) \times U(1)_R$-invariant solution 
in 4-dimensional supergravity 
is described in \cite{AW0908} (see also \cite{FKR}). 
The 11-dimensional description preserving ${\cal N}=2$ $SU(2) \times U(1)
\times U(1)_R$ symmetry is found in \cite{Ahn0909} and
the smaller ${\cal N}=2$ $U(1) \times U(1) \times U(1)_R$ symmetry flow is
discussed in \cite{Ahn0910}. Further study on \cite{AI} is 
done in \cite{AW1001} recently. 

When the 11-dimensional supergravity theory is reduced to 
4-dimensional ${\cal N}=8$ gauged supergravity, the 4-dimensional
spacetime is warped by a warp factor $\Delta$ which depends on both
4-dimensional coordinates and 7-dimensional internal coordinates.
The 7-dimensional internal metric of deformed 7-sphere is obtained
from the $AdS_4$ supergravity fields \cite{dWNW,dN87}. 
As they vary, the geometric structure of round 7-sphere changes.
We have the following
11-dimensional metric  
\bea
ds_{11}^2 =\Delta^{-1} \, \left(dr^2 +e^{2 A(r)}
\, \eta_{\mu\nu}\, dx^\mu dx^\nu \right)+ ds_7^2,
\label{11dmetric}
\eea
where the 3-dimensional metric is $\eta_{\mu \nu}=(-, +, +)$, the
radial coordinate is transverse to the domain wall, and the scale factor
$A(r)$ behaves linearly in $r$ at UV and IR regions.

The $AdS_4$ supergravity fields $(\rho,\chi)$ in 4-dimensional gauged
supergravity as well as the scale function $A$
satisfy the supersymmetric $SU(3) \times U(1)_R$-invariant 
RG flow equations \cite{AP} 
\bea
\frac{d \rho}{d r} & = & \frac{1}{8L \, \rho} \, \left[
  (\cosh 2\chi +1) + \rho^8\, (\cosh 2\chi-3) \right],
\nonu \\
\frac{d \chi}{d r} & = & \frac{1}{2L \, \rho^2} \,
(\rho^8-3) \,
\sinh 2\chi,
\nonu \\
\frac{d A}{d r} & = & \frac{1}{4L \, \rho^2} \, 
\left[ 3
(\cosh 2\chi+1) -\rho^8 (\cosh 2\chi-3)\right].
\label{flow}
\eea
There exist two critical points, ${\cal N}=8$ $SO(8)$ critical point 
at which $(\rho,\chi)=(1,0)$ and ${\cal N}=2$ $SU(3) \times U(1)_R$
critical point at which $(\rho,\chi)=(3^{\frac{1}{8}},\frac{1}{2}
\cosh^{-1} 2)$ \cite{Warner83}. The $L$ is the radius of the 
round 7-sphere ${\bf S}^7$.
We focus on the possible 11-dimensional lifts of the 
RG flows around ${\cal N}=2$ $SU(3) \times U(1)_R$ critical point 
in this paper and those around ${\cal N}=1$ $G_2$ critical point have
been described in \cite{AW1001} recently. 

For the specific 7-dimensional internal metric $ds_7^2$, how one can determine
the solution for  11-dimensional Einstein-Maxwell equations? 
For given 11-dimensional metric and 4-form field strengths, 
the 11-dimensional bosonic field equations are given by \cite{CJS}
\bea
R_{M}^{\;\;\;N} & = & \frac{1}{3} \,F_{MPQR} F^{NPQR}
-\frac{1}{36} \de^{N}_{M} \,F_{PQRS} F^{PQRS},
\nonu \\
\nabla_M F^{MNPQ} & = & -\frac{1}{576} \,E \,\ep^{NPQRSTUVWXY}
F_{RSTU} F_{VWXY},
\label{fieldequations}
\eea
where the covariant derivative is given by 
$ \nabla_M = \pa_M +
E^{-1} \, (\pa_M \, E) $ with the elfbein determinant $E = \sqrt{-
g_{11}}$. The 11-dimensional epsilon tensor with lower indices is 
purely numerical in our convention.

In order to  construct the various 
11-dimensional M-theory lifts of a supersymmetric RG flow (\ref{flow}),
we impose the nontrivial $AdS_4$ radial coordinate dependence of 
$AdS_4$ supergravity fields 
subject to the 4-dimensional RG flow equations. Then the
geometric parameters in the 7-dimensional metric at certain critical
point are controlled by the RG flow equations so that they can be
extrapolated from the critical points. After that,  an appropriate ansatz 
for the 4-form field strengths is made. 
Finally, the 11-dimensional Einstein-Maxwell bosonic equations \cite{CJS,DNP} 
can be checked in order to complete the M-theory
uplift.

At ${\cal N}=8$ $SO(8)$ critical point where $(\rho, \chi)=(1, 0)$,
the 7-dimensional internal space is round 7-sphere ${\bf S}^7$.
Either one can introduce 
the global coordinates as ${\bf S}^7$ appropriate for the base
round 6-sphere ${\bf S}^6$ or those as the Hopf fibration on ${\bf
  CP}^3$ where there are two $U(1)$ symmetries.
Each global coordinates share the common ${\bf CP}^2$ space. 
For the former compactification, we would like to find out the new
11-dimensional solutions by taking the other various Einstein-Kahler 2-folds
including the above ${\bf CP}^2$ space, 
inside of 6-dimensional manifold. For the latter compactification, 
the new 
11-dimensional solutions were found in
\cite{CPW,AW0908,Ahn0909,Ahn0910} where the 3-forms are fixed
by the unbroken $U(1)$ symmetry which is nothing but the $U(1)_R$ symmetry.

In this paper, 
we find out a new exact solution of ${\cal N}=1$ $SU(3)$-invariant flow 
(connecting from the ${\cal N}=8$ $SO(8)$ UV fixed point to 
${\cal N}=2$ $SU(3) \times U(1)_R$ IR fixed point) 
to the 11-dimensional Einstein-Maxwell
equations.  
This can be described in the 3-dimensional ${\cal N}=1$ 
Chern-Simons matter theory by introducing two mass terms for two
adjoint ${\cal N}=1$ superfields.
If the two masses  are equal to each other, then   
the previously known 
${\cal N}=2$ $SU(3) \times U(1)_R$-invariant flow \cite{Ahn0806n2} arises.
If both masses are nonzero but not necessarily 
equal, then the  
${\cal N}=1$ $SU(3)$-invariant flow occurs. 
We claim that we have found the 11-dimensional uplift of
the generic ${\cal N}=1$ flow with unequal mass
parameters found by \cite{BHPW}.

There exist various 4-dimensional Einstein-Kahler 2-folds which live
in 5-dimensional Sasaki-Einstein space. 
By replacing the above ${\bf CP}^2$ space with 
${\bf CP}^1 \times {\bf CP}^1$ space, Einstein-Kahler 2-fold, 
and other Einstein-Kahler 2-fold, respectively, 
we find out new 11-dimensional solutions 
of ${\cal N}=1$ $SU(2) \times SU(2)$-, $SU(2) \times U(1)$- and $U(1)
\times U(1)$-invariant RG flows connecting above two fixed points. 
The corresponding $U(1)$ bundles are replaced also.
We will present the middle one which can be generalized to the last
and leads to the first for particular limit. One can take the last one
because this includes the first two cases but there is no nonabelian
symmetry group. 

In section 2,  by changing only three coordinates among seven
internal coordinates characterized by previous 
parametrization \cite{CPW}, one keeps only $SU(3)$ symmetry inside of $SU(3)
\times U(1)_R$ symmetry.
Then, the 11-dimensional metric can be written in terms of these new
coordinates and it contains the Fubini-Study metric on ${\bf CP}^2$ space
\cite{AI02}.
The Ricci tensor can be expressed as a linear combination of the Ricci
tensor for $SU(3) \times U(1)_R$-invariant flow and 
similarly the 4-forms also are given by 
a linear combination of 4-forms for $SU(3) \times U(1)_R$-invariant flow.  
Then, we find out a new solution for 
the 11-dimensional Einstein-Maxwell equations  corresponding to the
11-dimensional lift of the ${\cal N}=1$ $SU(3)$-invariant RG flow
connecting from the ${\cal N}=8$ $SO(8)$ UV fixed point to 
${\cal N}=2$ $SU(3) \times U(1)_R$ IR fixed point.
In the 11-dimensional point
of view, both the metric and 4-forms preserve only the $SU(3)$
symmetry inside of $SU(3) \times U(1)_R$ symmetry. 
The possible
deformation in the gauge dual, ${\cal N}=1$ Chern-Simons matter theory, 
is discussed.

In section 3, 
by considering the 
the Einstein-Kahler 2-fold and $U(1)$ bundle 
living in the 5-dimensional Sasaki-Einstein
manifold $Y^{p,q}$, one keeps only $SU(2) \times U(1)$ 
symmetry inside of $SU(2) \times U(1)
\times U(1)_R$ symmetry.
We find out a new solution for the
11-dimensional lift of the ${\cal N}=1$ $SU(2) \times U(1)$-invariant RG flow.

In section 4, we summarize the results of this paper and present some
future directions.

In the Appendices, we present the detailed expressions for the 
Ricci tensor, 4-form field strengths, and
Maxwell equation.

The new 11-dimensional solutions 
of ${\cal N}=1$ $SU(2) \times SU(2)$-, and $U(1)
\times U(1)$-invariant RG flows 
connecting above two fixed points in 4-dimensions can be done similarly. 

\section{An ${\cal N}=1$ $SU(3)$-invariant 
supersymmetric flow}

$\bullet$ The 11-dimensional metric along the flow

Let us introduce the three orthogonal ${\bf R}^8$ vectors \cite{AI02}
\bea
U & = & (u^1, u^2, u^3, u^4, u^5, u^6, 0, 0) \, \sin \theta \, \sin \th_6, \nonu \\
V_1  & = & (0, 0, 0, 0, 0, 0, 1, 0) \, \cos \theta, \nonu \\
V_2  & = & (0, 0, 0, 0, 0, 0, 0, 1) \, \sin \theta \, \cos \th_6.
\label{UV}
\eea
The sum of these vectors 
is restricted on the round seven-sphere ${\bf S}^7$. 
That is, 
$\sum_{A=1}^{8} (X^A)^2 =1$ with $X= U + V_1 +V_2$. 
Note that $\sum_{A=1}^{6} (X^A)^2 +(X^8)^2 = \sin^2 \theta$ and the
ellipsoidal deformation arises along the remaining $V_1$($X^7$)-direction.
By looking at the inside of ${\bf S}^7$, 
the unit round five-sphere ${\bf S}^5$ with ${\bf CP}^2$-base 
can be  described by the following
six variables with the constraint $\sum_{i=1}^{6} (u^i)^2=1$,
or five angular variables $\th_i$($i =1, \cdots, 5$), 
\bea
u^1 + i \, u^2 & = & \sin \theta_1 \, \cos (\frac{\theta_2}{2})  
\, e^{\frac{i}{2}
(\theta_3 + \theta_4)}\, e^{i \th_5} \equiv z^1,
\nonu \\
u^3 + i \, u^4 & = & \sin \theta_1 \, \sin (\frac{\theta_2}{2})  
\, e^{\frac{i}{2}
(-\theta_3 + \theta_4)} \, e^{i \th_5} \equiv z^2,
\nonu \\
u^5 + i \, u^6 & = & \cos \theta_1
\,  e^{i \th_5} \equiv z^3.
\label{sixu}
\eea
The isometry of five-sphere ${\bf S}^5$ is given by $SU(3) \times U(1)$ where 
$SU(3)$ acts on three complex coordinates $z^i$ and  the $U(1)$ acts on 
each $z^i$ as the phase rotations. 
The 5-dimensional metric $(du)^2$ from (\ref{sixu}) can be rewritten 
as $(d u)^2 = ds_{FS(2)}^2 + (u, J \, d u)^2$ where $ds_{FS(2)}^2$ denotes 
the Fubini-Study metric on ${\bf CP}^2$-base of ${\bf S}^5$
characterized by four angular variables $\th_i$($i=1, \cdots, 4$)
\bea
 d s_{FS(2)}^2 = d \theta_1^2 + \frac{1}{4} \, \sin^2 \theta_1
 \left(\sigma_1^2 + \sigma_2^2 + \cos^2 \theta_1 \, \sigma_3^2 \right),
\label{FS}
\eea
and $(u, J \, d u) \equiv u^i \, J_{ij} \, u^j$ is the Hopf
fiber on it and is given by 
\bea
(u, J \, d u) = d \theta_5 + \frac{1}{2} \, \sin^2 \theta_1 \, \sigma_3. 
\label{fiber}
\eea 
The $J$ is the standard Kahler form: $J_{12} = J_{34} = J_{56} = J_{78}
= 1$.
The one-forms appearing in (\ref{FS}) are given by
$
\si_1  =  \cos \th_4 \, d \th_2 + \sin \th_2 \sin \th_4 \, d \th_3$, 
$\si_2   =  \sin \th_4 \, d \th_2 - \sin \th_2 \cos \th_4 \, d \th_3$, 
and $\si_3   =  d \th_4 + \cos \th_2  \, d \th_3$.
By extending to the 7-dimensional metric with the differentials 
$d \th$ and $d \th_6$, one sees 
the standard metric for round seven-sphere ${\bf S}^7$:
\bea
(d X)^2 = (dX,  dX) = d
  \theta^2 +
\sin^2 \theta  \, d \Omega_6^2, \qquad
d \Omega_6^2 =  d \theta_6^2 + \sin^2 \theta_6 \left[ d
    s_{FS(2)}^2 + (u, J \, d u)^2 \right]. 
\label{dxdx}
\eea
Inside of ${\bf S}^7$, the 6-dimensional metric is nothing but 
the metric for the unit round six-sphere ${\bf S}^6$.
So far, the background geometry provides the ${\cal N}=8$ maximally symmetric 
$AdS_4 \times {\bf S}^7$ solution \cite{DNP} characterized by
(\ref{11dmetric}) for $\rho=1$ and $\chi=0$
corresponding to $SO(8)$-invariant UV fixed point.
Then the Ricci tensor in (\ref{fieldequations}) has the form of
\bea
R_{M}^{\,N} =\frac{6}{L^2} \, \mbox{diag} \,
(-2, -2, -2, -2, 1, 1, 1, 1, 1, 1, 1),
\label{ricci}
\eea
where $L$ is the radius of ${\bf S}^7$, twice the $AdS_4$ radius and
the only nonzero 4-form field strength satisfying (\ref{fieldequations})
with (\ref{ricci}) 
is given by $F_{1234} = -\frac{18}{L}$, so-called Freund-Rubin
parametrization \cite{FR}. 

Now we turn on the $AdS_4$ supergravity scalar fields 
$(\rho, \chi)$ starting from the above $SO(8)$-invariant UV fixed
point 
$(\rho, \chi)=(1, 0)$. They develop a nontrivial profile as a function
of $r$ (\ref{flow}) becoming more significantly different from $(\rho,\chi)=(1,0)$
as one goes to the $SU(3) \times U(1)_R$ IR fixed point. 
Let us consider the deformation from the $\rho$-supergravity field first.
The deformation matrix $Q$ is given by \cite{CPW} 
\bea
Q = \mbox{diag}(\rho^{-2}, \rho^{-2}, \rho^{-2}, \rho^{-2}, \rho^{-2}, 
\rho^{-2}, \rho^6, \rho^6).
\label{Q}
\eea
The quadratic form $\xi^2 = (X, Q \, X) \equiv X^A Q_{AB} X^B$ from (\ref{UV})
and (\ref{Q})
can be calculated to be
\bea
\xi^2 = \rho^6 +(\rho^{-2}-\rho^6) \, \sin^2 \theta \,
  \sin^2 \th_6.
\label{xi}
\eea
One sees that the metric on the deformed ${\bf R}^8$ by ellipsoidal
squashing via $\rho$-field can be recombined  by
\bea
&& (dX, Q^{-1} \, dX)  =   \rho^2 \, (d U)^2 + \frac{1}{\rho^6}
  \left[ (d V_1)^2  +  (d V_2)^2 \right] 
\nonu \\
&& =  \rho^2 \, \sin^2 \theta \, \sin^2 \th_6 \,
\left[ ds_{FS(2)}^2 + (u, J \, d u)^2 \right] 
\label{dxQdx} 
\\
&& +  \rho^{-4} \, \xi^2 \, \left[\frac{\left(-
\cos \theta \, \sin \th_6 \, d \theta - \sin \theta \, \cos \th_6 \, d
\th_6 \right)^2}{ (1-\sin^2 \theta \,
    \sin^2 \th_6)} \right] 
 +
\rho^{-6} \, \left[
 \frac{ \left( \cos \th_6 \, d \theta
 -      \sin \theta \, \cos \theta \, 
\sin \th_6 \, d \th_6 \right)^2  }
{(1-\sin^2 \theta \,
    \sin^2 \th_6)} \right].
\nonu
\eea
The Fubini-Study metric on ${\bf CP}^2$ space and its Hopf fiber 
are given by (\ref{FS}) and (\ref{fiber}) respectively.
According to the (\ref{xi}), the $\rho^2$-terms from $(d U)^2$ in 
(\ref{dxQdx}) are
decomposed into two parts, the first two terms of (\ref{dxQdx}) and
some terms containing $\xi^2$, and moreover
the $\rho^{-6}$-terms from $(d V_1)^2$ and $(d V_2)^2$
are distributed into the remaining term in $\xi^2$ and the last term
of
(\ref{dxQdx}). 
At $\rho=1$(no deformation), this leads to the standard metric \cite{AI02}
for round
seven-sphere ${\bf S}^7$ charaterized by (\ref{dxdx}), 
by recollecting $d \theta^2$- and $d \th_6^2$-terms.
Although the full $SO(8)$ isometry is broken in (\ref{dxQdx}), 
the $SU(3) \times U(1)$
isometry is still preserved. 

The $U(1)$ Hopf fiber on 6-dimensional manifold can be written as
\bea
(X, J \, d X) & \equiv & X^A \, J_{AB} \, 
X^B = (U, J \,d U) + (V_1, J \, d V_2) + (
V_2, J \, d V_1) \nonu \\
& = & (u, J \, du) \sin^2 \theta \, \sin^2 \th_6 + 
 \left[\cos \th_6 \, d \theta -    \sin \theta \, \cos \theta \, 
\sin \th_6 \, d \th_6 \right],
\label{xjdx}
\eea 
where the square of last two terms is proportional to 
the last term of (\ref{dxQdx}). 
By recalling that the Fubini-Study metric on ${\bf CP}^3$ space can be
written in terms of the Fubini-Study metric on ${\bf CP}^2$ space, 
$ds_{FS(2)}^2$, that is given by (\ref{FS}) and other pieces, 
the metric on 6-dimensional manifold from 
(\ref{dxdx}) and 
(\ref{xjdx}) is rewritten as
\bea
&& (d X, d X) -(X, J\, d X)^2   =  \left[\frac{\left(-
\cos \theta \, \sin \th_6 \, d \theta - \sin \theta \, \cos \th_6 \, d
\th_6 \right)^2}{ (1-\sin^2 \theta \,
    \sin^2 \th_6)} \right]  
\label{6dmetric} 
\\
&& + \sin^2 \theta \sin^2 \th_6 
\left[ ds_{FS(2)}^2 + (1-\sin^2 \theta 
    \sin^2 \th_6)  \left( (u, J  d u) -  \frac{ \left( \cos \th_6  d \theta
 -      \sin \theta  \cos \theta  
\sin \th_6  d \th_6 \right)  }
{(1-\sin^2 \theta 
    \sin^2 \th_6)} \right)^2 \right].
\nonu
\eea
In the first term of the right hand side(RHS), 
we intentionally recollected $\rho^{-4} \, \xi^2$ terms in (\ref{dxQdx}) 
which will play the role of one of the frame basis
and then the other parts of (\ref{6dmetric}) (that will be other frame basis)
are automatically determined once we identify the Fubini-Study metric
on ${\bf CP}^2$ space. 

Let us consider the deformation from the $\chi$-supergravity field next.
Then the warped $SU(3)$-invariant 7-dimensional metric, by adding (\ref{dxQdx}) to 
the stretched fiber characterized by $\chi$-field, leads to \cite{AI02}
\bea
d s_7^2 = \sqrt{\Delta } \, L^2 \left[  (dX, Q^{-1} \, dX)  +
\frac{\sinh^2 \chi}{\xi^2} \,  (X, J \, d X)^2 \right],
\label{7d}
\eea
where the warp factor is given by \cite{CPW} 
\bea
\Delta = (\xi \, \cosh \chi)^{-\frac{4}{3}}.
\label{Delta}
\eea
Let us use the identity $\cosh^2 \chi-\sinh^2 \chi =1$ in the second
term and recollect the first term and the
second term with the coefficient $-\frac{1}{\xi^2}$ of (\ref{7d}).

By keeping the $\rho^{-4} \, \xi^2$-term in
(\ref{dxQdx})  and the $ds_{FS(2)}^2$ part 
as independent five-orthonormal frames and 
substituting (\ref{dxQdx}) and (\ref{xjdx}) into 
(\ref{7d}),
one obtains, with (\ref{fiber}), 
\bea
d s_7^2  & = & \sqrt{\Delta } \, L^2 \left( 
\frac{\xi^2}{\rho^4} \, \left[\frac{\left(-
\cos \theta \, \sin \th_6 \, d \theta - \sin \theta \, \cos \th_6 \, d
\th_6 \right)^2}{ (1-\sin^2 \theta \,
    \sin^2 \th_6)} \right] +  \rho^2 \, \sin^2 \theta \, \sin^2 \th_6 \,
ds_{FS(2)}^2 \right. 
\nonu \\
& + &  \left. \frac{\omega^2}{\xi^2} + \frac{\cosh^2 \chi}{\xi^2} \left[ 
 \cos \th_6 \, d \theta -    \sin \theta \, \cos \theta \, 
\sin \th_6 \, d \th_6 
+      \sin^2 \theta  \, 
 \sin^2 \th_6 (u, J \, d u )
\right]^2 \right),
\label{7d1}
\eea
where the remaining one-form is completely fixed as follows:
\bea
\omega   = 
\sin \theta  \sin \th_6 
 \sqrt{1-\sin^2 \theta 
 \sin^2 \th_6} \,
\left[ -\frac{1}{\rho^4} \, \frac{ \left( \cos \th_6  d \theta
 -      \sin \theta  \cos \theta  
\sin \th_6  d \th_6 \right)  }
{(1-\sin^2 \theta 
    \sin^2 \th_6)} 
 +     \rho^4  (u, J  d u ) \right].
\label{1-form}
\eea
Note that for no deformation ($\rho=1$), 
the square of this one-form (\ref{1-form}) becomes 
the last term in (\ref{6dmetric}).
Moreover, the last term in (\ref{7d}) without deformation ($\rho=1$
and $\chi=0$)  leads to the square of
(\ref{xjdx}) where $\frac{\cosh^2 \chi}{\xi^2}$ becomes one.

By combining the 4-dimensional metric and 7-dimensional metric
(\ref{7d1}) together with (\ref{11dmetric}),
one arrives at the following set of frames for the 11-dimensional
metric
with 11-dimensional coordinates $z^M= 
(x^1, x^2, x^3, r; \th, \th_1, \th_2,
\th_3, \th_4, \th_5, \th_6)$: 
\bea
e^1  & = & \Delta^{-\frac{1}{2}} \, e^A \, d x^1,
\qquad
e^2   =  \Delta^{-\frac{1}{2}} \, e^A \, d x^2,
\qquad
e^3   =  \Delta^{-\frac{1}{2}} \, e^A \, d x^3,
\qquad
e^4    =   \Delta^{-\frac{1}{2}}  \, d r,
\nonu \\
e^5 & = & {L} \,
\Delta^{\frac{1}{4}} \,  
\frac{\xi}{\rho^2} \,
\frac{\left[-
\cos \theta \, \sin \th_6 \, d \theta - \sin \theta \, \cos \th_6 \, d
\th_6 \right]}{ \sqrt{1-\sin^2 \theta \,
    \sin^2 \th_6}},
\nonu \\
 e^6 & = & {L} \,
\Delta^{\frac{1}{4}} \,  \rho \, \sin \theta \, \sin \th_6
\, d \theta_1,
\nonu \\
 e^7  &  = &  {L} \,
\Delta^{\frac{1}{4}} \,  \rho \,   \sin \theta \, 
\sin \theta_6 \, \frac{1}{2} \, \sin \theta_1 \, \sigma_1,
\nonu \\
 e^8 & = & {L} \,
\Delta^{\frac{1}{4}} \,  \rho \,  \sin \theta \, 
\sin \th_6 \, \frac{1}{2} \, \sin \theta_1 \, \sigma_2,
\nonu \\
 e^9  & = &  {L} \,
\Delta^{\frac{1}{4}} \, \rho \, \sin \theta \, 
\sin \th_6 \, \frac{1}{2} \, \sin \theta_1 \,\cos \theta_1 \, \sigma_3,
\nonu \\
 e^{10} & = & {L} \,
\Delta^{\frac{1}{4}} \,  \xi^{-1} \, \sin \theta \, \sin \th_6 \,
 \sqrt{1-\sin^2 \theta \,
    \sin^2 \th_6} \,
\left[ -\frac{1}{\rho^4} \, \frac{ \left( \cos \th_6 \, d \theta
 -      \sin \theta \, \cos \theta \, 
\sin \th_6 \, d \th_6 \right)  }
{(1-\sin^2 \theta \,
    \sin^2 \th_6)} \,  
 \right. \nonu \\
& + & \left.    \rho^4 \, 
   \left( d \th_5 +  \frac{1}{2}  \, \sin^2 \th_1  
 \,  \si_3\right) \right],
\nonu \\
e^{11}   & = &  {L} \,
\Delta^{\frac{1}{4}} \,  \xi^{-1}  \, \cosh \chi \,
\left[
 \cos \th_6 \, d \theta -    \sin \theta \, \cos \theta \, 
\sin \th_6 \, d \th_6 \right. \nonu \\
&+ &  \left.   \sin^2 \theta  \, 
 \sin^2 \th_6 \left( d \th_5 + \frac{1}{2}  \, \sin^2 \th_1  
 \,  \si_3  \right)
 \right],
\label{11d}
\eea
where the quadratic form $\xi^2$ is given by (\ref{xi}), 
the warp factor $\Delta$ is given by (\ref{Delta}) and 
the fiber on ${\bf CP}^2$ space (\ref{fiber}) is written in terms of 
the angular variables  $\th_i$($i=1, \cdots, 5$).
The $AdS_4$ supergravity fields $(\rho, \chi)$ and scale function $A$
have nontrivial $r$-dependence via (\ref{flow}). 
The $\xi^2$ and $\Delta$ depend on the radial coordinate  as well
as the internal coordinates $(\th, \th_6)$. The one forms $\si_i$ are
the same as previous expressions and are related to the angular
coordinates $(\th_2, \th_3,\th_4)$ as before. 
There are no $U(1)$ shifts in the coordinates $(\th, \th_6)$ except the $\th_5$.
The three coordinates \cite{CPW} where the base 6-dimensional 
manifold is ${\bf CP}^3$ space
are related to those in this paper or in \cite{AI02} as follows:
\bea
\mu  & = & \cos^{-1} \left( \sin \theta \, \sin \th_6 \right), \nonu \\
\phi  & = & \th_5- \cos^{-1} \left( 
\frac{\cos \theta}{\sqrt{1- \sin^2 \theta \, \sin^2
    \th_6 }} \right), 
\nonu \\
\psi  & = & \cos^{-1} \left(\frac{\cos \theta}{\sqrt{1- \sin^2 \theta \, \sin^2
    \th_6 }}\right). 
\label{mpp}
\eea
Through the transformation (\ref{mpp}),
one can easily see that the above 11-dimensional metric (\ref{11d})
becomes the one in \cite{CPW} exactly.
In order to use the Ricci tensor and 4-form field strengths for
$SU(3)$-invariant flow,
the partial differentiations between these coordinates are needed
and some of them are
\bea
\frac{\pa \mu}{\pa \theta} & = & -\frac{\cos \theta \, 
\sin \th_6}{\sqrt{1- \sin^2 \theta \, \sin^2
    \th_6}}, \qquad 
\frac{\pa \mu}{\pa \th_6} = 
 -\frac{\sin \theta \, 
\cos \th_6}{\sqrt{1- \sin^2 \theta \, \sin^2
    \th_6}},  
\nonu \\
\frac{\pa \phi}{\pa \theta} & = & -\frac{ 
\cos \th_6}{(1- \sin^2 \theta \, \sin^2
    \th_6)}, \qquad 
\frac{\pa \phi}{\pa \th_6} = 
 \frac{\sin \theta \, \cos \theta \, 
\sin \th_6}{(1- \sin^2 \theta \, \sin^2
    \th_6)}, \nonu \\
\frac{\pa \psi}{\pa \theta} & = & \frac{ 
\cos \th_6}{(1- \sin^2 \theta \, \sin^2
    \th_6)}, \qquad 
\frac{\pa \psi}{\pa \th_6} = -
 \frac{\sin \theta \, \cos \theta \, 
\sin \th_6}{(1- \sin^2 \theta \, \sin^2
    \th_6)}.
\label{diff}
\eea
Moreover, the partial differentiations of new variables $(\th, \th_5,
\th_6)$ 
with respect
to old variables $(\mu, \phi, \psi)$ from (\ref{mpp}) can be obtained also.  

$\bullet$ The Ricci tensor and the 4-form field strengths along the flow

Since the metric (\ref{11d}) is related to the metric given in \cite{CPW} via
the change of variables (\ref{mpp}), one can use the $SU(3) \times
U(1)_R$-invariant solution  summarized in the Appendix A and find
out the new solution which is invariant under the $SU(3)$ symmetry. 
The Ricci tensor can be obtained from (\ref{11d})
directly
or can be determined from the one preserving $SU(3) \times U(1)_R$ by using the
transformation on the coordinates between the two coordinate
systems (\ref{mpp}).
That is, the $SU(3)$-invariant Ricci tensor is given by 
\bea
\widetilde{R}_{M}^{\,\,N} =  \left(\frac{\pa z^P}{ \pa \widetilde{z}^{ M}} \right)
\left(\frac{\pa \widetilde{z}^{ N}}{\pa z^Q}
\right) R_{P}^{\,\,Q},
\label{rtilde}
\eea
where 
the 11-dimensional coordinates are given by
\bea
\widetilde{z}^M   =  (x^1, x^2, x^3, r; \theta, \theta_1, \theta_2, 
\theta_3, \theta_4, \theta_5, \theta_6), \qquad
z^M   =  (x^1, x^2, x^3, r; \mu, \theta_1, \theta_2, 
\theta_3, \theta_4, \phi, \psi).
\label{two11}
\eea 
Only three of them are distinct and
the new variables $(\th, \th_5,
\th_6)$ correspond to old variables $(\mu, \phi, \psi)$.
The Ricci tensor $R_{P}^{\,\,Q}$ for $SU(3)\times U(1)_R$-invariant flow
is presented in the Appendix $A$ (\ref{Ricci1}) explicitly where the
flow equations (\ref{flow}) are imposed.
The transformation rule between the two coordinate systems 
can be obtained from (\ref{mpp}).
The Ricci tensor $\widetilde{R}_{M}^{\,\,N}$ for $SU(3)$-invariant flow is 
given in the Appendix $B$ (\ref{Ricci2}) explicitly
and there exist off-diagonal components
$(4,5)$, $(4,11)$, $(5,4)$, $(5,10)$, $(5,11)$, 
$(8,5)$, $(8,10)$, $(8,11)$, $(9,5)$, $(9,10)$, $(9,11)$, 
$(10,5)$, $(10,11)$ $(11,4)$, $(11,5)$ and $(11,10)$.
At the IR critical point, the components of Ricci tensor $(4,5)$, $(4,11)$, $(5,4)$
and $(11,4)$ vanish.

For the 4-form field strengths, one has
\bea
\widetilde{F}_{MNPQ} = \left(\frac{\pa z^R}{ \pa \widetilde{z}^{M}} \right)
\left(\frac{\pa z^S}{\pa \widetilde{z}^{ N}}
\right) \left( \frac{\pa
  z^T}{\pa \widetilde{z}^{ P}}\right) \left(\frac{\pa z^U}{\pa
  \widetilde{z}^{ Q}}
\right) F_{RSTU},
\label{ftilde}
\eea
with (\ref{two11}), (\ref{mpp}) and (\ref{diff}).
These transformed 4-forms  
are given in
the Appendix $B$ (\ref{4-form3})  in terms of those (\ref{4-form1}) in
$SU(3) 
\times U(1)_R$-invariant flow and 
the transformed 4-forms with upper indices  are
described via (\ref{4-form4}) in terms of (\ref{4-form2}).
The 4-forms with upper indices  $\widetilde{F}^{MNPQ}$
can be obtained from $\widetilde{F}_{RSTU}$ (\ref{4-form3}) by using the 11-dimensional
inverse metric (\ref{11d}) 
or by multiplying the transformation matrices obtained from (\ref{mpp})
into the 4-forms with upper indices $F^{MNPQ}$ (\ref{4-form2}) of 
$SU(3) \times U(1)_R$-invariant flow as done in (\ref{ftilde}).
The 4-form $\widetilde{F}_{123\,11}$ is a new object for $SU(3)$-invariant flow.
At the IR critical point in 4-dimensions, 
the following 4-forms also vanish:
$
\widetilde{F}_{1235} = \widetilde{F}_{123\, 11} =
\widetilde{F}_{45mn} =\widetilde{F}_{4mnp}=0$.
For the $SU(3) \times U(1)_R$-invariant flow, 
the 4-forms $F_{1235}, F_{4mnp}$ and
$F_{45mn}$($m, n, p=6, \cdots, 11$)  become zero
at the IR critical point in 4-dimensions.
Once we suppose that 4-dimensional metric has the domain wall factor
(\ref{11dmetric}) which breaks the 4-dimensional conformal invariance, 
the mixed 4-forms occur along the whole RG flow.

Note that the internal 3-form corresponding to 
above 4-forms (\ref{ftilde})(the internal part of
$\widetilde{F}^{(4)}$ is given by 
$d \widetilde{C}^{(3)} + d \widetilde{C}^{(3)\ast} $) 
has the following expression 
\bea
\widetilde{C}^{(3)} =
\frac{L^3 \, \tanh \, \chi}{ 4[\rho^8 +(1-\rho^8) \, (1-|W|^2)] }
\left( 3 Z^{[1} \, d Z^2 \wedge d Z^{3]} \wedge
d W - \rho^8 \, W \, d Z^1 \wedge d Z^2 \wedge d Z^3 \right),
\label{c3}
\eea
where the rectangular coordinates with (\ref{UV}) and (\ref{sixu}) are given by
\bea
Z^i \equiv z^i \, \sin \th \, \sin \th_6 (i=1, 2, 3), 
\qquad W \equiv X^7 + i X^8 = \cos \th
+ i \sin \th \, \cos \th_6. 
\nonu
\eea
Then it is easy to see that the 3-form (\ref{c3}) has explicit $SU(3)$
symmetry 
because the ${\bf 3}$ of $SU(3)$ in $Z^i$ occurs as 
$\epsilon_{ijk} Z^i \wedge d Z^j \wedge d
Z^k$ or  $\epsilon_{ijk} d Z^i \wedge
d Z^j \wedge d Z^{k}$ 
with  the $SU(3)$-invariant epsilon tensor $\epsilon_{ijk}$.
The $X^7$ and $X^8$ are $SU(3)$ singlets.
Since the 3-form along the $(123)$-directions has $(\th, \th_6)$
dependence as well as $r$-dependence,
it is $SU(3)$-singlet. 

$\bullet$ Checking the Einstein equation

One checks the Einstein equation using the solution for
$SU(3) \times U(1)_R$-invariant flow. 
In \cite{CPW}, it was  shown that 
the Ricci tensor (\ref{Ricci1}), the 4-forms with lower indices 
(\ref{4-form1}) and the 4-forms with upper indices (\ref{4-form2})
satisfy the field equation (\ref{fieldequations}).  
Now one can
use the property of $SU(3) \times U(1)_R$-invariant flow. 
One replaces the Ricci tensor $ R_{M}^{\,\,N}$ in
terms of the quadratic 4-forms: $F^2$ or $F_{MPQR} F^{NPQR}$. 
After that one sees the transformed-Ricci tensor $ \widetilde{R}_{M}^{\,\,N}$ 
can be written in terms of the quadratic 4-forms 
for $SU(3) \times U(1)_R$-invariant flow from (\ref{rtilde}). 
Let us return to the right hand side of Einstein equation.      
Using (\ref{ftilde}) and $\widetilde{F}^{MNPQ}$, one can express the RHS 
in terms of quadratic 4-forms $F^2$ or $F_{MPQR} F^{NPQR}$
for $SU(3) \times U(1)_R$-invariant flow.

One can make the difference between the left hand side(LHS) and the 
RHS of Einstein equation and see whether this becomes zero or not. 
At first sight, some of the
components written in terms of quadratic 4-forms in $SU(3) \times U(1)_R$-invariant flow 
are not exactly vanishing. 
They contain the terms 
\bea
&& F_{MPQR} \, F^{NPQR}, \qquad M= 4, 5, \,\, N=10, 11, 
\qquad
F_{6PQR} \, F^{NPQR}, \qquad N= 4, 5, 9, 10, 11,
\nonu \\
&& 
F_{7PQR} \, F^{NPQR},  \qquad N = 4,5,6,8,9,10,11, \qquad
F_{8PQR} \, F^{NPQR}, \qquad N=4,5,6,7,9, \nonu \\
&& F_{9PQR} \, F^{NPQR}, \qquad N=4,5,6, \qquad
F_{MPQR} \, F^{NPQR}, \qquad M= 10, 11, \,\, N= 4, 5. 
\label{term}
\eea
However, 
after plugging the explicit solution $F_{MNPQ}$ (\ref{4-form1}) and 
$F^{MNPQ}$ (\ref{4-form2}) for 
$SU(3) \times U(1)_R$-invariant flow into (\ref{term}), 
then all of these (\ref{term}) 
are identically vanishing.  
Recall that these quadratic 4-forms above 
correspond to the off-diagonal terms of Einstein
equation for $SU(3) \times U(1)_R$-invariant 
flow which vanish identically except
$(4,5)$-, $(5,4)$-, $(8,10)$-, $(8,11)$-, $(9,10)$-, $(9,11)$-,
$(10,11)$-,
and $(11,10)$-components from the $ R_{M}^{\,\,N}$ (\ref{Ricci1}). Of course, 
the above extra piece (\ref{term}) does not possess
these nonzero off-diagonal terms. 
Therefore, we have shown the solution given by 
 the Ricci tensor 
$ \widetilde{R}_{M}^{\,\,N}$ (\ref{rtilde}), the 4-forms $\widetilde{F}_{MPQR}$ 
(\ref{ftilde}) and   the 4-forms $\widetilde{F}^{NPQR}$ 
indeed satisfies the 11-dimensional Einstein equation (\ref{fieldequations}). 

$\bullet$ Checking the Maxwell equation

Let us introduce the notation 
$
\frac{1}{2} \, \widetilde{E} \,  
\widetilde{\nabla}_M  \, \widetilde{F}^{MNPQ} \equiv
(\widetilde{N}\widetilde{P}\widetilde{Q})$ for simplicity,
and present all the nonzero components for the LHS of Maxwell
equations in terms of the 4-forms in $SU(3) \times U(1)_R$-invariant flow, using the
property of 11-dimensional solution.
According to the transformation rules, one can
express the transformed covariant derivative  
and transformed 4-forms in terms of those
for $SU(3) \times U(1)_R$-invariant flow together with 
$(\th,\th_6)$-dependent coefficient functions. In other words,
by using the Maxwell equation for $SU(3) \times U(1)_R$-invariant flow, 
one can replace the LHS of Maxwell equation with the quadratic 4-forms
$F_{RSTU} F_{VWXY}$.
The nonzero components are given in the Appendix (\ref{max}) explicitly.
The nonzero components of the Maxwell equation 
are characterized by the following indices 
$
(\widetilde{1}\widetilde{2}\widetilde{3})$, 
$(\widetilde{4}\widetilde{n}\widetilde{p})$, 
$(\widetilde{5}\widetilde{n}\widetilde{p})$, and 
$(\widetilde{m}\widetilde{n}\widetilde{p})$, 
with the number of components $1$, 
$8$ by choosing two out of six, 
$8$ by choosing two out of six and 
$13$ by choosing three out of six respectively. 
Other remaining components of the Maxwell equation become identically zero. 
Therefore, there exist 30-nonzero-components of the Maxwell equation. 
The  
$(\widetilde{1}\widetilde{2}\widetilde{3})$-component above
consists of 12-terms coming from the quadratic 4-forms $F_{RSTU} F_{VWXY}$.
For the other components, the right hand side contains
a single-term, three-terms or four-terms  in quadratic 4-forms.
 
Similarly, 
let us return to the RHS of Maxwell equation.      
Using (\ref{ftilde}) and 11-dimensional metric (\ref{11d}), one can express this 
in terms of quadratic 4-forms 
$F_{RSTU} F_{VWXY}$ for $SU(3) \times U(1)_R$-invariant flow.
We also transform the 11-dimensional determinant according to the
transformation rules appropriately. 
It turns out that
the difference between the LHS obtained from previous paragraph and the RHS 
in this paragraph of Maxwell equation becomes zero. Therefore, we have
shown that the solution (\ref{ftilde}) for given 11-dimensional metric
(\ref{11d}) indeed satisfies the Maxwell
equation (\ref{fieldequations}).

Now we have shown that the solution (\ref{ftilde})
together with the Appendices $A$ and $B$
consists of an exact solution to 11-dimensional supergravity by
bosonic field equations (\ref{fieldequations}) as long as the
deformation parameters $(\rho ,\chi)$ of the 7-dimensional internal
space and the domain wall amplitude $A$ develop in the $AdS_4$
radial direction along the ${\cal N}=1$ $SU(3)$-invariant RG flow (\ref{flow})
connecting from ${\cal N}=8$ $SO(8)$ UV fixed point to ${\cal N}=2$
$SU(3) \times U(1)_R$ IR fixed point in 4-dimensions.
Compared with the previous solution for $SU(3) \times
U(1)_R$-invariant 
flow \cite{CPW}, 
they share the common ${\bf CP}^2$
space inside 7-dimensional internal space but three remaining
coordinates
are different from each other. This provides the symmetry
breaking of $SU(3) \times U(1)_R$ into 
its subgroup $SU(3)$ along the whole RG flow in 11-dimensions.

$\bullet$ The mass deformation in dual gauge theory

The mass deformation of BL theory \cite{BL} has the fermion mass term in
the Lagrangian \cite{mass}.
For ${\cal N}=1$ supersymmetry, the bosonic mass terms consist of two
independent terms when the 4-form $F_3^{+}$ vanishes 
in 4-dimensional gauged supergravity \cite{Ahn0905}.
One writes down the mass-deformed superpotential in ${\cal N}=1$
superfield notation    
\bea
\Delta W = \frac{1}{2} \, m_7  \Tr \Phi_7^2 + \frac{1}{2} \, m_8 \Tr
\Phi_8^2.  
\label{desuper}
\eea 
The original ${\cal N}=1$ superpotential $W$
has quartic terms in $\Phi_I$ and comes from the D-term and F-term
of the ${\cal N}=2$ action \cite{BKKS,BHPW} and the superpotential $W$ has 
terms not having $(\Phi_7, \Phi_8)$, terms in linear in $\Phi_7$,
terms in linear in $\Phi_8$ and terms that depend on $\Phi_7$ and
$\Phi_8$. 
When we integrate out $(\Phi_7,\Phi_8)$ in the deformed $W+\Delta W$
with (\ref{desuper})
at low energy
scale, 
we obtain
the quartic terms coming from the original $W$ which do not contain
mass parameters and two kinds of sextic terms with two independent
parameters which depend on $(m_7, m_8)$
by solving the equations of motion for $(\Phi_7, \Phi_8)$ 
in $W+\Delta W$. This should flow to 
the superconformal field theory in the IR and
the gravity dual shows that 
it will flow to the ${\cal N}=2$ 
$SU(3) \times U(1)_R$-invariant fixed point.
In the IR, when the two parameters are equal to each other in the resulting
superpotential $\widehat{W}$, 
it will flow to the ${\cal N}=2$ $SU(3) \times U(1)_R$-invariant fixed point.
At the IR critical point, the supersymmetry 
should be enhanced from ${\cal N}=1$ to ${\cal N}=2$ and there exists 
$U(1)_R$ symmetry. 

Thus, we have found ${\cal N}=1$ superconformal Chern-Simons matter
theory with global $SU(3)$ symmetry and expect that $SU(3)$-invariant 
$U(N) \times U(N)$ Chern-Simons matter theory for $N > 2$ with $k=1,
2$ is dual to the background of this paper with $N$ unit of flux.
Namely, we have described the 11-dimensional uplift of 
the generic ${\cal N}=1$ flow with unequal mass parameters found by 
\cite{BHPW} \footnote{Since the 4-dimensional 
supersymmetric flow in \cite{BHPW,AW} involves four supergravity fields
rather than two we considered in this paper, one might ask what is the
exact meaning of 11-dimensional uplift of \cite{BHPW}? In
the context of 
4-dimensional ${\cal N}=8$ gauged supergravity, the $SU(3)$-invariant
sector can be realized by four real supergravity fields. The
11-dimensional metric with common $SU(3)$-invariance 
is constructed in \cite{AI02} where the 7-dimensional internal metric
is more complicated than the one in (\ref{7d}) due to the presence of
four fields denoted by $(a,b,c,d)$ rather than two by $(a,c)$. The
definition for $(a,b,c,d)$ \cite{AI02} is given in terms of the
fields in \cite{Warner83}. Along the 
constraints $b=\frac{1}{a} \equiv \rho^{-4}$ and 
$d=c \equiv \cosh \chi$, the supersymmetric flow
characterized by four fields becomes the one in
(\ref{flow}). Therefore, we have considered the 4-dimensional
superpotential
on a restricted 2-dimensional slice (rather than 4) 
of the scalar manifold and the differential equations for the fields
are the gradient flow equations of this superpotential. 

What happens
for 11-dimensional point of view? Since the 11-dimensional metric is
given in terms of four fields, the Ricci tensor can be obtained 
straightforwardly. This becomes the Ricci tensor 
(\ref{Ricci1}) along the above constraints. What about 4-forms? One
expects that the internal 4-forms looks like as (\ref{c3}) but the
dependence on the four fields arises in various way. 
For the 3-form with membrane indices, one should generalize the
4-dimensional superpotential given in \cite{AW} to the geometric
superpotential where the dependence on the internal coordinates occurs.   
Along the
constraints, this 4-forms should be the same as the one in (\ref{c3}).
Our work in this paper is the first task to complete the
11-dimensional uplift of the supersymmetric flow invloving the four
supergravity fields which is still an open problem.  

In this sense, we have found the 11-dimensional solution for the 
4-dimensional supersymmetric flow presented in \cite{BHPW,AW}
along the constraints and 
the corresponding $SU(3)$-invariant 
RG flow to the IR point of 3-dimensional dual gauge theory is the
curve ($m_1 \neq m_2$) 
connecting the $SO(8)$ point to the $SU(3) \times U(1)_R$ point
in Figure 1 of \cite{BHPW}. 
Recall that the $SU(3) \times U(1)_R$-invariant flow  is 
the straight line $m_1=m_2$ (with the horizontal axis $m_1$
and the vertical axis $m_2$) and its 11-dimensional uplift is found
in \cite{CPW}. The coordinates $(m_1,m_2)$ of \cite{BHPW} correspond 
to the previous $(m_7,m_8)$ in (\ref{desuper}).
Of course, at the IR(or UV) critical point, our
solution is an exact 11-dimensional uplift of \cite{BHPW} because the
values of four supergravity fields at the IR(or UV) critical point are located at 
the above constraints.       }.

\section{An ${\cal N}=1$ $SU(2) \times U(1)$-invariant 
supersymmetric flow}

Let us consider the 5-dimensional Sasaki-Einstein space $Y^{p,q}$
used in \cite{GMSW} and it consists of
the Einstein-Kahler 2-fold and the $U(1)$ bundle. Let us replace 
the ${\bf CP}^2$ metric (\ref{FS}) and the fiber (\ref{fiber}) with 
the Einstein-Kahler 2-fold and $U(1)$ bundle of 
$Y^{p,q}$ space respectively. Then 
the following set of frames for the 11-dimensional
metric can be written as 
\bea
e^1  & = & \Delta^{-\frac{1}{2}} \, e^A \, d x^1,
\qquad
e^2   =  \Delta^{-\frac{1}{2}} \, e^A \, d x^2,
\qquad
e^3   =  \Delta^{-\frac{1}{2}} \, e^A \, d x^3,
\qquad
e^4    =   \Delta^{-\frac{1}{2}}  \, d r,
\nonu \\
e^5   & = &  L \,
\Delta^{\frac{1}{4}} \, 
\frac{\xi}{\rho^2} \,
\frac{\left[-
\cos \theta \, \sin \th_6 \, d \theta - \sin \theta \, \cos \th_6 \, d
\th_6 \right]}{ \sqrt{1-\sin^2 \theta \,
    \sin^2 \th_6}},
\nonu \\
 e^6  &  = &  L \,
\Delta^{\frac{1}{4}} \, 
 \rho \, \sin \theta \, \sin \th_6
\, \sqrt{\frac{1-y}{6}} \, d \theta_1,
\nonu \\
 e^7 & = & L \,
\Delta^{\frac{1}{4}} \,
 \rho \, \sin \theta \, \sin \th_6
\, \sqrt{\frac{1-y}{6}} \, \sin \theta_1
\, d \phi_1,
\nonu \\
 e^8  & = &  L \,
\Delta^{\frac{1}{4}} \, 
 \rho \, \sin \theta \, \sin \th_6
\, \frac{1}{\sqrt{w\,q}} \, d y,
\nonu \\
 e^9 & = & L \,
\Delta^{\frac{1}{4}} \, 
 \rho \, \sin \theta \, \sin \th_6
\, \frac{1}{6} \, \sqrt{w\,q} \, 
(d\, \beta + \cos \theta_1 \, d \phi_1),
\nonu
\\
e^{10} & = & {L} \,
\Delta^{\frac{1}{4}} \,  \xi^{-1} \, \sin \theta \, \sin \th_6 \,
 \sqrt{1-\sin^2 \theta \,
    \sin^2 \th_6} \,
\left[ -\frac{1}{\rho^4} \, \frac{ \left( \cos \th_6 \, d \theta
 -      \sin \theta \, \cos \theta \, 
\sin \th_6 \, d \th_6 \right)  }
{(1-\sin^2 \theta \,
    \sin^2 \th_6)} \,  
 \right. \nonu \\
& + & \left.    \rho^4 \, 
\frac{1}{3} \,   \left[ (d \th_5 -\cos \theta_1 \, d \phi_1) +  y
    \, ( d \beta + 
\cos \theta_1 \, d \phi_1) \right]  \right],
\nonu \\ 
e^{11}   & = &  {L} \,
\Delta^{\frac{1}{4}} \,  \xi^{-1}  \, \cosh \chi \,
\left[
 \cos \th_6 \, d \theta -    \sin \theta \, \cos \theta \, 
\sin \th_6 \, d \th_6 \right. \nonu \\
&+ &  \left.   \sin^2 \theta  \, 
 \sin^2 \th_6 \, \frac{1}{3}\,
\left[ (d \th_5 -\cos \theta_1 \, d \phi_1) + y\, ( d \beta + 
\cos \theta_1 \, d \phi_1) \right] 
 \right]
\label{11d2}
\eea
where the $y$-dependent functions with some parameter $a$ are given by
\bea
w \equiv \frac{2(a-y^2)}{1-y}, \qquad
q \equiv \frac{a -3 y^2 + 2y^3}{a -y^2}, \qquad
a \equiv \frac{1}{2} -\frac{(p^2-3q^2)}{4p^3} \sqrt{4p^2-3q^2}.
\nonu
\eea
The quadratic form $\xi^2$ is given by (\ref{xi}), 
the warp factor $\Delta$ is given by (\ref{Delta}) and 
the fiber on the Einstein-Kahler 2-fold  is written in terms of 
the angular variables  $\th_5, \th_1, \phi_1, y$ and $\beta$. See the 
frame $e^{10}$ and the frame $e^{11}$.
There are $U(1)$ symmetries in $\beta$ and $\th_5$.
The $AdS_4$ supergravity fields $(\rho, \chi)$ and scale function $A$
have nontrivial $r$-dependence via (\ref{flow}) as before. 
The three coordinates \cite{Ahn0909}
are related to those in this paper: 
$(\mu, \alpha)$ correspond to $(\th, \th_6)$ of (\ref{mpp}) 
and $\psi =\th_5$. That is,
 $\cos \mu = \sin \th \, \sin \th_6$
and $\cos \alpha = 
\frac{\cos \th}{\sqrt{1-\sin^2 \th \, \sin^2 \th_6}}$.
Through the transformation,
one can easily see that the above 11-dimensional metric (\ref{11d2})
becomes the one in \cite{Ahn0909} exactly.
In order to obtain the Ricci tensor and 4-form field strengths for 
$SU(2) \times U(1)$-invariant flow, one needs
the partial differentiations between these coordinates.

$\bullet$ The Ricci tensor and the 4-form field strengths along the flow

The Ricci tensor can be obtained from (\ref{11d2})
directly
or can be determined from the one preserving $SU(2) \times U(1) \times U(1)_R$ by using the
transformation on the coordinates between the two coordinate
systems.
That is, the $SU(2) \times U(1) $-invariant Ricci tensor is given by 
$\widetilde{R}_{M}^{\,\,N}$ (\ref{rtilde})
where 
the 11-dimensional coordinates are given by
$
\widetilde{z}^M$ in (\ref{two11})
and $ z^M   =  (x^1, x^2, x^3, r; \mu,  \theta_1, \phi_1, 
y, \beta, \psi, \alpha)$.
Only two of them are distinct and
the Ricci tensor $R_{P}^{\,\,Q}$ for $SU(2) \times U(1) \times U(1)_R$-invariant flow
is presented in the Appendix $C$ (\ref{Ricci5}) explicitly.
The Ricci tensor (\ref{rtilde}) for $SU(2) \times U(1)$-invariant flow is 
given in the Appendix $D$ (\ref{Ricci6}) explicitly
and there exist off-diagonal components
$(4,5)$, $(4,11)$, $(5,4)$, $(5,10)$, $(5,11)$, 
$(7,5)$, $(7,10)$, $(7,11)$, $(9,5)$, $(9,10)$, $(9,11)$, 
$(10,5)$, $(10,11)$ $(11,4)$, $(11,5)$ and $(11,10)$.
At the IR critical point in 4-dimensions, 
the components $(4,5)$, $(4,11)$, $(5,4)$
and $(11,4)$ vanish.

For the 4-form field strengths, one can use
(\ref{ftilde}).
These transformed 4-forms  
are given, in terms of those (\ref{4-form9}) in $SU(2) \times U(1) 
\times U(1)_R$-invariant flow, in
the Appendix $D$ (\ref{4-form11}) and 
the transformed 4-forms with upper indices, in terms of (\ref{4-form10}), are
described by (\ref{4-form12}).
At the IR critical point in 4-dimensions, 
the following 4-forms also vanish:
$
\widetilde{F}_{1235} = \widetilde{F}_{123\, 11} =
\widetilde{F}_{45mn} =\widetilde{F}_{4mnp}=0$.
For the $SU(2) \times U(1) \times U(1)_R$-invariant flow, 
the 4-forms $F_{1235}, F_{4mnp}$ and
$F_{45mn}$($m, n, p=6, \cdots, 11$) become zero
at the IR critical point.

$\bullet$ Checking the Einstein equation

In \cite{Ahn0909}, 
it is  known that the solution by
the Ricci tensor (\ref{Ricci5}), the 4-forms with lower indices
(\ref{4-form9}) and the 4-forms with upper indices (\ref{4-form10}) 
satisfies the field equation (\ref{fieldequations}).  
The Ricci tensor can be written in
terms of the quadratic 4-forms. This implies that 
the transformed-Ricci tensor 
can be written in terms of the quadratic 4-forms for 
$SU(2) \times U(1) \times U(1)_R$-invariant flow. 
One can express the RHS of Einstein equation 
in terms of quadratic 4-forms for 
$SU(2) \times U(1) \times U(1)_R$-invariant flow from the relation (\ref{ftilde}).

One can make the difference between the LHS and the 
RHS of Einstein equation and see whether this difference is zero or not. 
Some of the
components written in terms of quadratic 4-forms in 
$SU(2) \times U(1) \times U(1)_R$-invariant flow 
are not exactly vanishing. 
They contain the terms 
\bea
&& F_{MPQR} \, F^{NPQR}, \qquad M= 4, 5, \,\, N=10, 11, 
\qquad
F_{6PQR} \, F^{NPQR}, \qquad N= 4, 5, 7, 8, 9, 10, 11,
\nonu \\
&& 
F_{7PQR} \, F^{NPQR},  \qquad N = 4, 5, 6, 8, 9, \qquad
F_{8PQR} \, F^{NPQR}, \qquad N=4, 5, 9, 10, 11, \nonu \\
&& F_{9PQR} \, F^{NPQR}, \qquad N=4, 5, 8, \qquad
F_{MPQR} \, F^{NPQR}, \qquad M= 10, 11, \,\, N= 4, 5. 
\label{term2}
\eea
After plugging the explicit solution for the lower 4-forms 
(\ref{4-form9}) and the upper 4-forms (\ref{4-form10}) of 
$SU(2) \times U(1) 
\times U(1)_R$-invariant flow, all of these (\ref{term2}) 
are vanishing identically.  
Recall that these quadratic 4-forms 
correspond to the off-diagonal terms of Einstein
equation for $SU(2) \times U(1) \times U(1)_R$-invariant 
flow which vanish identically except
$(4,5)$-, $(5,4)$-, $(7,10)$-, $(7,11)$-, $(9,10)$-, $(9,11)$-,
$(10,11)$-,
and $(11,10)$-components from the Ricci tensor (\ref{Ricci5}).  
The extra piece (\ref{term2}) does not possess
these nonzero off-diagonal terms. 
Therefore, we have shown the solution characterized by 
(\ref{rtilde}) and (\ref{ftilde}) for $SU(2) \times U(1)$-invariant flow
indeed satisfies the 11-dimensional Einstein equation. 

$\bullet$ Checking the Maxwell equation

By using the Maxwell equation for $SU(2) \times U(1) \times U(1)_R$-invariant flow, 
one can replace the LHS of Maxwell equation with the quadratic 4-forms.
Eventually,
the nonzero components are given in the Appendix (\ref{max2}) explicitly.
They
are characterized by the following indices 
$
(\widetilde{1}\widetilde{2}\widetilde{3})$, 
$ (\widetilde{4}\widetilde{n}\widetilde{p})$, 
$(\widetilde{5}\widetilde{n}\widetilde{p})$, 
$(\widetilde{m}\widetilde{n}\widetilde{p})$, 
where $\widetilde{m}, \widetilde{n} ,\widetilde{p} =6,
\cdots, 11$,
with the number of components $1$, 
$9$ by choosing two out of six, 
$9$ by choosing two out of six and 
$14$ by choosing three out of six respectively. 
Other remaining components of the Maxwell equation become identically zero. 
There exist 33-nonzero-components of the Maxwell equation. 
The RHS of Maxwell equation for the 
$(\widetilde{1}\widetilde{2}\widetilde{3})$-component above
consists of 12-terms coming from the quadratic 4-forms.
For the other components, the RHS contains
a single-term, two-terms or three-terms  in quadratic 4-forms.
Similarly, 
one can express the RHS 
in terms of quadratic 4-forms for 
$SU(2) \times U(1) \times U(1)_R$-invariant flow via (\ref{ftilde}). 
The difference between the LHS and the RHS 
of Maxwell equation becomes zero. Therefore, we have
shown that the solution by the 4-forms (\ref{ftilde}) with 
11-dimensional metric (\ref{11d2}) and the 4-forms with upper indices 
indeed satisfies the Maxwell
equation.

Therefore, we have shown that the solution by the Ricci tensor (\ref{Ricci6}),
the 4-forms with lower indices (\ref{4-form11}) 
and the 4-forms with upper indices (\ref{4-form12})
consists of an exact solution to 11-dimensional supergravity by
bosonic field equations (\ref{fieldequations}) and  the
deformation parameters $(\rho ,\chi)$ and 
the domain wall amplitude $A$ develop in the $AdS_4$
radial direction along the ${\cal N}=1$ $SU(2) \times U(1)$-invariant RG flow.
Compared with the previous solution \cite{Ahn0909} for $SU(2) \times U(1) 
\times U(1)_R$-invariant flow, 
they share the common
Einstein-Kahler 2-fold
inside 7-dimensional internal space but two remaining
coordinates
are different from each other. This provides the symmetry
breaking of $SU(2) \times U(1) \times U(1)_R$ into 
its subgroup $SU(2) \times U(1)$ along the whole RG flow.

\section{Conclusions and outlook }

We have found out four new 11-dimensional solutions 
of ${\cal N}=1$ $SU(3)$-invariant  flow with 
the Ricci tensor (\ref{Ricci2}), the 4-forms (\ref{4-form3}) and
the 4-forms (\ref{4-form4}), 
${\cal N}=1$ 
$SU(2) \times U(1)$-invariant  flow with the Ricci tensor 
(\ref{Ricci6}), the 4-forms (\ref{4-form11}) and
the 4-forms (\ref{4-form12}). 
In 4-dimensional sense, they have common RG flows given by (\ref{flow}). 
Inside of 7-dimensional internal metric, they have their own 
Einstein-Kahler 2-fold and its $U(1)$ bundle living in   
5-dimensional Sasaki-Einstein space (${\bf S}^5$,
$Y^{p,q}$) and share other two coordinates characterized
by $(\th, \th_6)$. Although the 4-forms are obtained from (\ref{rtilde}),
one can determine them by constructing the corresponding 3-forms
in frame basis directly described in \cite{AW0908,Ahn0909,Ahn0910}.
 
$\bullet$ 
One can analyze the similar RG flow descriptions around ${\cal N}=1$ $G_2$ critical
point.
According to the branching rules of $G_2$ into its subgroups, 
one expects that the 11-dimensional solution should
preserve $SU(3)$ or $SU(2) \times SU(2)$ symmetries. 
It would be interesting to
find out the correct 4-forms for given 11-dimensional 
metric.
Furthermore, the most general 5-dimensional Sasaki-Einstein space can
be considered and the global symmetries become the smaller $SU(2)
\times U(1)$ symmetry or $U(1) \times U(1)$ symmetry.

$\bullet$ It is an open problem to find 
the 11-dimensional lifts of holographic ${\cal N}=1$
supersymmetric RG flows \cite{BHPW} 
connecting from ${\cal N}=1$ $G_2$ critical point to 
${\cal N}=2$ $SU(3) \times U(1)_R$ critical
point. One can think of 
$SU(3)$-, $SU(2) \times SU(2)$-, $SU(2) \times U(1)$- and $U(1) \times
U(1)$-invariant flows with 
the global coordinates for ${\bf S}^7$ appropriate for the base
round 6-sphere ${\bf S}^6$ or  those as the Hopf fibration on ${\bf CP}^3$ space. 
Along the supersymmetric ${\cal N}=1$ RG flows, one 
expects that there exist nontrivial 4-form field strengths
and the work of \cite{dN87} will be 
useful to obtain these 4-forms explicitly.

$\bullet$ It is an open problem to 
consider the case where there exist four supergravity fields
preserving $SU(3)$ symmetry. 
In particular limit, one has
11-dimensional lift \cite{CPW} of  ${\cal N}=2$ $SU(3) \times U(1)_R$-invariant flow and 
for other limit, one obtains the 11-dimensional lift \cite{AI} of
${\cal N}=1$ $G_2$-invariant flow. The decoding of the 4-forms written as
the $SU(3)$-singlet vacuum expectation values, the covariant
derivative in round 7-sphere ${\bf S}^7$ and the Killing vectors in
\cite{dN87} will be useful also.

\vspace{.7cm}

\centerline{\bf Acknowledgments}

This work was supported by the 
National Research Foundation of Korea(NRF) grant 
funded by the Korea government(MEST)(No. 2009-0084601).
CA acknowledges the warm hospitality by the 
School of Physics, at Korea Institute for Advanced Study(KIAS)
where this work was initiated.

\newpage

\appendix

\renewcommand{\thesection}{\large \bf \mbox{Appendix~}\Alph{section}}
\renewcommand{\theequation}{\Alph{section}\mbox{.}\arabic{equation}}

\section{The $SU(3) \times U(1)_R$-invariant flow }

In this Appendix, we summarize the Ricci tensor and the 4-form field
strengths for $SU(3) \times U(1)_R$-invariant flow \cite{CPW}.
 
\subsection{The Ricci tensor }

The nonzero Ricci tensor in the coordinate basis from the 11-dimensional
metric \cite{CPW}, after imposing the flow
equations (\ref{flow}), can be written as follows:
\bea
R_1^{\, 1} & = & -\frac{1}{3 \, L^2\, u^{\frac{2}{3}} \,
  v^{\frac{4}{3}} \, (c_{\mu}^2  + u^2\, s^2_{\mu} )^{\frac{8}{3}}} 
2\left[ 2 u^8 v^2 (v^2-1)s_{\mu}^4 +
2 v^2 (v^2 +3) c_{\mu}^4 \right. \nonu \\
& + &   u^6 
\left[-2(-5+c_{2\mu})+ v^2(-11+c_{2\mu}) + 4v^4 c_{\mu}^2 \right]s_{\mu}^2
\nonu \\
&+ & \left. u^2 \left[12c_{\mu}^2 +v^2(9-13c_{2\mu})+ 
4 v^4 s_{\mu}^2\right]c_{\mu}^2
+u^4\left[6s_{2\mu}^2+v^2(5-8c_{2\mu}+5c_{4\mu}) +
v^4 s_{2\mu}^2 \right] \right]
\nonu \\
 & = & R_2^{\, 2} =R_3^{\, 3}
= -2 R_6^{\, 6} = -2 R_7^{\, 7} = -2
R_8^{\, 8}=-2 R_9^{\, 9},
\nonu \\
R_4^{\, 4} & = &
\frac{1}{6 \, L^2\, u^{\frac{2}{3}} \,
  v^{\frac{10}{3}} \, (c_{\mu}^2  + u^2\, s^2_{\mu} )^{\frac{8}{3}}} 
\left[ -4 v^2 (2v^4-21v^2+27)c_{\mu}^4 -
4u^8 v^2 (2v^4-5v^2 +3) s_{\mu}^4 \right. \nonu \\
& + &   2u^2 v^2 
\left[-48+60c_{2\mu}+ v^2(15-43c_{2\mu}) + 4v^4 s_{\mu}^2 \right]c_{\mu}^2
 \nonu \\
& - &  2 u^6 \left[24c_{\mu}^2 -4v^2(7+4c_{2\mu})+
v^4 (17+ 5c_{2\mu})-4v^6 c_{\mu}^2 \right] s_{\mu}^2
\nonu \\
&- & \left. u^4 v^2 \left(33-48c_{2\mu}+27c_{4\mu} +4v^2
\left[2+4c_{2\mu}-7c_{4\mu}+v^2(-1+c_{4\mu})\right] \right) \right],  
\nonu \\
R_4^{\, 5} & = &  
\frac{1}{{2  \, L^3 \,
  (c^2_{\mu} + u^2  s^2_{\mu} )^{\frac{8}{3}}}}
u^{\frac{5}{6}}  \, v^{\frac{2}{3}}\,
\left(-2c_{\mu}^2(v^2-3) \right. \nonu \\
& + & \left. u^2\left[-5-11c_{2\mu}+14v^2c_{\mu}^2+
u^2\left(-11+9c_{2\mu}+
2s_{\mu}^2\left[u^2 -v^2(u^2-7)\right]\right)\right] \right) s_{2\mu}, 
\nonu \\
R_5^{\, 4} & = &  
\frac{1}{{2  \, L \, u^{\frac{1}{6}} \, v^{\frac{4}{3}}\,
  (c^2_{\mu} + u^2  s^2_{\mu} )^{\frac{8}{3}}}}
\left(-2c_{\mu}^2(v^2-3) \right. \nonu \\
& + & \left. u^2\left[-5-11c_{2\mu}+14v^2c_{\mu}^2+
u^2\left(-11+9c_{2\mu}+
2s_{\mu}^2\left[u^2 -v^2(u^2-7)\right]\right)\right] \right) s_{2\mu}, \nonu \\ 
R_5^{\, 5} & = & \frac{1}{6  \,L^2 \, u^{\frac{2}{3}} \,
  v^{\frac{4}{3}} \, (c_{\mu}^2  + u^2\, s^2_{\mu} )^{\frac{8}{3}}} 
\left[ 4u^8 v^2 (v^2-1)s_{\mu}^4 +
4v^2 (v^2+3) c_{\mu}^4 \right. \nonu \\
& + &   u^4 \left[6 - 6c_{4\mu} +v^2 
(19-16c_{2\mu}+ c_{4\mu})+ 5v^4(-1+c_{4\mu})\right] \nonu \\
& + & \left. 4 u^2 \left[6c_{\mu}^2 +v^2(3-5c_{2\mu})+
2v^4 s_{\mu}^2\right]c_{\mu}^2
+4 u^6\left[-1-v^2+v^4+(-7+5v^2 +v^4)c_{2\mu} \right] s_{\mu}^2 \right], 
\nonu \\
R_8^{\, 10} &= & -
\frac{1}{4  \,L^2 \,
  v^{\frac{10}{3}} \, (c_{\mu}^2  + u^2\, s^2_{\mu} )^{\frac{11}{3}}} 
\left[ c_{\th_2} c_{\mu}^2 s_{\th_1}^2 
u^{\frac{4}{3}} (v^2-1) \left( -8 s_{\mu}^4 u^8 
-24 c_{\mu}^4  v^2 \right. \right. \nonu \\
& - &   12 s_{\mu}^2 u^6 ( -3 + c_{2\mu} + 2c_{\mu}^2 v^2 ) 
+ 8 c_{\mu}^2 u^2 (-3s_{\mu}^2- 5s_{\mu}^2 v^2 + 
2c_{\mu}^2 v^4 ) \nonu \\
& + & \left.  \left. u^4 
\left[4 ( -1 + 5c_{2\mu})s_{\mu}^2 +
(28 c_{2\mu}-5(3+c_{4\mu})) v^2 + 8
(5+c_{2\mu}) s_{\mu}^2 v^4 \right] \right) \right],
\nonu \\
R_8^{\, 11} &= & 
\frac{1}{L^2 \,
  v^{\frac{10}{3}} \, (c_{\mu}^2  + u^2\, s^2_{\mu} )^{\frac{11}{3}}} 
\left[ 2c_{\th_2} c_{\mu}^2 s_{\th_1}^2 
u^{\frac{10}{3}} (v^2-1) \left( - s_{\mu}^4 u^6 
+2 c_{\mu}^4  v^4 \right. \right. \nonu \\
& - & \left. \left. s_{\mu}^2 u^4 (-3 +(1+2c_{2\mu})v^2) 
 +  c_{\mu}^2 u^2( 3s_{\mu}^2 +
(-1 + 2c_{2\mu})v^2 + 6s_{\mu}^2 v^4) \right) \right],
\nonu \\
R_9^{\, 10} &= & 
 -
\frac{1}{4  \,L^2 \,
  v^{\frac{10}{3}} \, (c_{\mu}^2  + u^2\, s^2_{\mu} )^{\frac{11}{3}}} 
\left[ c_{\mu}^2 s_{\th_1}^2 
u^{\frac{4}{3}} (v^2-1) \left( -8 s_{\mu}^4 u^8 
-24 c_{\mu}^4  v^2 \right. \right. \nonu \\
& - &   12 s_{\mu}^2 u^6 ( -3 + c_{2\mu} + 2c_{\mu}^2 v^2 ) 
+ 8 c_{\mu}^2 u^2 (-3s_{\mu}^2- 5s_{\mu}^2 v^2 + 
2c_{\mu}^2 v^4 ) \nonu \\
& + & \left.  \left. u^4 
\left[4 ( -1 + 5c_{2\mu})s_{\mu}^2 +
(28 c_{2\mu}-5(3+c_{4\mu})) v^2 + 8
(5+c_{2\mu}) s_{\mu}^2 v^4 \right] \right) \right],
\nonu \\
R_9^{\, 11} &= & 
\frac{ 2c_{\mu}^2 s_{\th_1}^2 
u^{\frac{10}{3}} (v^2-1)}{L^2 \,
  v^{\frac{10}{3}} \, (c_{\mu}^2  + u^2\, s^2_{\mu} )^{\frac{11}{3}}} 
\left[  - s_{\mu}^4 u^6 
+2 c_{\mu}^4  v^4  \right. \nonu \\
& - & \left.  s_{\mu}^2 u^4 (-3 +(1+2c_{2\mu})v^2) 
 +  c_{\mu}^2 u^2( 3s_{\mu}^2 +
(-1 + 2c_{2\mu})v^2 + 6s_{\mu}^2 v^4)  \right],
\nonu \\
R_{10}^{\, 10} & = & \frac{\left[-1+2v(
- 3 + 4v^2)(-\sqrt{v^2-1} +v(-3+4v(v+\sqrt{v^2-1}))) \right]}
{48 \, L^2 \, u^{\frac{2}{3}} \,
  v^{\frac{11}{3}} \, (c_{\mu}^2  + u^2\, s^2_{\mu} )^{\frac{22}{3}}\,
(v+ \sqrt{v^2-1})^6} 
\left[ \right. \nonu \\
&& 32c_{\mu}^6 v^4 (v^2+3) + 32 s_{\mu}^4
  u^{10}(v^2-1)(6c_{\mu}^2+
s_{\mu}^2 v^4)   \nonu \\
& + &   32c_{\mu}^4 u^2 v^2  
\left[-12 c_{\mu}^2+ (15 + c_{2\mu}) v^2 + 
3s_{\mu}^2 v^4 \right] -4 c_{\mu}^2 u^4 \left[
36 s_{2\mu}^2 + 6(-1+c_{4\mu}) v^2  \right. 
\nonu \\
&+ & \left. (28c_{2\mu}-3(39+5c_{4\mu})) v^4 
+4(7+14c_{2\mu}+3c_{4\mu})v^6 \right] \nonu \\
& - & u^6\left[
24(1-5c_{2\mu})s_{2\mu}^2 + 32c_{\mu}^4(1-7c_{2\mu}) v^2 
\right. 
\nonu \\
&+& (-182+237c_{2\mu}+134c_{4\mu}+3c_{6\mu}) v^4+
16(14+3c_{2\mu})s_{2\mu}^2 v^6) -8s_{\mu}^2 u^8
\left[36c_{\mu}^2(-3+c_{2\mu}) \right.
\nonu \\
& - & \left.   \left. v^2 \left[-61-84c_{2\mu}+c_{4\mu}+
3v^2(5+16c_{2\mu}+3c_{4\mu}+s_{2\mu}^2 v^2) \right]  
\right]\right], 
\nonu \\
R_{10}^{\, 11}  & = &  \frac{4 \, c_{\mu}^2 \, u^{\frac{10}{3}} 
(v^2-1)  }
{L^2\,
  v^{\frac{10}{3}}\, (c^2_{\mu} + u^2 \, 
s^2_{\mu} )^{\frac{11}{3}}}
\left[-s_{\mu}^4 u^6 + 2 c_{\mu}^4 v^4 -s_{\mu}^2 u^4
(-3 +(1+ 2 c_{2\mu}) v^2) \right. \nonu \\
& + & \left. c_{\mu}^2 u^2 (3s_{\mu}^2 +(-1+2c_{2\mu})
v^2 + 6s_{\mu}^2 v^4) \right],  
\nonu \\
R_{11}^{\,\,10} & = &-
\frac{\left[1  +  
4 v ( -\sqrt{v^2-1} + 2 v( -1 +v^2 + v\sqrt{v^2-1})) \right]}{4 \, L^2 \,
  v^{\frac{10}{3}} \, (c_{\mu}^2  + u^2\, s^2_{\mu} )^{\frac{11}{3}}\,
(v+ \sqrt{v^2-1})^4} 
\left[ \right. \nonu \\
& & u^{\frac{4}{3}} (v^2-1) (-16 c_{\mu}^2 s_{\mu}^4 u^8 + 
16 c_{\mu}^4 ( 3 s_{\mu}^2 +(-4+c_{2\mu}) v^2)  \nonu \\
& + & 4s_{\mu}^2 u^6 ( 9 + 6c_{2\mu}- 3c_{4\mu} + (1-14c_{2\mu}+
c_{4\mu}) v^2 ) \nonu \\
& + & 16 c_{\mu}^2 u^2 ( -( 2 + 5 c_{2\mu}) s_{\mu}^2 +
(-4 + 3c_{2\mu})s_{\mu}^2 v^2 + 2c_{2\mu} v^4) \nonu \\
& + & \left. u^4 ( 4 ( -5 +6 c_{2\mu} ) s_{2\mu}^2 + (( 53-40c_{2\mu})
c_{2\mu} + 3 c_{6\mu}) v^2 + 32( 2 + 
c_{2\mu})s_{\mu}^2 v^4))
\right], 
\nonu \\
R_{11}^{\, 11} & = & 
 \frac{\left[-1+2v(- 3 + 4v^2)(-\sqrt{v^2-1}
     +v(-3+4v(v+\sqrt{v^2-1})))
\right]}
{48 \, L^2 \, u^{\frac{2}{3}} \,
  v^{\frac{10}{3}} \, 
(c_{\mu}^2  + u^2\, s^2_{\mu} )^{\frac{11}{3}}\,
(v+ \sqrt{v^2-1})^6} 
\left[ \right. \nonu \\
&& 32c_{\mu}^6 v^4 (v^2+3) + 32 s_{\mu}^4
  u^{10}(v^2-1)(-6c_{\mu}^2+
s_{\mu}^2 v^4)   \nonu \\
& + &   32c_{\mu}^4 u^2 v^2  
\left[6 c_{\mu}^2+ (6 -8 c_{2\mu}) v^2 + 
3s_{\mu}^2 v^4 \right] -4 c_{\mu}^2 u^4 \left[
-36 s_{2\mu}^2 + 48 s_{2\mu}^2 v^2  \right. 
\nonu \\
&+ & \left. (-39 + 124c_{2\mu}+3c_{4\mu}) v^4 
-4(11+10c_{2\mu}+3c_{4\mu})v^6 \right] \nonu \\
& + & 8s_{\mu}^2 u^8 \left[
36c_{\mu}^2(-3+c_{2\mu}) + 8(7+12c_{2\mu}-c_{4\mu}) v^2 
\right. 
\nonu \\
&-& \left. 12(1+5c_{2\mu}) v^4+
3s_{2\mu}^2 v^6 \right] + u^6
\left[24(1-5c_{2\mu}) s_{2\mu}^2 \right.
\nonu \\
& + &   32 c_{\mu}^2 (-7+6c_{2\mu}
-5c_{4\mu})v^2 +
(26-63c_{2\mu}+214c_{4\mu}+15c_{6\mu}) v^4
\nonu \\
&+& \left. \left. 16(16+3c_{2\mu})s_{2\mu}^2 v^6 \right]  
\right],
\label{Ricci1}
\eea
where we introduce 
\bea
 u \equiv \rho^4, \qquad v \equiv \cosh \chi.
\label{uv}
\eea
There are additional
nonzero Ricci tensor components ($R_{8}^{\,10}, R_{8}^{\,11}, R_{9}^{\,10}$
and $R_{9}^{\,11}$) that depend on the internal coordinates $\th_1$ or
$\th_2$, 
compared to the Ricci tensor in the frame basis \cite{CPW,Ahn0909}. 
One also obtains (\ref{Ricci1}) directly from the Ricci tensor in the
frame basis with the help of vielbeins.  
At the IR fixed point($u=\sqrt{3}$ and $v=\sqrt{\frac{3}{2}}$) in 4-dimensions, 
the off-diagonal components $R_{4}^{\,5}$ and
$R_{5}^{\,4}$
vanish.
We use a simplified notation for the trigonometric function as 
in $s_{\mu} \equiv \sin \mu$ and so on.

\subsection{The 4-form field strengths }

The nonzero 4-form field strengths satisfying (\ref{fieldequations})
for given Ricci tensor (\ref{Ricci1}) and 11-dimensional metric
presented in \cite{CPW}
are summarized as follows:
\bea
F_{1234} & = & 
\left[\frac{3 e^{3A}}{L \, \rho^4} \, \cosh^2 \chi\right] \left[ 
c^2_{\mu} (-5 + \cosh 2\chi)+ 2\rho^8 (-2 +c_{2\mu} + \rho^8 \, s^2_{\mu} \,
\sinh^2 \chi)\right], 
\nonu \\
F_{1235} & = &  \left[\frac{3 e^{3A}}{2 \, \rho^2} \right] \left[
1+ \cosh2\chi + \rho^8 \, (-3 +\cosh2\chi)
\right] \,  s_{2\mu}, 
\nonu \\
F_{4567} & = &  \left[ \frac{3 \, L^2 (-3 + \rho^8) \,
 \tanh \chi }{2 \, \rho^2}  \right] \, c^2_{\mu} \, c_{\th_4+3\phi
+4\psi}\, s_{\th_1} = -2 s_{\th_1}^{-1} \, 
c_{\th_1}^{-1} \, s_{\th_2}^{-1} \, F_{4589},
\nonu \\
F_{4568} & = &  \left[ \frac{3 \, L^2 \, (-3 + \rho^8) \,
\tanh \chi}{2 \, \rho^2}  \right] \, c^2_{\mu} \, s_{\th_4+3\phi
+4\psi}\, s_{\th_1} \, s_{\th_2}=2 s_{\th_1}^{-1} \, 
s_{\th_2} \, c_{\th_1}^{-1} \, c_{\th_2}^{-1} \, F_{4578}
\nonu \\
& = & 2 s_{\th_1}^{-1} \, c_{\th_1}^{-1} \, 
s_{\th_2} \, F_{4579}, \nonu \\
F_{4678} & = & 
\left[\frac{ 3 L^2 \,
 \rho^6 \,
\mbox{sech}^2 \chi \, \sinh 2\chi}
{16\,(c^2_{\mu} +\rho^8 \, s^2_{\mu})^2}\right]  
 \left[ 2c_{\mu}^2(-2+\cosh2\chi) +
\rho^8 (-5 + 2c_{\mu}^2 \, \cosh2\chi + 2\rho^8 \, 
s_{\mu}^2) \right] \nonu \\
& \times & s^3_{\th_1} \,
c_{\th_2} \, 
c^3_{\mu} \,
s_{\mu} \, s_{\th_4 + 3\phi + 4\psi} =  c_{\th_2} \,F_{4679} =\frac{1}{2}
s_{\th_1}^2\, c_{\th_2} \,  F_{467\,10} = -
s_{\th_1} \, c_{\th_1}^{-1} \, s_{\th_2}^{-1} \, c_{\th_2}\,
F_{489\,10},
\nonu    \\
F_{467\,11} & = & 
\left[ \frac{3 L^2 \,
\tanh \chi}{4 \, \rho^2 \, (c^2_{\mu} +\rho^8 \, s^2_{\mu})^2} \right]
\nonu \\
&\times & \left[ 6c_{\mu}^2 +
\rho^8 (1-7 c_{2\mu} + 2\cosh2\chi + \rho^8 (-9 + 5 c_{2\mu} +
2\cosh2\chi + 2 \, \rho^8\, s^2_{\mu}) \right] \nonu \\
& \times &  s_{\th_1} \, 
c^3_{\mu} \,
s_{\mu} \, s_{\th_4 + 3\phi + 4\psi} = - 2 s_{\th_1}^{-1} \,
c_{\th_1}^{-1} \, s_{\th_2}^{-1} \, F_{489\,11},
\nonu \\
F_{4689} & = & 
- \left[\frac{ 3 L^2 \,
 \rho^6 \,
\mbox{sech}^2 \chi \, \sinh 2\chi}
{16(c^2_{\mu} +\rho^8 \, s^2_{\mu})^2}\right] \nonu \\
& \times &  \left[ 2c_{\mu}^2(-2+\cosh2\chi) +
\rho^8 (-5 + c_{2\mu} + 2c_{\mu}^2 \, 
\cosh2\chi + 2\rho^8 \, s_{\mu}^2) \right] \nonu \\
& \times &   s^3_{\th_1} \, s_{\th_2} \,
c^3_{\mu} \,
s_{\mu} \, c_{\th_4 + 3\phi + 4\psi}
= \frac{1}{2} \, s_{\th_1}^2 \, F_{468\,10}
= c_{\th_1}^{-1} \, s_{\th_1}\, c_{\th_2}^{-1} \, s_{\th_2} \,
F_{478\,10} = c_{\th_1}^{-1} \, s_{\th_1} \, s_{\th_2} \,
F_{479\,10},
\nonu \\
F_{468\,11} & = & 
 -\left[\frac{ 3 L^2 \,
\tanh \chi }
{4  \rho^2  (c^2_{\mu} +\rho^8  s^2_{\mu})^2} \right]
\nonu \\
& \times &  \left[ 6c_{\mu}^2 +
\rho^8 (1-7 c_{2\mu} + 2\cosh2\chi + \rho^8 (-9 + 5 c_{2\mu} +
2\cosh2\chi + 2 \rho^8  s^2_{\mu})) \right] \nonu \\
& \times & s_{\th_1} \, s_{\th_2} \, 
c^3_{\mu} \,
s_{\mu} \, c_{\th_4 + 3\phi + 4\psi}= 2 \, s_{\th_1}^{-1} \, 
s_{\th_2} \, c_{\th_1}^{-1} \, c_{\th_2}^{-1} \, F_{478\,11} = 2
s_{\th_1}^{-1} \, c_{\th_1}^{-1} \, s_{\th_2} \, F_{479\,11},
\nonu \\
F_{5678} & = & \left[
 \frac{ 3 L^3 \,
\tanh \chi}{8  \, (c^2_{\mu} +\rho^8 \, s^2_{\mu})^3}
\right]
\left[ 6c_{\mu}^4 +
\rho^8 \, c_{\mu}^2 \, (9-7 c_{2\mu}) 
+ 10 \, \rho^{16} \, s_{\mu}^4 -
2\, \rho^{24}\, s^4_{\mu} \right] \nonu \\
& \times & s_{\th_1}^3 \, c_{\th_2} \, 
c^4_{\mu} \,
 s_{\th_4 + 3\phi + 4\psi} =  c_{\th_2} \, F_{5679}=
\frac{1}{2} \, s^2_{\th_1} \, c_{\th_2} \, 
 F_{567\,10} = -s_{\th_1} \, c_{\th_1}^{-1} \, 
s_{\th_2}^{-1} \, c_{\th_2} \, F_{589\,10},
\nonu \\
F_{567\,11} & = & 
\left[ \frac{ 3 L^3 \,
\tanh \chi}{8  \, (c^2_{\mu} +\rho^8 \, s^2_{\mu})^3} \right]
\nonu \\
& \times & \left[ 4c_{\mu}^4 (2+c_{2\mu}) +
\rho^8  c_{\mu}^2  (13-6 c_{2\mu}-3c_{4\mu}) 
+ 12  \rho^{16}  (2+ c_{2\mu} )  s_{\mu}^4 -
4 \rho^{24} s^4_{\mu} c_{2\mu} \right] \nonu \\
& \times & s_{\th_1}  \, 
c^2_{\mu} \, 
 s_{\th_4 + 3\phi + 4\psi} = -2 
s_{\th_1}^{-1} \, c_{\th_1}^{-1} \, s_{\th_2}^{-1} \, F_{589\,11},
\nonu \\
F_{5689} & = & 
 -\left[\frac{3 L^3 \,
\tanh \chi}{8  \, (c^2_{\mu} +\rho^8 \, s^2_{\mu})^3} 
\right]
\left[ 6c_{\mu}^4 +
\rho^8 \, c_{\mu}^2 \, (9-7 c_{2\mu}) 
+ 10 \, \rho^{16} \, s_{\mu}^4 -
2\, \rho^{24}\, s^4_{\mu} \right] \nonu \\
& \times & s_{\th_1}^3  \, s_{\th_2} \, 
c^4_{\mu} \,
 c_{\th_4 + 3\phi + 4\psi} = \frac{1}{2} s_{\th_1}^2 \, F_{568\,10}
= s_{\th_1} \, c_{\th_1}^{-1} \, s_{\th_2} \, c_{\th_2}^{-1} \, 
F_{578\,10} = s_{\th_1}\, c_{\th_1}^{-1} \, 
s_{\th_2} \, F_{579\,10},
\nonu \\
F_{568\,11} & = & 
\left[ \frac{ 3 L^3 \,
\tanh \chi}{4  \, (c^2_{\mu} +\rho^8 \, s^2_{\mu})^2}
\right]
 \left[ -2 c_{\mu}^2 (2+c_{2\mu}) +
\rho^8 \, ( -5 + 2 c_{2\mu} + c_{4\mu} +
2 \rho^8\, c_{2\mu}\, s^2_{\mu}) \right] \nonu \\
& \times &  s_{\th_1}  \, s_{\th_2} \, 
c^2_{\mu}  \,
 c_{\th_4 + 3\phi + 4\psi},
\nonu \\
F_{578\,11} & = & 
 \left[ \frac{ 3 L^3 \,
\tanh \chi}{16  \, (c^2_{\mu} +\rho^8 \, s^2_{\mu})^3} \right]
\, s_{\th_1}^2 \,  c_{\th_1}  \, c_{\th_2} \, 
c^2_{\mu}  \,
 c_{\th_4 + 3\phi + 4\psi} \nonu \\
& \times & \left[ -4 c_{\mu}^4 (2+c_{2\mu}) +
\rho^8 \, c_{\mu}^2 \, ( -13 + 6 c_{2\mu} + 3 c_{4\mu}) -12
\rho^{16} \, ( 2+ c_{2\mu}) \, s_{\mu}^4 + 4 \rho^{24} \, 
c_{2\mu} \, s^4_{\mu}) \right]
\nonu  \\
& = &  c_{\th_2} \,  
F_{579\,11},
\nonu \\
F_{678\,11} & = &  - \left[\frac{3 \, L^3 \, (3 + \rho^8) \,
 \tanh \chi}{4 \,(c^2_{\mu} +\rho^8 \,
  s^2_{\mu}) } \right] 
 \, c^3_{\mu} \, s_{\mu} \, c_{\th_4+3\phi
+4\psi}\, s_{\th_1}^3 \, c_{\th_2} = c_{\th_2} \, F_{679\,11}
=\frac{1}{2} s_{\th_1}^2 \, c_{\th_2} \, F_{67\,10\,11}
\nonu \\
& = & - s_{\th_1} \, c_{\th_1}^{-1} \, 
c_{\th_2} \, s_{\th_2}^{-1} \, F_{89\,10\,11}, 
\nonu \\
F_{689\,11} & = &  - \left[\frac{3 \, L^3 \, (3 + \rho^8) \,
 \tanh \chi}{4 \,(c^2_{\mu} +\rho^8 \,
  s^2_{\mu}) } \right]
 \, c^3_{\mu} \, s_{\mu} \, s_{\th_4+3\phi
+4\psi}\, s_{\th_1}^3 \, s_{\th_2} =\frac{1}{2} 
s_{\th_1}^2\, F_{68\,10\,11} \nonu \\
& = & 
s_{\th_1} \,  c_{\th_1}^{-1} \, s_{\th_2}\,
c_{\th_2}^{-1} \, F_{78\,10\,11} 
 =  s_{\th_1} \, c_{\th_1}^{-1} \, s_{\th_2} \, F_{79\,10\,11}.
\label{4-form1} 
\eea
One sees that the 4-forms (\ref{4-form1}) has the dependence on 
$(\th_4 + 3\phi + 4\psi)$.
According to the shifts $\phi \rightarrow \frac{4}{3} \gamma $ and $\psi
\rightarrow \psi - \gamma$, corresponding to the $U(1)_R$ charge, 
it is evident that 4-forms 
preserve this $U(1)_R$ charge. Note that $\th_4$ is one of the Euler
angles on ${\bf S}^3$ in (\ref{sixu}).
At the IR fixed point in 4-dimensions, the components $F_{1235}, F_{4mnp}$ and
$F_{45mn}$
vanish.
By using the 4-form field strengths in the
frame basis \cite{CPW,AW0908} and vielbeins, one also obtains (\ref{4-form1}).

The 4-form field strengths with upper indices can be obtained from
those with lower indices (\ref{4-form1}) by multiplying the
11-dimensional inverse metric and they are given by as follows:    
\bea
F^{1234} & = & -\left[\frac{3 e^{-3A} \, \rho^{\frac{4}{3}}}
{L \, \cosh^{\frac{10}{3}} \chi \, (c^2_{\mu} +\rho^8 \,
  s^2_{\mu})^{\frac{8}{3}}} 
\right] \left[ 
c^2_{\mu} (-5 + \cosh 2\chi)+ 2\rho^8 (-2 +c_{2\mu} + \rho^8 \, s^2_{\mu} \,
\sinh^2 \chi)\right], 
\nonu \\
F^{1235} & = &  -\left[\frac{3  \, e^{-3A} \, 
\rho^{\frac{22}{3}}\,  \mbox{sech}^{\frac{10}{3}} \chi}
{2 \,L^2 \, (c^2_{\mu} +\rho^8 \,
  s^2_{\mu})^{\frac{8}{3}}} \right]  \left[
1+ \cosh2\chi + \rho^8 (-3 +\cosh2\chi)
\right] \,  s_{2\mu}, 
\nonu \\
F^{4567} & = &  \left[ \frac{6  \, (-3 + \rho^8) \,
 \tanh \chi \,  }{L^4   \, \rho^{\frac{2}{3}}
\, \mbox{sech}^{\frac{2}{3}} 
\chi  \, (c^2_{\mu} +\rho^8 \,
  s^2_{\mu})^{\frac{2}{3}} }  \right]\, c_{\th_4+3\phi
+4\psi} \, s_{\th_1}^{-1} \, c_{\mu}^{-2} 
= -\frac{1}{2} s_{\th_1} \, 
c_{\th_1} \, s_{\th_2} \, F^{4589}
\nonu \\
& = & s_{\th_1}^{-1} \, c_{\th_1} \, s_{\th_2}\, 
F^{458\,10} = - s_{\th_1}^{-1} \, c_{\th_1}\,
c_{\th_2}^{-1} \, s_{\th_2} \, F^{459\,10},
\nonu \\
F^{4568} & = &  
 \left[ \frac{6  \, (-3 + \rho^8) \,
 \tanh \chi \,  }{L^4   \, \rho^{\frac{2}{3}}
\, \mbox{sech}^{\frac{2}{3}} 
\chi  \, (c^2_{\mu} +\rho^8 \,
  s^2_{\mu})^{\frac{2}{3}} }  \right]\, s_{\th_4+3\phi
+4\psi} \, s_{\th_1}^{-1} \, s_{\th_2}^{-1}\, 
c_{\mu}^{-2} = 
- c_{\th_2}^{-1} \, F^{4569} 
\nonu \\
& =& \frac{1}{2}\, s_{\th_1} \, c_{\th_1} \, 
s_{\th_2}^{-1} \, F^{4579} =
- s_{\th_1}^{-1} \, c_{\th_1} \, s_{\th_2}^{-1} \, 
F^{457\,10}, 
\nonu \\
F^{467\,10} & = & 
\left[ \frac{3}
{2 \, L^4 \, \rho^{\frac{2}{3}} \, \cosh^{\frac{4}{3}} \chi \,
(c^2_{\mu} +\rho^8 \, s^2_{\mu})^{\frac{5}{3}}} \right]
\nonu \\
&\times & \left[ -s_{2\mu}^2 \, (\rho^8-1) \, \tanh \chi\, (1+
\cosh 2\chi + \rho^8(-3 + \cosh 2\chi )) \right.
\nonu \\
&+ & \left.  2 \,  (\rho^8-3) \,
\sinh 2\chi \, (c_{\mu}^2 + \rho^8 s_{\mu}^2) \right]  
\,   s_{\th_1}^{-1} \, s_{\mu}^{-1} \, 
c_{\mu}^{-3} \, s_{\th_4 + 3\phi + 4\psi} = 
-\frac{1}{2} s_{\th_1}\, c_{\th_1}\,
s_{\th_2}\, F^{489\,10},
\nonu \\
F^{467\,11} & = & \left[
 \frac{6  \, \mbox{sech}^{\frac{4}{3}} \chi }
{L^4 \, \rho^{\frac{2}{3}} \,
(c^2_{\mu} +\rho^8 \, s^2_{\mu})^{\frac{5}{3}}} 
\right]
\left[ 3 c_{\mu} \, s_{\mu}^{-1} \, \cosh \chi \, \sinh \chi 
+ \rho^8 (-3 c_{2\mu} \, s_{2\mu}^{-1} + s_{\mu}^{-1} \, 
c_{\mu}^{-1}) \sinh 2\chi \right.
\nonu \\
&+ & \left. \rho^{16}\, (-3 +
\sinh^2 \chi) \, s_{\mu} \, c_{\mu}^{-1} \, \tanh \chi \right]  
\,   s_{\th_1}^{-1} \, s_{\th_4 + 3\phi + 4\psi} = 
s_{\th_1}^{-1} \, 
c_{\th_1} \, s_{\th_2} \, F^{48\,10\,11} \nonu \\
& = &
- 2 s_{\th_1}^{-1} \, 
c_{\th_1} \, s_{\th_2} \, c_{\th_2}^{-1} \, 
F^{49\,10\,11} = -\frac{1}{2} s_{\th_1} \, c_{\th_1} \, s_{\th_2}  
\, F^{489\,11},
\nonu \\
F^{468\,10} & = & -
\left[ \frac{3}
{2 \, L^4 \, \rho^{\frac{2}{3}} \, \cosh^{\frac{4}{3}} \chi \,
(c^2_{\mu} +\rho^8 \, s^2_{\mu})^{\frac{5}{3}}} \right]
\nonu \\
& \times & 
\left[ -s_{2\mu}^2 \,  (\rho^8-1)\, \tanh \chi \, (1+
\cosh 2\chi + \rho^8(-3 + \cosh 2\chi )) \right.
\nonu \\
&+ & \left.  2 \, 
(\rho^8-3) \, \sinh 2\chi \, (c_{\mu}^2 + \rho^8 s_{\mu}^2) \right]  
\,   s_{\th_1}^{-1} \, s_{\th_2}^{-1} \, s_{\mu}^{-1} \, 
c_{\mu}^{-3} \, c_{\th_4 + 3\phi + 4\psi} = - c_{\th_2}^{-1}\,
F^{469\,10} \nonu \\
& = & \frac{1}{2} s_{\th_1} \, c_{\th_1}\, s_{\th_2}^{-1} \,
F^{479\,10},
\nonu \\
F^{468\,11} & = & -
\left[ \frac{3  \, \mbox{sech}^{\frac{4}{3}} \chi }
{L^4 \, \rho^{\frac{2}{3}} \,
(c^2_{\mu} +\rho^8 \, s^2_{\mu})^{\frac{5}{3}}} 
\right]
\left[ 3 c_{\mu} \, s_{\mu}^{-1} \, \sinh 2\chi 
+ \rho^8 (2-3 c_{2\mu})\, s_{\mu}^{-1} \, 
c_{\mu}^{-1}) \sinh 2\chi \right.
\nonu \\
&+ & \left. \rho^{16}\, (-7 +
\cosh 2\chi) \, s_{\mu} \, c_{\mu}^{-1} \, \tanh \chi \right]  
\,   s_{\th_1}^{-1} \, s_{\th_2}^{-1} \, 
c_{\th_4 + 3\phi + 4\psi} =
- c_{\th_2}^{-1} \, F^{469\,11} \nonu \\
& = & \frac{1}{2} s_{\th_1}\, c_{\th_1} \, s_{\th_2}^{-1} \,
F^{479\,11} =-c_{\th_1}\, s_{\th_1}^{-1} \, 
s_{\th_2}^{-1}\, F^{47\,10\,11},
\nonu \\
F^{567\,10} & = & 
- \left[\frac{6 \, (\rho^8-1) \,
 \sinh \chi \, \cosh^{\frac{4}{3}} \chi}
{L^5 \, \rho^{\frac{8}{3}} \, (c^2_{\mu} +\rho^8 \, 
s^2_{\mu})^{\frac{5}{3}}} 
\right]
\left[ 3c_{\mu}^2 \mbox{sech}^{\frac{5}{3}} \chi 
+ \rho^8 (2 \cosh^{\frac{1}{3}} \chi \, 
\right. \nonu \\
& - & \left. ( -2 + c_{2\mu}) \mbox{sech}^{\frac{5}{3}} \chi + 
\rho^8 s_{\mu}^2 \, \mbox{sech}^{\frac{5}{3}} \chi) \right]  
\,
s_{\th_1}^{-1}\, c_{\mu}^{-2} \, s_{\th_4 + 3\phi + 4\psi}= -
\frac{1}{2} s_{\th_1}\, c_{\th_1} \, s_{\th_2} \, F^{589\,10},
\nonu \\
F^{567\,11} & = & 
 \left[\frac{6  \, \rho^{\frac{16}{3}} \,
 \sinh \chi \, \cosh^{\frac{1}{3}} \chi}
{L^5 \,  \, 
(c^2_{\mu} +\rho^8 \, s^2_{\mu})^{\frac{5}{3}}} 
\right]
\left[ -2 \cosh^{\frac{4}{3}} \chi +
3 \mbox{sech}^{\frac{2}{3}} \chi + \rho^8 ( 2 \cosh^{\frac{4}{3}} \chi
+ ( -2  \right.
\nonu \\
&+ & \left.  3 c_{\mu}^{-2} ) \mbox{sech}^{\frac{2}{3}} \chi +
\rho^8 \, s_{\mu}^2 \, c_{\mu}^{-2} \, \mbox{sech}^{\frac{2}{3}} \chi  \right]  
\, s_{\th_1}^{-1} \, s_{\th_4 + 3\phi + 4\psi} = -\frac{1}{2}
s_{\th_1} \, c_{\th_1} \, s_{\th_2} \,
F^{589\,11}
\nonu \\
& = & c_{\th_1} \, s_{\th_1}^{-1} \, s_{\th_2} \, 
F^{58\,10\,11} = 
- s_{\th_1}^{-1} \, c_{\th_1} 
\, s_{\th_2} \, c_{\th_2}^{-1} \, F^{59\,10\,11},
\nonu \\
F^{568\,10} & = & 
\left[\frac{6 \, (\rho^8-1) \,
 \sinh \chi \, \cosh^{\frac{4}{3}} \chi}
{L^5 \, \rho^{\frac{8}{3}} \, (c^2_{\mu} +\rho^8 \, 
s^2_{\mu})^{\frac{5}{3}}} 
\right]
\left[ 3c_{\mu}^2 \mbox{sech}^{\frac{5}{3}} \chi 
+ \rho^8 (2 \cosh^{\frac{1}{3}} \chi \, 
\right. \nonu \\
& - & \left. ( -2 + c_{2\mu}) \mbox{sech}^{\frac{5}{3}} \chi + 
\rho^8 s_{\mu}^2 \, \mbox{sech}^{\frac{5}{3}} \chi) \right]  
\,
s_{\th_1}^{-1}\, s_{\th_2}^{-1} \,
c_{\mu}^{-2} \, c_{\th_4 + 3\phi + 4\psi} = - c_{\th_2}^{-1} \, 
 F^{569\,10}
\nonu \\
&= & \frac{1}{2} s_{\th_1} \, c_{\th_1} \, s_{\th_2}^{-1} \, 
F^{579\,10}, 
\nonu \\
F^{568\,11} & = & -\left[
 \frac{6  \, \rho^{\frac{16}{3}} \,
 \sinh \chi \, \cosh^{\frac{1}{3}} \chi}
{L^5 \,  \, 
(c^2_{\mu} +\rho^8 \, s^2_{\mu})^{\frac{5}{3}}} 
\right]
\left[ -2 \cosh^{\frac{4}{3}} \chi +
3 \mbox{sech}^{\frac{2}{3}} \chi + \rho^8 ( 2 \cosh^{\frac{4}{3}} \chi 
+ ( -2  \right.
\nonu \\
&+ & \left.  3 c_{\mu}^{-2} ) \mbox{sech}^{\frac{2}{3}} \chi +
\rho^8 \, s_{\mu}^2 \, c_{\mu}^{-2} \, \mbox{sech}^{\frac{2}{3}} \chi  \right]  
\, s_{\th_1}^{-1} \, s_{\th_2}^{-1} \, 
c_{\th_4 + 3\phi + 4\psi} = - c_{\th_2}^{-1} \, F^{569\,11}
\nonu  \\
& = & \frac{1}{2} s_{\th_1} \, c_{\th_1} \,
s_{\th_2}^{-1} \, F^{579\,11} = - c_{\th_1} \,
s_{\th_1}^{-1} \, s_{\th_2}^{-1} \, F^{57\,10\,11},
\nonu \\
F^{67\,10\,11} & = &-\left[
 \frac{6\, (\rho^8+3)}{L^5 \, \rho^{\frac{8}{3}}} \, \sinh \chi
 \, \mbox{sech}^{\frac{1}{3}} \chi \, 
 (c^2_{\mu} +\rho^8 \, s^2_{\mu})^{\frac{1}{3}} \right] 
s_{\th_1}^{-1}\, s_{\mu}^{-1} \, c_{\mu}^{-3} \, c_{\th_4 + 3\phi +
  4\psi} \nonu \\
& = & -\frac{1}{2} s_{\th_1}\, c_{\th_1}\, s_{\th_2} \,
F^{89\,10\,11}, 
\nonu \\
F^{68\,10\,11} & = &-
\left[ \frac{6\, (\rho^8+3)}{L^5 \, \rho^{\frac{8}{3}}} \, \sinh \chi
 \, \mbox{sech}^{\frac{1}{3}} \chi \, 
 (c^2_{\mu} +\rho^8 \, s^2_{\mu})^{\frac{1}{3}} \right]
s_{\th_1}^{-1}\, s_{\th_2}^{-1} \, s_{\mu}^{-1} \, 
c_{\mu}^{-3} \, s_{\th_4 + 3\phi + 4\psi} \nonu \\
&= & - c_{\th_2}^{-1} \, F^{69\,10\,11} = 
\frac{1}{2} s_{\th_1} \, c_{\th_1} \, s_{\th_2}^{-1} 
\, F^{79\,10\,11}.
\label{4-form2} 
\eea
One also obtains (\ref{4-form2}) from
the 4-form field strengths in the
frame basis \cite{CPW,AW0908} and vielbeins.
At the IR fixed point in 4-dimensions, the components $F^{1235}, F^{4mnp}$ and
$F^{45mn}$
vanish as before.
It turns out that after computing the RHS of
(\ref{fieldequations}) using both (\ref{4-form1}) and (\ref{4-form2}),
the dependence on the combination $(\th_4 + 3\phi + 4\psi)$ disappears
completely.
This coincides with the fact that the Ricci tensor (\ref{Ricci1})
does not depend on these variables. 

\section{The $SU(3)$-invariant flow }

In this Appendix, we describe the Ricci tensor and the 4-form field
strengths for $SU(3)$-invariant flow that can be written in terms of 
corresponding Ricci tensor and 4-form field strengths respectively for $SU(3)
\times U(1)_R$-invariant flow.

\subsection{The Ricci tensor  }

The Ricci tensor in the coordinate basis from the 11-dimensional
metric (\ref{11d}), after imposing the flow
equations (\ref{flow}), can be written in terms of those (\ref{Ricci1}) in the
Appendix A as follows:
\bea
\widetilde{R}_{1}^{\,\,1} & = & R_{1}^{\,\,1}, \qquad
\widetilde{R}_{2}^{\,\,2}  =  R_{2}^{\,\,2}, \qquad
\widetilde{R}_{3}^{\,\,3} =  R_{3}^{\,\,3},  \qquad
\widetilde{R}_{4}^{\,\,4}    =  R_{4}^{\,\,4}, \nonu \\
\widetilde{R}_{4}^{\,\,5} & = & -\left[\frac{
\cos \theta \, \sin
  \th_6}
{\sqrt{1-\sin^2 \theta \, \sin^2
    \th_6} } \right] R_{4}^{\,\,5}, \qquad
\widetilde{R}_{4}^{\,\,11}   =  - 
\left[\frac{ \csc \theta \, \cos
  \th_6}
{\sqrt{1-\sin^2 \theta \, \sin^2
    \th_6} } \right] R_{4}^{\,\,5},
\nonu \\
\widetilde{R}_{5}^{\,\,4} & = & -\left[
\frac{\cos \theta \,\sin \th_6}{\sqrt{1-\sin^2 \theta \, \sin^2
    \th_6} } \right] R_{5}^{\,\,4},
 \nonu \\
\widetilde{R}_{5}^{\,\,5} & = & \left[\frac{1 }
{1+\cot^2 \th_6 \, \sec^2
    \theta } \right] R_{5}^{\,\,5} - 
\left[\frac{\cos^2 \th_6}{1-\sin^2 \theta \, \sin^2
    \th_6} \right] \left(R_{10}^{\,\,11}-R_{11}^{\,\,11}\right), \nonu \\
\widetilde{R}_{5}^{\,\,10} & = & -\left[\frac{\cos \th_6 }
{1-\sin^2 \theta \, \sin^2
    \th_6 } \right] \left( R_{10}^{\,\,10} + R_{10}^{\,\,11}-
  R_{11}^{\,\,10}-
R_{11}^{\,\,11} \right),  
\nonu \\
\widetilde{R}_{5}^{\,\,11} & = & \left[\frac{2 \cot \theta \,  \, \sin
    2\th_6 }
{3 + \cos 2\theta +  2 \cos 2\th_6
    \sin^2 \theta } \right] \left( R_{5}^{\,\,5} + R_{10}^{\,\,11} 
- R_{11}^{\,\,11}\right),
\nonu \\
\widetilde{R}_{6}^{\,\,6} & = & R_{6}^{\,\,6}, \qquad 
\widetilde{R}_{7}^{\,\,7}  =  R_{7}^{\,\,7}, \nonu \\
\widetilde{R}_{8}^{\,\,5}  & = & \left[\cos \th_6 \right] R_{8}^{\,\,11}, \qquad
\widetilde{R}_{8}^{\,\,8} =   R_{8}^{\,\,8}, \qquad
\widetilde{R}_{8}^{\,\,10}  =  R_{8}^{\,\,10} + R_{8}^{\,\,11}, \qquad
\widetilde{R}_{8}^{\,\,11}   = -
\left[\cot \theta \, \sin \th_6 \right] R_{8}^{\,\,11},  \nonu \\ 
\widetilde{R}_{9}^{\,\,5}  & = & \left[\cos \th_6 \right] R_{9}^{\,\,11}, \qquad
\widetilde{R}_{9}^{\,\,9}    =    R_{9}^{\,\,9}, 
\qquad
\widetilde{R}_{9}^{\,\,10}   =  R_{9}^{\,\,10} + R_{9}^{\,\,11}, \qquad
\widetilde{R}_{9}^{\,\,11}    =   -\left[\cot \theta \, \sin \th_6 \right]  
R_{9}^{\,\,11},
\nonu \\
\widetilde{R}_{10}^{\,\,5}  & = & \left[\cos \th_6 \right] R_{10}^{\,\,11},
\qquad \widetilde{R}_{10}^{\,\,10}  =   R_{10}^{\,\,10} +
R_{10}^{\,\,11}, \qquad
\widetilde{R}_{10}^{\,\,11}   = -\left[\cot \theta \, \sin \th_6 \right]
 R_{10}^{\,\,11},
\nonu \\
\widetilde{R}_{11}^{\,\,4} & = & -\left[
\frac{\sin \theta \,\cos \th_6}{\sqrt{1-\sin^2 \theta \, \sin^2
    \th_6} } \right] R_{5}^{\,\,4},
 \nonu \\
\widetilde{R}_{11}^{\,\,5} & = & \left[\frac{ \sin 2\theta \,  \, \sin
    2\th_6 }
{3 + \cos 2\theta +  2 \cos 2\th_6
    \sin^2 \theta } \right] \left( R_{5}^{\,\,5} + R_{10}^{\,\,11} 
- R_{11}^{\,\,11}\right), \nonu \\
\widetilde{R}_{11}^{\,\,10} & = & \left[\frac{\cos \th \, \sin \theta
  \, \sin \th_6}
{1-\sin^2 \theta \, \sin^2
    \th_6 } \right] \left( R_{10}^{\,\,10} + R_{10}^{\,\,11}-
  R_{11}^{\,\,10}-
R_{11}^{\,\,11} \right),  
\nonu \\
\widetilde{R}_{11}^{\,\,11} & = & \left[\frac{1 }
{1+\tan^2 \th_6 \, \cos^2
    \theta } \right] R_{5}^{\,\,5} - 
\left[\frac{\cos^2 \theta \, \sin^2 \th_6}{1-\sin^2 \theta \, \sin^2
    \th_6} \right] \left(R_{10}^{\,\,11}-R_{11}^{\,\,11}\right).
\label{Ricci2}
\eea
There are extra
nonzero off-diagonal Ricci tensor components
($\widetilde{R}_{4}^{\,11}, 
\widetilde{R}_{5}^{\,10},
\widetilde{R}_{5}^{\,11}, \widetilde{R}_{8}^{\,5}, 
\widetilde{R}_{9}^{\,5}, \widetilde{R}_{10}^{\,5}, \widetilde{R}_{11}^{\,4}$
and $\widetilde{R}_{11}^{\,5}$) for $SU(3)$-invariant flow, 
compared to the Ricci tensor (\ref{Ricci1}) for $SU(3)\times U(1)_R$-invariant flow. 
Also the 11-dimensional metric (\ref{11d}) generates (\ref{Ricci2})
directly. At the IR fixed point in 4-dimensions, 
the components $\widetilde{R}_4^{\,5},
\widetilde{R}_4^{\,11}, \widetilde{R}_5^{\,4}$ and $\widetilde{R}_{11}^{\,4}$
vanish. 

\subsection{The 4-form field strengths}

The 4-form field strengths satisfying (\ref{fieldequations})
for given Ricci tensor (\ref{Ricci2}) and 11-dimensional metric (\ref{11d}),
in terms of those (\ref{4-form1}) for $SU(3) \times U(1)_R$-invariant flow,  
are summarized as follows:
\bea
\widetilde{F}_{1234} & = & F_{1234}, \qquad
\widetilde{F}_{1235}   =  -\left[\frac{\cos \th \, \sin \th_6}
{\sqrt{1-\sin^2 \th \, \sin^2
    \th_6}} \right] F_{1235}, \nonu \\
\widetilde{F}_{123\,11}  & = &  -\left[\frac{\cos \th_6 \, \sin \th}
{\sqrt{1-\sin^2 \th \, \sin^2
    \th_6}} \right] F_{1235},
\nonu \\
\widetilde{F}_{45mn} & = & -
\left[\frac{\cos \th \, \sin \th_6}
{\sqrt{1-\sin^2 \th \, \sin^2
    \th_6}} \right] F_{45mn} - \left[ \frac{\cos \th_6}
{1-\sin^2 \th \, \sin^2
    \th_6} \right] \left( F_{4mn\,10}- F_{4mn\, 11} \right), \nonu \\
& & (m, n=6, \cdots,  10),
\nonu \\
\widetilde{F}_{4mnp}  & = &  F_{4mnp}, (m, n, p =6, \cdots, 10),
\nonu \\
\widetilde{F}_{4mn\,11} & = & 
-\left[\frac{\sin \th\, \cos \th_6}
{\sqrt{1-\sin^2 \th \, \sin^2
    \th_6}} \right] F_{45mn} + 
\left[ \frac{\sin \th\, \, \cos \th\, \sin \th_6}
{1-\sin^2 \th \, \sin^2
    \th_6} \right] \left( F_{4mn\,10}- F_{4mn\, 11} \right), \nonu \\
& & (m, n=6, \cdots,  10),
\nonu \\
\widetilde{F}_{5mnp} & = &  -\left[\frac{\cos \th \, \sin \th_6}
{\sqrt{1-\sin^2 \th \, \sin^2
    \th_6}} \right] F_{5mnp} -   \left[\frac{\cos \th_6}{1-\sin^2 \th \, \sin^2
    \th_6} \right] F_{mnp\,11}, \nonu \\
& & (m, n, p =6, \cdots, 10),
\label{4-form3}
\\
\widetilde{F}_{5mn\,11} & = & -\left[\frac{5\sin \th +4 \cos 2\th_6 \,
  \sin^3 \th + \sin 3\th}
{8(1-\sin^2 \th \, \sin^2
    \th_6)^{\frac{3}{2}}} \right] \left( F_{5mn\,10} - F_{5mn\, 11}
\right), 
\quad (m, n =6, \cdots, 9),
\nonu \\
\widetilde{F}_{mnp\,11} & = & \left[\frac{\sin \th \, \cos \th_6}
{\sqrt{1-\sin^2 \th \, \sin^2
    \th_6}} \right] F_{5mnp} -   \left[\frac{\sin \th_6 \, \cos
 \th \, \sin \th}
{1-\sin^2 \th \, \sin^2
    \th_6} \right] F_{mnp\,11}, \quad (m, n, p =6, \cdots, 10).
\nonu
\eea
The $\widetilde{F}_{123\,11}$ is new, compared to the $SU(3)\times
U(1)_R$-invariant flow.
At the IR fixed point in 4-dimensions, the components $\widetilde{F}_{1235},
\widetilde{F}_{123\,11},
 \widetilde{F}_{4mnp}$ and
$\widetilde{F}_{45mn}$
vanish.

The 4-form field strengths with upper indices can be obtained from
those with lower indices (\ref{4-form3}) by multiplying the
11-dimensional inverse metric (\ref{11d}) and they are given by as follows: 
\bea
\widetilde{F}^{1234} & = & F^{1234}, \qquad
\widetilde{F}^{1235}   =  -\left[\frac{\cos \th \, \sin \th_6}
{\sqrt{1-\sin^2 \th \, \sin^2
    \th_6}} \right] F^{1235}, \nonu \\
\widetilde{F}^{123\,11}  & = &  -\left[\frac{\csc \th \, \cos \th_6}
{\sqrt{1-\sin^2 \th \, \sin^2
    \th_6}} \right] F^{1235}, 
\nonu \\
\widetilde{F}^{45mn} & = & 
\left[\cos \th_6
 \right] F^{4mn\, 11}
-\left[\frac{\cos \th \, \sin \th_6}
{\sqrt{1-\sin^2 \th \, \sin^2
    \th_6}} \right] F^{45mn}, \qquad (m, n=6, \cdots,  9),
\nonu \\
\widetilde{F}^{45m\,10} & = & 
 \left[\cos \th_6 \right] F^{4m\,10\,11}
-\left[\frac{\cos \th \, \sin \th_6}
{\sqrt{1-\sin^2 \th \, \sin^2
    \th_6}} \right] \left( F^{45m\, 10} + F^{45m\, 11} \right), \nonu \\
&& (m =6, \cdots, 9),
\nonu \\
\widetilde{F}^{45m\,11} & = & \left[
\csc \th \, \sqrt{1-\sin^2 \th \, \sin^2
    \th_6} \right] F^{45m\, 11}, \qquad (m =6, \cdots, 10),
\nonu \\
\widetilde{F}^{4mn\,10} & = & F^{4mn\, 10} +  F^{4mn\, 11}, 
\qquad (m, n =6, \cdots, 9),
\nonu \\
\widetilde{F}^{4mn\,11} & = & 
- \left[\sin
  \th_6 \, \cot \th  \right] F^{4mn\,11}
-\left[\frac{\csc \th \, \cos \th_6}
{\sqrt{1-\sin^2
    \th \, \sin^2 \th_6}} 
\right] F^{45mn}, \qquad (m, n =6, \cdots, 10),
\nonu \\
\widetilde{F}^{5mn\,10} & = & 
 -  \left[\cos \th_6 
\right] F^{mn\,10\,11}
-\left[\frac{\cos \th \, \sin \th_6}
{\sqrt{1-\sin^2 \th \, \sin^2
    \th_6}} \right] \left( F^{5mn\, 10} + F^{5mn\, 11} \right), \nonu \\
&& (m =6, \cdots, 9),
\nonu \\
\widetilde{F}^{5mn\,11} & = & \left[\csc \th \, 
\sqrt{1-\sin^2 \th \, \sin^2 \th_6} 
\right] F^{5mn\, 11}, \qquad (m, n =6, \cdots, 10),
\nonu \\
\widetilde{F}^{mn\,10\,11} & = & - \left[\cot \th \, 
\sin \th_6 \right] F^{mn\, 10\,11}  
 +  \left[\frac{\csc \th \,  \cos \th_6}
{\sqrt{1-\sin^2 \th \, \sin^2 \th_6}}  \right] \left( F^{5mn\,10} + F^{5mn\,11} \right), 
\nonu \\ 
&& (m, n =6, \cdots, 9).
\label{4-form4}
\eea
There exists a new 4-form  
$
\widetilde{F}^{123\,11}$, 
compared to the 4-form field strengths (\ref{4-form2}) for $SU(3) \times
U(1)_R$-invariant flow.
At the IR fixed point in 4-dimensions, the components $\widetilde{F}^{1235},
\widetilde{F}^{123\,11}, \widetilde{F}^{4mnp}$ and
$\widetilde{F}^{45mn}$
vanish.
By substituting the Ricci tensor (\ref{Ricci2}) and 
4-form field strengths (\ref{4-form3}) and (\ref{4-form4}) into the
Einstein equation in (\ref{fieldequations}), one can check that 
the LHS coincides with the RHS exactly. In doing this, 
one uses the fact that the solution characterized by (\ref{Ricci1}),
(\ref{4-form1}) 
and (\ref{4-form2}) for $SU(3) \times U(1)_R$-invariant flow
really satisfies the Einstein equation. Therefore one concludes that 
the Einstein
equation for $SU(3)$-invariant flow is satisfied.  

\subsection{The left hand side of Maxwell equation}

By introducing the notation 
\bea
\frac{1}{2} \, E \,  
\widetilde{\nabla}_M  \, \widetilde{F}^{MNPQ} \equiv (NPQ),
\nonu
\eea
let us present all the nonzero components of left hand side of Maxwell
equations, in terms of 4-forms (\ref{4-form1}), as follows:
\bea
(123) & = & 
-F_{489\,11} F_{567\,10} + F_{489\,10} F_{567\,11}
+  F_{479\,11} F_{568\,10}- F_{479\,10} F_{568\,11} \nonu \\
& + & F_{468\,11} F_{579\,10} - 
F_{468\,10} F_{579\,11} - F_{467\,11} F_{589\,10}   
+  F_{467\,10} F_{589\,11} \nonu \\
& -& F_{4589} F_{67\,10\,11} + F_{4579}
F_{68\,10\,11}  
+ F_{4568} F_{79\,10\,11} - F_{4567} F_{89\,10\,11}, \nonu \\
(467)  & = &  
F_{1235} \, F_{89\,10\,11}, \qquad
(468)  =  -
F_{1235} \, F_{79\,10\,11}, \qquad
(469)  =  
F_{1235} \, F_{78\,10\,11}, \nonu \\
(479)  & = &  -
F_{1235} \, F_{68\,10\,11}, \qquad
(47\,10)  =  
F_{1235}  F_{689\,11}, \qquad
(489)    =   
F_{1235} \, F_{67\,10\,11}, \nonu \\
(48\,10) & = & -
F_{1235}   F_{679\,11}, \qquad
(49\,10)  = 
F_{1235}  F_{678\,11}, \nonu \\
(567) & = & \left[\cos \th_6 \right] 
( F_{1235}  F_{489\,10} - F_{1234}  F_{589\,10}) 
 + \left[
\frac{
 \cos \th \, \sin \th_6}{\sqrt{1-\sin^2 \th \, \sin^2 \th_6}} 
\right] F_{1234} \, F_{89\,10\,11}, \nonu \\
(568) & = & -\left[\cos \th_6 \right] 
( F_{1235}  F_{479\,10} - F_{1234}  F_{579\,10}) 
 - \left[
\frac{
 \cos \th \, \sin \th_6}{\sqrt{1-\sin^2 \th \, \sin^2 \th_6}} \right] 
F_{1234} \, F_{79\,10\,11}, \nonu \\
(569) & = & \left[\cos \th_6 \right] 
( F_{1235}  F_{478\,10} - F_{1234}  F_{578\,10}) 
 + \left[
\frac{
 \cos \th \, \sin \th_6}{\sqrt{1-\sin^2 \th \, \sin^2 \th_6}} \right]
F_{1234} \, F_{78\,10\,11}, \nonu \\
(579) & = & -\left[\cos \th_6 \right] 
(F_{1235}  F_{468\,10} - F_{1234}  F_{568\,10}) 
 - \left[ \frac{
 \cos \th \, \sin \th_6}{\sqrt{1-\sin^2 \th \, \sin^2 \th_6}}\right] 
F_{1234}  \, F_{68\,10\,11},
\nonu \\
(57\,10) & = &  \left[\cos \th_6 \right] (F_{1235}\, F_{4689}
-F_{1234}\, F_{5689})
+\left[
   \frac{\cos \th \, \sin \th_6}{ \sqrt{1-\sin^2 \th \,
  \sin^2 \th_6}}
\right] F_{1234} \, F_{689\,11},  
\nonu \\
(589) & = & \left[\cos \th_6 \right] 
(F_{1235}  F_{467\,10} - F_{1234}  F_{567\,10}) 
 +  \left[\frac{
 \cos \th \, \sin \th_6}{\sqrt{1-\sin^2 \th \, \sin^2 \th_6}} 
\right] F_{1234} \,  F_{67\,10\,11},
\nonu \\
(58\,10) & = &  -     
\left[\cos \th_6 \right] (F_{1235}\, F_{4679} -F_{1234}\, F_{5679})
   -\left[\frac{\cos \th \, \sin \th_6}{ \sqrt{1-\sin^2 \th \,
  \sin^2 \th_6}}
\right] F_{1234} \, F_{679\,11},  
\nonu \\
(59\,10) & = & 
\left[\cos \th_6 \right] (F_{1235}\, F_{4678} -F_{1234}\, F_{5678})+
  \left[\frac{\cos \th \, \sin \th_6}{ \sqrt{1-\sin^2 \th \,
  \sin^2 \th_6}}
\right] F_{1234} \, F_{678\,11},  
\nonu \\
(67\,10) & = & 
F_{1235} 
\left( F_{489\,10} - F_{489\,11}\right) + F_{1234} 
\left( -F_{589\,10} 
+ F_{589\,11} \right),
\nonu \\
(67\,11) & = & \left[\cot \th \,  \sin \th_6 \right] 
\left(-F_{1235}  F_{489\,10} + F_{1234}  F_{589\,10}\right) 
 +  \left[\frac{
 \csc \th \, \cos \th_6}{\sqrt{1-\sin^2 \th \, \sin^2 \th_6}} 
\right] F_{1234} \,  F_{89\,10\,11},
\nonu \\
(68\,10) & = & 
F_{1235} \left(- F_{479\,10} + F_{479\,11} \right) 
+ F_{1234} \left( F_{579\,10} -  F_{579\,11}\right),
\nonu \\
(68\,11) & = & \left[\cot \th  \sin
\th_6  \right]
(F_{1235}  F_{479\,10} - F_{1234}  F_{579\,10}) 
-  \left[\frac{
 \csc \th \, \cos \th_6}{\sqrt{1-\sin^2 \th \, \sin^2 \th_6}} 
\right] F_{1234} \,  F_{79\,10\,11},
\nonu \\
(69\,10) & = & 
F_{1235} ( F_{478\,10} - F_{478\,11}) + F_{1234} 
( -F_{578\,10}
+ F_{578\,11} ),
\nonu \\
(69\,11) & = & \left[\cot \th  \sin \th_6 \right] 
(-F_{1235}  F_{478\,10} + F_{1234}  F_{578\,10}) 
 +  \left[\frac{
 \csc \th \, \cos \th_6}{\sqrt{1-\sin^2 \th \, \sin^2 \th_6}} 
\right] F_{1234}  \, F_{78\,10\,11},
\nonu \\
(79\,10) & = & 
F_{1235} (- F_{468\,10} + F_{468\,11}) + F_{1234} ( F_{568\,10} - 
F_{568\,11} ),
\nonu \\
(79\,11) & = & \left[\cot \th  \sin \th_6 \right] 
(F_{1235}  F_{468\,10} - F_{1234}  F_{568\,10}) 
 - \left[\frac{
 \csc \th \, \cos \th_6}{\sqrt{1-\sin^2 \th \,  \sin^2 \th_6}} 
\right] F_{1234}\,  F_{68\,10\,11},
\nonu \\
(7\,10\,11) & = & \left[\cot \th  \sin \th_6 \right]
(-F_{1235}  F_{4689} + F_{1234}  F_{5689}) 
 +  \left[\frac{
 \csc \th  \, \cos \th_6}{\sqrt{1-\sin^2 \th \, \sin^2 \th_6}} 
\right] F_{1234} \, F_{689\,11},
\nonu \\
(89\,10) & = & 
F_{1235} ( F_{467\,10} -F_{467\,11} ) + 
F_{1234} ( -F_{567\,10} + F_{567\,11} ),
\nonu \\
(89\,11) & = & \left[\cot \th  \sin \th_6 \right] 
(-F_{1235}  F_{467\,10} + F_{1234}  F_{567\,10}) 
 +  \left[\frac{
 \csc \th \, \cos \th_6}{\sqrt{1-\sin^2 \th \, \sin^2 \th_6}} 
\right] F_{1234} \,  F_{67\,10\,11},
\nonu \\
(8\,10\,11) & = & \left[\cot \th  \sin \th_6 \right] 
(F_{1235}  F_{4679} - F_{1234}  F_{5679}) 
 -  \left[\frac{
 \csc \th \, \cos \th_6}{\sqrt{1-\sin^2 \th \, \sin^2 \th_6}} 
\right] F_{1234}\,  F_{679\,11},
\label{max}
\\
(9\,10\,11) & = & \left[\cot \th  \sin \th_6 \right] 
(-F_{1235}  F_{4678} + F_{1234}  F_{5678}) 
 +  \left[\frac{
 \csc \th \, \cos \th_6}{\sqrt{1-\sin^2 \th \, \sin^2 \th_6}} 
\right] F_{1234} \,  F_{678\,11}.
\nonu
\eea
One can easily check that 
the RHS of Maxwell equation (\ref{fieldequations})
with (\ref{4-form1}) and (\ref{11d}) is exactly coincident with the
above
LHS of Maxwell equation (\ref{max}) as we explained in the section 2.

\section{$SU(2) \times U(1) \times U(1)_R$-invariant flow}

In this Appendix, we summarize the Ricci tensor and the 4-form field
strengths for $SU(2) \times U(1) \times U(1)_R$-invariant flow 
\cite{Ahn0909}.

\subsection{The Ricci tensor}

The nonzero Ricci tensor in the coordinate basis from the 11-dimensional
metric \cite{Ahn0909}, after imposing the flow
equations (\ref{flow}), can be written as follows:
\bea
R_1^{\, 1} & = & -\frac{1}{L^2\, u^{\frac{2}{3}} \,
  v^{\frac{4}{3}} \, (c_{\mu}^2  + u^2\, s^2_{\mu} )^{\frac{8}{3}}} 
24 \left[ 2 u^8 v^2 (v^2-1)s_{\mu}^4 +
2 v^2 (v^2 +3) c_{\mu}^4 \right. \nonu \\
& + &   u^6 
\left[-2(-5+c_{2\mu})+ v^2(-11+c_{2\mu}) + 4v^4 c_{\mu}^2 \right]s_{\mu}^2
\nonu \\
&+ & \left. u^2 \left[12c_{\mu}^2 +v^2(9-13c_{2\mu})+ 
4 v^4 s_{\mu}^2\right]c_{\mu}^2
+u^4\left[6s_{2\mu}^2+v^2(5-8c_{2\mu}+5c_{4\mu}) +
v^4 s_{2\mu}^2 \right] \right]
\nonu \\
 & = & R_2^{\, 2} =R_3^{\, 3}
= -2 R_6^{\, 6} = -2 R_7^{\, 7} = -2
R_8^{\, 8}=-2 R_9^{\, 9},
\nonu \\
R_4^{\, 4} & = &
\frac{6}{ \, L^2\, u^{\frac{2}{3}} \,
  v^{\frac{10}{3}} \, (c_{\mu}^2  + u^2\, s^2_{\mu} )^{\frac{8}{3}}} 
\left[ -4 v^2 (2v^4-21v^2+27)c_{\mu}^4 -
4u^8 v^2 (2v^4-5v^2 +3) s_{\mu}^4 \right. \nonu \\
& + &   2u^2 v^2 
\left[-48+60c_{2\mu}+ v^2(15-43c_{2\mu}) + 4v^4 s_{\mu}^2 \right]c_{\mu}^2
 \nonu \\
& - &  2 u^6 \left[24c_{\mu}^2 -4v^2(7+4c_{2\mu})+
v^4 (17+ 5c_{2\mu})-4v^6 c_{\mu}^2 \right] s_{\mu}^2
\nonu \\
&- & \left. u^4 v^2 \left(33-48c_{2\mu}+27c_{4\mu} +4v^2
\left[2+4c_{2\mu}-7c_{4\mu}+v^2(-1+c_{4\mu})\right] \right) \right],  
\nonu \\
R_4^{\, 5} & = &  
\frac{18}{{ L^3 \,
  (c^2_{\mu} + u^2  s^2_{\mu} )^{\frac{8}{3}}}}
u^{\frac{5}{6}}  \, v^{\frac{2}{3}}\,
\left(-2c_{\mu}^2(v^2-3) \right. \nonu \\
& + & \left. u^2\left[-5-11c_{2\mu}+14v^2c_{\mu}^2+
u^2\left(-11+9c_{2\mu}+
2s_{\mu}^2\left[u^2 -v^2(u^2-7)\right]\right)\right] \right) s_{2\mu}, 
\nonu \\
R_5^{\, 4} & = &  
\frac{18}{{ L \, u^{\frac{1}{6}} \, v^{\frac{4}{3}}\,
  (c^2_{\mu} + u^2  s^2_{\mu} )^{\frac{8}{3}}}}
\left(-2c_{\mu}^2(v^2-3) \right. \nonu \\
& + & \left. u^2\left[-5-11c_{2\mu}+14v^2c_{\mu}^2+
u^2\left(-11+9c_{2\mu}+
2s_{\mu}^2\left[u^2 -v^2(u^2-7)\right]\right)\right] \right) s_{2\mu}, 
\nonu  \\
R_5^{\, 5} & = & \frac{6}{  \,L^2 \, u^{\frac{2}{3}} \,
  v^{\frac{4}{3}} \, (c_{\mu}^2  + u^2\, s^2_{\mu} )^{\frac{8}{3}}} 
\left[ 4u^8 v^2 (v^2-1)s_{\mu}^4 +
4v^2 (v^2+3) c_{\mu}^4 \right. \nonu \\
& + &   u^4 \left[6 - 6c_{4\mu} +v^2 
(19-16c_{2\mu}+ c_{4\mu})+ 5v^4(-1+c_{4\mu})\right] \nonu \\
& + & \left. 4 u^2 \left[6c_{\mu}^2 +v^2(3-5c_{2\mu})+
2v^4 s_{\mu}^2\right]c_{\mu}^2
+4 u^6\left[-1-v^2+v^4+(-7+5v^2 +v^4)c_{2\mu} \right] s_{\mu}^2 \right], 
\nonu \\
R_7^{\, 10} & = &
\frac{36(y-1) c_{\th_1} \, c_{\mu}^2 \, u^{\frac{4}{3}}\, 
(v^2-1)}{  L^2 \,
v^{\frac{10}{3}} \, (c_{\mu}^2  + u^2\, s^2_{\mu} )^{\frac{11}{3}}}
\left[ 12 c_{\mu}^4 v^2 + u^2 ( s_{2\mu}^2(3+ 5 v^2) + 2s_{\mu}^2 u^2 (4-2c_{2\mu} \right.   
\nonu \\
&+ & \left.
(3-9c_{2\mu}) v^2 + 2s_{\mu}^2 ( -4 v^4 + u^2(-3 +v^2)))) \right],
\nonu \\
R_7^{\, 11} & = &
\frac{48(y-1) c_{\th_1} \, c_{\mu}^2 \, u^{\frac{10}{3}}\, 
(v^2-1)}{  L^2 \,
v^{\frac{10}{3}} \, (c_{\mu}^2  + u^2\, s^2_{\mu} )^{\frac{11}{3}}}
\left[ - s_{\mu}^4 u^6 + 2c_{\mu}^4 v^4 - s_{\mu}^2 u^4 
(-3+ ( 1+ 2c_{2\mu}) v^2) \right. \nonu \\
&+ & \left.
c_{\mu}^2  
u^2 ( 3s_{\mu}^2 + (-1 +2c_{2\mu}) v^2 + 6s_{\mu}^2 v^4) \right],
\nonu \\
R_9^{\, 10} & = &
\frac{36 y \, c_{\mu}^2 \, u^{\frac{4}{3}}\, 
(v^2-1)}{  L^2 \,
v^{\frac{10}{3}} \, (c_{\mu}^2  + u^2\, s^2_{\mu} )^{\frac{11}{3}}}
\left[ 12 c_{\mu}^4 v^2 + u^2 ( s_{2\mu}^2 ( 3+ 5 v^2) + 
2s_{\mu}^2 u^2 ( 4-2c_{2\mu}  \right. \nonu \\
&+ & \left.
 (3-9 c_{2\mu}) v^2 + 2s_{\mu}^2 ( -4 v^4 + u^2 ( -3 + v^2)) ))
\right],
\nonu \\
R_9^{\, 11} & = &
\frac{48 y \, c_{\mu}^2 \, u^{\frac{10}{3}}\, 
(v^2-1)}{  L^2 \,
v^{\frac{10}{3}} \, (c_{\mu}^2  + u^2\, s^2_{\mu} )^{\frac{11}{3}}}
\left[ 
- s_{\mu}^4 u^6 + 2c_{\mu}^4 v^4 - s_{\mu}^2 u^4 
(-3+ ( 1+ 2c_{2\mu}) v^2) \right. \nonu \\
&+ & \left.
c_{\mu}^2  
u^2 ( 3s_{\mu}^2 + (-1 +2c_{2\mu}) v^2 + 6s_{\mu}^2 v^4)
\right],
\nonu \\
R_{10}^{\, 10} & = & \frac{3\left[-1+2v(
- 3 + 4v^2)(-\sqrt{v^2-1} +v(-3+4v(v+\sqrt{v^2-1}))) \right]}
{L^2 \, u^{\frac{2}{3}} \,
  v^{\frac{10}{3}} \, (c_{\mu}^2  + u^2\, 
s^2_{\mu} )^{\frac{11}{3}}\,
(v+ \sqrt{v^2-1})^6} 
\left[ \right. \nonu \\
&& 8s_{\mu}^6 u^{10} v^4 (v^2-1) + 8 c_{\mu}^6
  v^4 (v^2+3) + 8c_{\mu}^4 u^2 v^2 ( -12 c_{\mu}^2 +
(15 + c_{2\mu}) v^2 + 3 s_{\mu}^2 v^4 )   \nonu \\
& + &   8 s_{\mu}^4 u^8   
(18 c_{\mu}^2 - (7 + 13c_{2\mu}) v^2 - 
6 s_{\mu}^2 v^4 + 3 c_{\mu}^2 v^6)  
\nonu \\
& + & 
s_{\mu}^2 u^6 ( 48 c_{\mu}^2 (-2 + c_{2\mu})+
224 c_{\mu}^4 v^2 +(5-124c_{2\mu} -57c_{4\mu}) v^4 +
8(-2+c_{2\mu} + 3c_{4\mu})v^6)
\nonu \\
&+ & \left.
c_{\mu}^2 u^4(-36 s_{2\mu}^2 + 12s_{2\mu}^2 v^2 + (81-
76 c_{2\mu} + 3c_{4\mu}) v^4 + 16 s_{\mu}^2 v^6) \right], 
\nonu \\
R_{10}^{\, 11} & = &
\frac{48 \, c_{\mu}^2 \, u^{\frac{10}{3}}\, 
(v^2-1)}{  L^2 \,
v^{\frac{10}{3}} \, (c_{\mu}^2  + u^2\, s^2_{\mu} )^{\frac{11}{3}}}
\left[ - s_{\mu}^4 u^6 + 2c_{\mu}^4 v^4 - s_{\mu}^2 u^4 
(-3+ ( 1+ 2c_{2\mu}) v^2) \right. \nonu \\
&+ & \left.
c_{\mu}^2  
u^2 ( 3s_{\mu}^2 + (-1 +2c_{2\mu}) v^2 + 6s_{\mu}^2 v^4) \right],
\nonu \\
R_{11}^{\, 10} & = &
\frac{108 \, s_{\mu}^2 \, u^{\frac{4}{3}}\, 
(v^2-1)}{  L^2 \,
v^{\frac{10}{3}} \, (c_{\mu}^2  + u^2\, s^2_{\mu} )^{\frac{11}{3}}}
\left[ - 4 s_{\mu}^4 u^6 v^2 + 4c_{\mu}^4(-3 +5 v^2) +
2s_{\mu}^2 u^4 
(10 c_{\mu}^2 +(1-11c_{2\mu} ) v^2 \right. \nonu \\
&- & \left.
4 s_{\mu}^2 v^4) +  
u^2 ( 2(c_{2\mu} + c_{4\mu}) + s_{2\mu}^2 v^2 ( 7 + 2v^2)) 
\right],
\nonu \\
R_{11}^{\, 11} & = & \frac{3\left[-1+2v(
- 3 + 4v^2)(-\sqrt{v^2-1} +v(-3+4v(v+\sqrt{v^2-1}))) \right]}
{L^2 \, u^{\frac{2}{3}} \,
  v^{\frac{10}{3}} \, (c_{\mu}^2  + u^2\, 
s^2_{\mu} )^{\frac{11}{3}}\,
(v+ \sqrt{v^2-1})^6} 
\left[ \right. \nonu \\
&& 8s_{\mu}^6 u^{10} v^4 (v^2-1) + 8 c_{\mu}^6
  v^4 (v^2+3) + 8c_{\mu}^4 u^2 v^2 (6 c_{\mu}^2 +
(6 - 8 c_{2\mu}) v^2 + 3 s_{\mu}^2 v^4 )   
\nonu \\
& + &   8 s_{\mu}^4 u^8   
(-18 c_{\mu}^2 + 4(2 + 5c_{2\mu}) v^2 - 
12 c_{2\mu} v^4 + 3 c_{\mu}^2 v^6)  
\nonu \\
&+ & 
s_{\mu}^2 u^6 (-48 c_{\mu}^2 (-2 + c_{2\mu}) - 8 c_{\mu}^2
(19 + c_{2\mu}) v^2 \nonu \\
&+ & ( 47 + 20 c_{2\mu} + 45 c_{4\mu}) v^4 + 8 ( 4 + c_{2\mu}-3
c_{4\mu} ) v^6) \nonu \\
& + & \left.
c_{\mu}^2 u^4 ( 36 s_{2\mu}^2 - 48 s_{2\mu}^2 v^2 + (75
-76 c_{2\mu} + 9 c_{4\mu} ) v^4 + 16 s_{\mu}^2 v^6) \right], 
\label{Ricci5}
\eea
where we use a simplified notation (\ref{uv}).
There are additional
nonzero Ricci tensor components ($R_{7}^{\,10}, R_{7}^{\,11}, R_{9}^{\,10}$
and $R_{9}^{\,11}$) that depend on the internal coordinates $\th_1$ or
$y$, as
compared to the Ricci tensor in the frame basis \cite{Ahn0909}. 
One also obtains (\ref{Ricci5}) directly from the Ricci tensor in the
frame basis with the help of the vielbeins.  
At the IR fixed point in 4-dimensions, the off-diagonal components $R_{4}^{\,5}$ and
$R_{5}^{\,4}$
vanish.

\subsection{The 4-form field strengths}

The nonzero 4-form field strengths satisfying (\ref{fieldequations})
for given Ricci tensor (\ref{Ricci5}) and 11-dimensional metric
presented in \cite{Ahn0909}
are summarized as follows:
\bea
F_{1234} & = & 
\left[\frac{3 \, e^{3A}\, \cosh^2 \chi}{L \, \rho^4} \right]
\left[ c_{\mu}^2 (-5 + \cosh 2\chi) + 2\rho^8 (-2 +
c_{2\mu}+\rho^8 \, s_{\mu}^2 \, \sinh^2 \chi) \right],
\nonu \\
F_{1235} & = & \left[
\frac{3 \, e^{3A} }{2\, \rho^2} \right]
\left[ 1+ \cosh 2 \chi + \rho^8 \, (-3 + \cosh 2 \chi)  
\right]  \, s_{2\mu},
\nonu \\
F_{4567} & = & \left[
\frac{L^2 \,  
(\rho^8 -3) \, \tanh \chi }{ 6\rho^2}
\right] \sqrt{\frac{3}{2} \,(1-y) \, w \, q} \,\, c_{\th_1} \, c_{\mu}^2 \, 
s_{\alpha+\psi} = c_{\th_1} \, 
F_{4569} 
 =   \frac{w \, q}{6} \, 
s_{\th_1}^{-1} \, c_{\th_1} \, F_{4578}, 
\nonu \\
F_{4568} & = &- 
\left[\frac{L^2 \,  
(\rho^8 -3) \, \tanh \chi }{ \rho^2}
\right] \sqrt{\frac{3(1-y)}{2 \,w \, q}}  \, c_{\mu}^2 \, 
c_{\alpha+\psi} = -  \frac{6}{w \, q} \, 
s_{\th_1}^{-1} \, F_{4579}, 
\nonu \\
F_{4678} & = &- 
\left[\frac{L^2 \, \rho^6 \, \mbox{sech}^2 \chi \, \sinh 2\chi}{8 \, 
 (c_{\mu}^2 +
\rho^8 \, s_{\mu}^2)^2} \right] 
\nonu \\
&\times & \left[
 2c_{\mu}^2 (-2 + \cosh 2\chi) + \rho^8 ( -5 + c_{2\mu} +
2c_{\mu}^2 \, \cosh 2 \chi + 2 \rho^8 \, s_{\mu}^2) 
\right] \, 
\frac{(1-y)^{\frac{3}{2}}}{\sqrt{\frac{3}{2} \, w \, q} } \, 
c_{\th_1} \, c_{\mu}^3 \, s_{\mu} \, s_{\alpha +\psi} \nonu 
\\
& = &
\frac{(1-y)}{y} \, c_{\th_1} \,
F^{4689} = (1-y) \, c_{\th_1} \,
F^{468\,10} = -\frac{6(1-y)}{w\,q} \, s_{\th_1}^{-1} \,
c_{\th_1} \,
F^{479\,10},
\nonu \\
F_{4679} & = &-
\left[\frac{L^2 \, \rho^6 \, \tanh \chi }
{24  \, (c_{\mu}^2 +
\rho^8 \, s_{\mu}^2)^2} \right] \nonu \\
& \times & \left[
 2c_{\mu}^2 (-2 + \cosh 2\chi) + \rho^8 ( -5 + c_{2\mu} +
2c_{\mu}^2 \, \cosh 2 \chi + 2 \rho^8 \, s_{\mu}^2) 
\right] \nonu \\
& \times & \sqrt{\frac{1}{6}(1-y) \, w\, q} \, 
c_{\th_1} \, c_{\mu}^2\, s_{2\mu} \, c_{\alpha +\psi} = 
F_{467\,10} = c_{\th_1}\, F_{469\,10} = \frac{1}{6} 
y^{-1} \, w \, q\, c_{\th_1}
\, s_{\th_1}^{-1}\, F_{4789} \nonu \\
& = & \frac{1}{6} 
\, w \, q\, c_{\th_1}
\, s_{\th_1}^{-1}\, F_{478\,10},
\nonu \\
F_{467\,11} & = & \left[
\frac{L^2 \, \rho^6 \, \tanh \chi }
{4 \, \rho^2 \, (c_{\mu}^2 +
\rho^8 \, s_{\mu}^2)^2}\right] \nonu \\
& \times & \left[
 -3c_{\mu}^2  + \rho^8 ( c_{\mu}^2 -(4+
\cosh 2\chi )s_{\mu}^2 - \rho^8 \, (-4 + \cosh 2 \chi)
\, s_{\mu}^2 )
\right] \nonu \\
& \times & \sqrt{\frac{1}{6} (1-y) \, w\, q} \, 
c_{\th_1} \, s_{2\mu} \, c_{\mu}^2 \, c_{\alpha +\psi} =
c_{\th_1} \, F_{469\,11} = \frac{1}{6} w\, q\, 
c_{\th_1} \, s_{\th_1}^{-1} \, F_{478\,11},
\nonu \\
F_{468\,11} & = &
\left[\frac{L^2 \, \tanh \chi }
{4 \, \rho^2 \, (c_{\mu}^2 +
\rho^8 \, s_{\mu}^2)^2} \right] \nonu \\
& \times & \left[
 -3c_{\mu}^2  + \rho^8 ( c_{\mu}^2 -(4+
\cosh 2\chi )s_{\mu}^2 - \rho^8 \, (-4 + \cosh 2 \chi)
\, s_{\mu}^2 )
\right] \nonu \\
& \times & \sqrt{\frac{6(1-y)}{w\,q}} \, 
s_{2\mu} \, c_{\mu}^2 \, s_{\alpha +\psi} =  -\frac{6}{w\, q} \,
s_{\th_1}^{-1} \,
F_{479\,11},
\nonu \\
F_{5678} & = &- \left[
\frac{L^3 \, \tanh \chi }
{4  \, (c_{\mu}^2 +
\rho^8 \, s_{\mu}^2)^3} \right] \left[
 6c_{\mu}^4  + \rho^8 c_{\mu}^2 (9-
7 c_{2\mu} ) + 10  \rho^{16} \, s_{\mu}^4-2 \rho^{24}
\, s_{\mu}^4 
\right] \nonu \\
& \times & \sqrt{\frac{2}{3}} \, \frac{(1-y)^{\frac{3}{2}}}{\sqrt{w\,q}} \, 
c_{\mu}^4 \, c_{\th_1} \, s_{\alpha +\psi} = 
y^{-1} \, (1-y) \, c_{\th_1}\,
 F_{5689} \nonu \\
& = &  
(1-y) \, c_{\th_1}  \, F_{568\,10} = 
-\frac{6}{w\,q} (1-y)  \, c_{\th_1} \,
s_{\th_1}^{-1} \, F_{579\,10},
\nonu \\
F_{5679} & = &- \left[
\frac{L^3 \, \tanh \chi }
{36   \, (c_{\mu}^2 +
\rho^8 \, s_{\mu}^2)^3}\right] \left[
 6c_{\mu}^4  +  \rho^8 c_{\mu}^2 (9-
7 c_{2\mu} ) + 10  \rho^{16} \, s_{\mu}^4 - 2 \rho^{24}
\, s_{\mu}^4 
\right] \nonu \\
& \times & \sqrt{\frac{3}{2}\, (1-y)\, w \,q} \, 
c_{\mu}^4 \, c_{\th_1} \, c_{\alpha +\psi} = F_{567\,10}=
 \, c_{\th_1} \, F_{569\,10} \nonu \\
& = & \frac{1}{6}  y^{-1} \,
w\, q \, s_{\th_1}^{-1} \, c_{\th_1} \, F_{5789} = 
\frac{1}{6} w\, q\, s_{\th_1}^{-1} \, c_{\th_1}\, F_{578\,10},
\nonu \\
F_{567\,11} & = &- \left[
\frac{L^3 \, \tanh \chi }
{6  \, (c_{\mu}^2 +
\rho^8 \, s_{\mu}^2)^2} \right] \left[
c_{\mu}^2  +  3 \rho^8 + \rho^{16} s_{\mu}^2
\right] \, \sqrt{\frac{3}{2}\, (1-y)\, w \,q} \, 
s_{\mu}^2 \, c_{\mu}^2 \, c_{\th_1} \, 
c_{\alpha +\psi} \nonu \\
& = & c_{\th_1} \, F_{569\,11} = 
\frac{1}{6} w\,q \, c_{\th_1} \,
s_{\th_1}^{-1} \, F_{578\,11},
\nonu \\
F_{568\,11} & = &- \left[
\frac{L^3 \, \tanh \chi }
{  (c_{\mu}^2 +
\rho^8 \, s_{\mu}^2)^2}\right] \left[
c_{\mu}^2  +  3 \rho^8 + \rho^{16} s_{\mu}^2
\right] \, \sqrt{\frac{3(1-y)}{2 w\,q }} \, 
s_{\mu}^2 \, c_{\mu}^2  
\, s_{\alpha +\psi} = -\frac{6}{w\,q} 
\, s_{\th_1}^{-1}\, F_{579\,11},
\nonu \\
\nonu \\
F_{678\,11} & = & -\left[
\frac{L^3 \, (\rho^8+3) \, \tanh \chi }{ \,  
(c_{\mu}^2 + \rho^8 \, s_{\mu}^2)} \right]
\frac{(1-y)^{\frac{3}{2}}}{\sqrt{6
    w\, q} }  \, c_{\th_1} \,
c_{\mu}^3 \, s_{\mu} \, c_{\alpha +\psi} = \frac{(1-y)}{y} \,
c_{\th_1} \, F_{689\,11} \nonu \\
& = & (1-y) \, c_{\th_1} \, F_{68\,10\,11}=
-\frac{6(1-y)}{w\,q} \, 
c_{\th_1} \, s_{\th_1}^{-1} \, F_{79\,10\,11},  
\nonu \\
F_{679\,11} & = &  -\left[\frac{L^3 \, (\rho^8+3) \, \tanh \chi }{ 18  \, 
(c_{\mu}^2 + \rho^8 \, s_{\mu}^2)} \right]
\sqrt{\frac{3}{2} (1-y)\, w \, q}
\, 
(1-y)^{\frac{3}{2}}\, c_{\th_1} \,
c_{\mu}^3 \, s_{\mu} \, s_{\alpha +\psi} 
 =  F_{67\,10\,11} \nonu \\
&= & c_{\th_1} \, F_{69\,10\,11}=
 \frac{1}{6} c_{\th_1} \, s_{\th_1}^{-1} \, y^{-1} 
\, w\, q\, F_{789\,11}  =  \frac{1}{6} c_{\th_1} \, s_{\th_1}^{-1} \, w\, q\, 
F_{78\,10\,11}.
\label{4-form9}
\eea
One sees that the 4-forms (\ref{4-form9}) has the dependence on 
$(\al + \psi)$.
According to the shifts $\al \rightarrow \al + \gamma $ and $\psi
\rightarrow \psi - \gamma$, corresponding to the $U(1)_R$ charge, 
it is evident that 4-forms 
preserve this $U(1)_R$ charge. 
At the IR fixed point in 4-dimensions, the components $F_{1235}, F_{4mnp}$ and
$F_{45mn}$
vanish.
By using the 4-form field strengths in the
frame basis \cite{Ahn0909} and vielbeins, one also obtains (\ref{4-form9}).

The 4-form field strengths with upper indices can be obtained from
those with lower indices (\ref{4-form9}) by multiplying the
11-dimensional inverse metric \cite{Ahn0909} and they are given by as follows:  
\bea
F^{1234} & = & -\left[\frac{3 e^{-3A} \, \rho^{\frac{4}{3}}}
{L \, \cosh^{\frac{10}{3}} \chi \, (c^2_{\mu} +\rho^8 \,
  s^2_{\mu})^{\frac{8}{3}}} 
\right] \left[ 
c^2_{\mu} (-5 + \cosh 2\chi)+ 2\rho^8 (-2 +c_{2\mu} + \rho^8 \, s^2_{\mu} \,
\sinh^2 \chi)\right], 
\nonu \\
F^{1235} & = &  -\left[\frac{3  \, e^{-3A} \, 
\rho^{\frac{22}{3}}\,  \mbox{sech}^{\frac{10}{3}} \chi}
{2 \,L^2 \, (c^2_{\mu} +\rho^8 \,
  s^2_{\mu})^{\frac{8}{3}}} \right]  \left[
1+ \cosh2\chi + \rho^8 (-3 +\cosh2\chi)
\right] \,  s_{2\mu}, 
\nonu \\
F^{4568} & = &  
 -\left[ \frac{6   
\, (-3 + \rho^8) \,
 \tanh \chi \,  }{L^4    \, \rho^{\frac{2}{3}}
\, \mbox{sech}^{\frac{2}{3}} 
\chi  \, (c^2_{\mu} +\rho^8 \,
  s^2_{\mu})^{\frac{2}{3}} }  \right]\, 
\sqrt{\frac{3\, w\,q}{2(1-y)}} \, c_{\al
+\psi} \, 
c_{\mu}^{-2} = - \frac{1}{6} w\, q\, s_{\th_1} \, F^{4579},
\nonu \\ 
& = & \frac{1}{6y} w\, q\, s_{\th_1} \, F^{457\,10}=
\frac{1}{6(1-y)} s_{\th_1} \, c_{\th_1}^{-1} \, F^{459\,10},
\nonu \\
F^{4569} & = &  
 \left[ \frac{18  \, (-3 + \rho^8) \,
 \tanh \chi \,  }{L^4   \, \rho^{\frac{2}{3}}
\, \mbox{sech}^{\frac{2}{3}} 
\chi  \, (c^2_{\mu} +\rho^8 \,
  s^2_{\mu})^{\frac{2}{3}} }  \right]\, 
\sqrt{\frac{6}{(1-y) \, w\,q}} 
\, s_{\al +\psi} \, 
c_{\mu}^{-2} = - y^{-1} \, F^{456\,10} 
\nonu \\
& = & \frac{6}{ w\,q} 
s_{\th_1} \, F^{4578} = 
\frac{6}{ w \, q} s_{\th_1} \, c_{\th_1}^{-1} \,
F^{4589} = - \frac{6}{ w\,q} 
s_{\th_1} \, c_{\th_1}^{-1} \, F^{458\,10},
\nonu \\
F^{468\,10} & = & 
 - 
\left[ \frac{9  }{L^4   \, \rho^{\frac{2}{3}}
\, \cosh^{\frac{4}{3}} 
\chi  \, (c^2_{\mu} +\rho^8 \,
  s^2_{\mu})^{\frac{5}{3}} }  \right]\, 
\left[-\sinh 2\chi + \rho^8 ( \sinh 2\chi ( 1- 3 s_{\mu}^2
\, c_{\mu}^{-2}) \right. \nonu \\
& + &  \left. 
2(-3 +\cosh 2\chi) \tanh \chi + \rho^8 \,
\sinh 2\chi
\, s_{\mu}^2 \, c_{\mu}^{-2} )  \right] \,
\sqrt{\frac{3 \, w\,q }{2(1-y)}} 
\, s_{\al +\psi} \, 
s_{\mu} \, c_{\mu}^{-1}  \nonu \\
& = &   - 
\frac{1}{6} \, w\, q\, s_{\th_1} \, F^{479\,10},
\nonu \\
F^{468\,11} & = & 
- \left[ \frac{3 \, \mbox{sech}^{\frac{4}{3}}
\chi }{L^4   \, \rho^{\frac{2}{3}}
\, (c^2_{\mu} +\rho^8 \,
  s^2_{\mu})^{\frac{5}{3}} }  \right]\, 
\left[3 c_{\mu} \, s_{\mu}^{-1} \,
\sinh 2\chi + \rho^8 \, s_{\mu}^{-1} \,
c_{\mu}^{-1} \, \sinh 2\chi 
( 2- 3 c_{2\mu}) \right. \nonu \\
& + &  \left. \rho^{16} \,
(-7 +\cosh 2\chi) s_{\mu} \, c_{\mu}^{-1} \,
\tanh \chi    \right] \,
\sqrt{\frac{3 \, w\,q }{2(1-y)}} 
\, s_{\al +\psi} \, 
 \nonu \\
& = & - \frac{1}{6} w\, q\, s_{\th_1} \, 
 F^{479\,11} = 
\frac{1}{6 y} w\, q\, s_{\th_1} \,
F^{47\,10\,11} = \frac{w\, q}{6(1-y)}\, s_{\th_1} \, c_{\th_1}^{-1}  
\, F^{49\,10\,11},
\nonu \\
F^{469\,10} & = & 
-\left[ \frac{27  }{L^4   \, \rho^{\frac{2}{3}}
\, \cosh^{\frac{4}{3}} 
\chi  \, (c^2_{\mu} +\rho^8 \,
  s^2_{\mu})^{\frac{5}{3}} }  \right]\, 
\left[-\sinh 2\chi + \rho^8 ( \sinh 2\chi ( 1- 3 s_{\mu}^2
\, c_{\mu}^{-2}) \right. \nonu \\
& + &  \left. 
2(-3 +\cosh 2\chi) \tanh \chi + \rho^8 \,
\sinh 2\chi
\, s_{\mu}^2 \, c_{\mu}^{-2} )  \right] \,
\sqrt{\frac{6 }{(1-y)\, w\, q}} 
\, c_{\al +\psi} \, 
s_{\mu} \, c_{\mu}^{-1} \nonu \\
& =& \frac{6}{w\,q} \, s_{\th_1} \, F^{478\,10} =
\frac{6}{w\,q} \, s_{\th_1} \, c_{\th_1}^{-1} \, F^{489\,10},
\nonu \\
F^{469\,11} & = & -
\left[\frac{18  \, \mbox{sech}^{\frac{4}{3}} \chi }
{L^4 \, \rho^{\frac{2}{3}} \,
(c^2_{\mu} +\rho^8 \, s^2_{\mu})^{\frac{5}{3}}} 
\right] 
\left[ 3 c_{\mu} \, s_{\mu}^{-1} \, \cosh \chi \, \sinh \chi 
+ \rho^8 (-3 c_{2\mu} \, s_{2\mu}^{-1} + s_{\mu}^{-1} \, 
c_{\mu}^{-1} )
\, \sinh 2\chi \right.
\nonu \\
&+ & \left. \rho^{16}\, (-3 +
\sinh^2 \chi) \, s_{\mu} \, 
c_{\mu}^{-1} \, \tanh \chi \right]  
\, \sqrt{\frac{6 }{(1-y)\, w\,q}} \,
c_{\al + \psi} \nonu \\
& = & - \frac{1}{y}  \, 
F^{46\,10\,11}= \frac{6}{w\,q} s_{\th_1} \, 
F^{478\,11} = \frac{6}{w\,q} s_{\th_1} \, 
c_{\th_1}^{-1} \, F^{489\,11} = - \frac{6}{w\,q} s_{\th_1} \,
c_{\th_1}^{-1} F^{48\,10\,11},
\nonu \\
F^{568\,10} & = &-
\left[\frac{18  \, \cosh^{\frac{4}{3}} \chi \, \sinh \chi }
{L^5 \, \rho^{\frac{8}{3}} \,
(c^2_{\mu} +\rho^{\frac{8}{3}} \, s^2_{\mu})^{\frac{5}{3}}} 
\right]
\left[ 3 c_{\mu}^2  \, \mbox{sech}^{\frac{5}{3}} \chi 
+
\rho^8 (-(-2+ c_{2\mu}) \, \mbox{sech}^{\frac{5}{3}} \chi \right. 
\nonu \\
& + &  \left. 2 \cosh^{\frac{1}{3}} \chi  \, s_{\mu}^2 + 
\rho^8
(-2 \cosh^{\frac{1}{3}} \chi + \mbox{sech}^{\frac{5}{3}} 
\chi ) s_{\mu}^2) \right]  
\, \sqrt{\frac{3 w\,q}{2(1-y)}} \,
c_{\mu}^{-2} \, s_{\al + \psi}
\nonu \\
& = &  -\frac{w\,q}{6} s_{\th_1} \, F^{579\,10},
\nonu \\
F^{568\,11} & = & -
\left[\frac{6 \, \rho^{\frac{16}{3}} \, 
\cosh^{\frac{1}{3}} \chi \, \sinh \chi }
{L^5  \,
(c^2_{\mu} +\rho^{\frac{8}{3}} \, s^2_{\mu})^{\frac{5}{3}}} 
\right] 
\left[-2 \cosh^{\frac{4}{3}} \chi 
+ 3 \mbox{sech}^{\frac{2}{3}} \chi \right. 
\nonu \\
& + &  \left. \rho^8
(2 \cosh^{\frac{4}{3}} \chi +(-2 + 3c_{\mu}^{-2}) 
\mbox{sech}^{\frac{2}{3}} 
\chi + \rho^8 \,
\mbox{sech}^{\frac{2}{3}} \chi  s_{\mu}^2 \,
c_{\mu}^{-2}) \right]  
\, \sqrt{\frac{3 w\,q}{2(1-y)}} \,
 s_{\al + \psi} \nonu \\
&= & - \frac{w\,q}{6 y} \, w\, q\, s_{\th_1} \, 
 F^{579\,11} = 
\frac{w\,q}{6y} s_{\th_1} \, F^{57\,10\,11}=
\frac{1}{6(1-y)} \, s_{\th_1} \, c_{\th_1}^{-1} \,
w\, q\, F^{59\,10\,11},
\nonu \\
F^{569\,10} & = &
\left[ \frac{54  \, \cosh^{\frac{1}{3}} \chi \, \sinh \chi }
{L^5 \, \rho^{\frac{8}{3}} \,
(c^2_{\mu} +\rho^{\frac{8}{3}} \, s^2_{\mu})^{\frac{5}{3}}} 
\right] 
\left[ -3 \mbox{sech}^{\frac{2}{3}} \chi 
+
\rho^8 ((-2+ c_{2\mu}) \, c_{\mu}^{-2} \, 
\mbox{sech}^{\frac{2}{3}} \chi \right. 
\nonu \\
& - &  \left. 2 \cosh^{\frac{4}{3}} \chi  s_{\mu}^2 \,
c_{\mu}^{-2} + 
\rho^8
(2 \cosh^{\frac{4}{3}} \chi -\mbox{sech}^{\frac{2}{3}} 
\chi ) s_{\mu}^2 \, c_{\mu}^{-2}) \right]  
\, \sqrt{\frac{6}{(1-y) w\,q}} \,
 c_{\al + \psi} \nonu \\
&= & \frac{6}{w\,q} \, s_{\th_1} 
\,  F^{578\,10} = \frac{6}{w\,q}  s_{\th_1} \, c_{\th_1}^{-1}\,
F^{589\,10},
\nonu \\
F^{569\,11} & = & -
\left[ \frac{18  \, \rho^{\frac{16}{3}} \,
 \sinh \chi \, \cosh^{\frac{1}{3}} \chi}
{L^5 \,  \, 
(c^2_{\mu} +\rho^8 \, s^2_{\mu})^{\frac{5}{3}}} 
\right] 
\left[ -2 \cosh^{\frac{4}{3}} \chi +
3 \mbox{sech}^{\frac{2}{3}} 
\chi + \rho^8 ( 2 \cosh^{\frac{4}{3}} 
+ ( -2  \right.
\nonu \\
&+ & \left.  3 c_{\mu}^{-2} ) \mbox{sech}^{\frac{2}{3}} \chi +
\rho^8 \, s_{\mu}^2 \, c_{\mu}^{-2} \, 
\mbox{sech}^{\frac{2}{3}} \chi  \right]  
\,   \sqrt{\frac{6}{(1-y)\, w\,q}} \,
c_{\al + \psi} = -\frac{1}{y} \, 
F^{56\,10\,11} \nonu \\
& = & \frac{6}{w\,q} s_{\th_1} \, c_{\th_1}^{-1} \, 
F^{589\,11} = \frac{6}{w\,q} s_{\th_1} \, F^{578\,11}=
- \frac{6}{w\,q} s_{\th_1} \, c_{\th_1}^{-1}\, F^{58\,10\,11},
\nonu \\
F^{68\,10\,11} & = &
\left[ \frac{18\, (\rho^8+3)}{L^5 \, \rho^{\frac{8}{3}}} \, \sinh \chi
 \, \mbox{sech}^{\frac{1}{3}} \chi \, 
 (c^2_{\mu} +\rho^8 \, s^2_{\mu})^{\frac{1}{3}} \right] 
\sqrt{\frac{3 w\,q}{2(1-y)}} \, s_{\mu}^{-1} \, c_{\mu}^{-3} \, c_{\al +
  \psi} \nonu \\
& = &  
-\frac{1}{6} w\, q\, s_{\th_1} \, F^{79\,10\,11}, 
\nonu \\
F^{69\,10\,11} & = &-\left[
 \frac{54\, (\rho^8+3)}{L^5 \, \rho^{\frac{8}{3}}} \, \sinh \chi
 \, \mbox{sech}^{\frac{1}{3}} \chi \, 
 (c^2_{\mu} +\rho^8 \, s^2_{\mu})^{\frac{1}{3}} \right] 
\sqrt{\frac{6 }{(1-y)w\,q}} \, 
s_{\mu}^{-1} \, c_{\mu}^{-3} \, s_{\al +
  \psi}\nonu \\ 
 & = & \frac{6}{w\,q} \, s_{\th_1}  \, F^{78\,10\,11} =
\frac{6}{w\,q} 
c_{\th_1}^{-1} \, s_{\th_1} \, F^{89\,10\,11}. 
\label{4-form10}
\eea
One also obtains (\ref{4-form10}) from
the 4-form field strengths in the
frame basis \cite{Ahn0909} and vielbeins.
At the IR fixed point in 4-dimensions, the components $F^{1235}, F^{4mnp}$ and
$F^{45mn}$
vanish.
It turns out that after computing the right hand side of
(\ref{fieldequations}) using both (\ref{4-form9}) and (\ref{4-form10}),
the dependence on the combination $(\al + \psi)$ disappears
completely.
This coincides with the fact that the Ricci tensor (\ref{Ricci5})
does not depend on these variables. 

\section{$SU(2) \times U(1)$-invariant flow}

In this Appendix, we describe the Ricci tensor and the 4-form field
strengths for $SU(2) \times U(1)$-invariant flow that can be written in terms of 
corresponding Ricci tensor and 4-form field strengths respectively for
$SU(2) \times U(1) \times U(1)_R$-invariant flow.

\subsection{The Ricci tensor}

The Ricci tensor in the coordinate basis from the 11-dimensional
metric (\ref{11d2}), after imposing the flow
equations (\ref{flow}), can be written in terms of those (\ref{Ricci5}) in the
Appendix E as follows:
\bea
\widetilde{R}_{1}^{\,\,1} & = & R_{1}^{\,\,1}, \qquad
\widetilde{R}_{2}^{\,\,2}  =  R_{2}^{\,\,2}, \qquad
\widetilde{R}_{3}^{\,\,3} =  R_{3}^{\,\,3},  \qquad
\widetilde{R}_{4}^{\,\,4}    =  R_{4}^{\,\,4}, \nonu \\
\widetilde{R}_{4}^{\,\,5} & = & -\left[\frac{
\cos \theta \, \sin
  \th_6}
{\sqrt{1-\sin^2 \theta \, \sin^2
    \th_6} } \right] R_{4}^{\,\,5}, \qquad
\widetilde{R}_{4}^{\,\,11}   =  - 
\left[\frac{ \csc \theta \, \cos
  \th_6}
{\sqrt{1-\sin^2 \theta \, \sin^2
    \th_6} } \right] R_{4}^{\,\,5},
\nonu \\
\widetilde{R}_{5}^{\,\,4} & = & -\left[
\frac{\cos \theta \,\sin \th_6}{\sqrt{1-\sin^2 \theta \, \sin^2
    \th_6} } \right] R_{5}^{\,\,4},
 \nonu \\
\widetilde{R}_{5}^{\,\,5} & = & \left[\frac{1 }
{1+\cot^2 \th_6 \, \sec^2
    \theta } \right] R_{5}^{\,\,5} + 
\left[\frac{\cos^2 \th_6}{1-\sin^2 \theta \, \sin^2
    \th_6} \right] R_{11}^{\,\,11}, \nonu \\
\widetilde{R}_{5}^{\,\,10} & = & \left[\frac{\cos \th_6 }
{1-\sin^2 \theta \, \sin^2
    \th_6 } \right] 
  R_{11}^{\,\,10},  \qquad
\widetilde{R}_{5}^{\,\,11}  =  \left[\frac{ \cot \theta   \, \sin
    \th_6 \, \cos \th_6 }
{1-\sin^2 \theta \, \sin^2
    \th_6} \right] \left( R_{5}^{\,\,5} 
- R_{11}^{\,\,11}\right),
\nonu \\
\widetilde{R}_{6}^{\,\,6} & = & R_{6}^{\,\,6}, \nonu \\ 
\widetilde{R}_{7}^{\,\,5}  & = & \left[ \cos \th_6 \right] R_{7}^{\,\,11}, \qquad
\widetilde{R}_{7}^{\,\,7}  =  R_{7}^{\,\,7}, \qquad
\widetilde{R}_{7}^{\,\,10}  =  R_{7}^{\,\,10}, \qquad
\widetilde{R}_{7}^{\,\,11}  =  - \left[\cot \th \, \sin \th_6 \right]  R_{7}^{\,\,11},
\nonu \\
\widetilde{R}_{8}^{\,\,8}  & = &   R_{8}^{\,\,8},   \nonu \\ 
\widetilde{R}_{9}^{\,\,5}  & = & \left[\cos \th_6 \right] R_{9}^{\,\,11}, \qquad
\widetilde{R}_{9}^{\,\,9}    =    R_{9}^{\,\,9}, 
\qquad
\widetilde{R}_{9}^{\,\,10}   =  R_{9}^{\,\,10}, 
\qquad
\widetilde{R}_{9}^{\,\,11}    =   -\left[\cot \theta \, \sin \th_6 \right]  
R_{9}^{\,\,11},
\nonu \\
\widetilde{R}_{10}^{\,\,5}  & = & \left[\cos \th_6 \right] R_{10}^{\,\,11},
\qquad \widetilde{R}_{10}^{\,\,10}  =   R_{10}^{\,\,10}, \qquad
\widetilde{R}_{10}^{\,\,11}   = -\left[\cot \theta \, \sin \th_6 \right]
 R_{10}^{\,\,11},
\nonu \\
\widetilde{R}_{11}^{\,\,4} & = & -\left[
\frac{\sin \theta \,\cos \th_6}{\sqrt{1-\sin^2 \theta \, \sin^2
    \th_6} } \right] R_{5}^{\,\,4},
 \qquad
\widetilde{R}_{11}^{\,\,5}  =  \left[\frac{ \sin \theta \, \cos \th \,    \sin
    \th_6 \, \cos \th_6}
{1-\sin^2 \theta \, \sin^2
    \th_6 } \right] \left( R_{5}^{\,\,5}  
- R_{11}^{\,\,11}\right), \nonu \\
\widetilde{R}_{11}^{\,\,10} & = & -\left[\frac{\cos \th \, \sin \theta
  \, \sin \th_6}
{1-\sin^2 \theta \, \sin^2
    \th_6 } \right] 
  R_{11}^{\,\,10},  
\nonu \\
\widetilde{R}_{11}^{\,\,11} & = & \left[\frac{1 }
{1+\tan^2 \th_6 \, \cos^2
    \theta } \right] R_{5}^{\,\,5} + 
\left[\frac{\cos^2 \theta \, \sin^2 \th_6}{1-\sin^2 \theta \, \sin^2
    \th_6} \right] R_{11}^{\,\,11}.
\label{Ricci6}
\eea
There are extra
nonzero off-diagonal Ricci tensor components
($\widetilde{R}_{4}^{\,11}, 
\widetilde{R}_{5}^{\,10},
\widetilde{R}_{5}^{\,11}, \widetilde{R}_{7}^{\,5}, 
\widetilde{R}_{9}^{\,5}, \widetilde{R}_{10}^{\,5}, \widetilde{R}_{11}^{\,4}$
and $\widetilde{R}_{11}^{\,5}$) for $SU(2) \times U(1)$-invariant flow, 
compared to the Ricci tensor (\ref{Ricci5}) for $SU(2) \times U(1) 
\times U(1)_R$-invariant flow. 
Also the 11-dimensional metric (\ref{11d2}) generates (\ref{Ricci6})
directly. At the IR fixed point in 4-dimensions, the components $\widetilde{R}_4^{\,5},
\widetilde{R}_4^{\,11}, \widetilde{R}_5^{\,4}$ and $\widetilde{R}_{11}^{\,4}$
vanish. 

\subsection{The 4-form field strengths}

The 4-form field strengths satisfying (\ref{fieldequations})
for given Ricci tensor (\ref{Ricci6}) and 
11-dimensional metric (\ref{11d2}),
in terms of those (\ref{4-form9}) for $SU(2) \times U(1) \times U(1)_R$-invariant flow,  
are summarized as follows:
\bea
\widetilde{F}_{1234} & = & F_{1234}, \qquad
\widetilde{F}_{1235}   =  -\left[\frac{\cos \th \, \sin \th_6}
{\sqrt{1-\sin^2 \th \, \sin^2
    \th_6}} \right] F_{1235}, \nonu \\
\widetilde{F}_{123\,11}  & = &  -\left[\frac{\cos \th_6 \, \sin \th}
{\sqrt{1-\sin^2 \th \, \sin^2
    \th_6}} \right] F_{1235},
\nonu \\
\widetilde{F}_{45mn} & = & -\left[\frac{\cos \th \, \sin \th_6}
{\sqrt{1-\sin^2 \th \, \sin^2
    \th_6}} \right] F_{45mn} + \left[ \frac{\cos \th_6}
{1-\sin^2 \th \, \sin^2
    \th_6} \right] F_{4mn\, 11}, \nonu \\
& & (m, n) =(6,7), (6,8),(6,9), (7,8), (7,9),
\nonu \\
\widetilde{F}_{46mn}  & = &  F_{46mn}, \qquad 
(m, n) =(7,8), (7,9), (7,10), (8,9), (8, 10), (9,10),
\nonu \\
\widetilde{F}_{4mn\,11} & = & 
-\left[\frac{\sin \th\, \cos \th_6}
{\sqrt{1-\sin^2 \th \, \sin^2
    \th_6}} \right] F_{45mn} - \left[ \frac{\sin \th\, \, \cos \th\, \sin \th_6}
{1-\sin^2 \th \, \sin^2
    \th_6} \right] F_{4mn\, 11}, \nonu \\
& & (m, n)=(6,7), (6,8), (6,9), (7,8), (7,9),
\nonu \\
\widetilde{F}_{47mn}  & = &  F_{47mn}, \qquad 
(m, n) =(8,9), (8, 10), (9,10),
\nonu \\
\widetilde{F}_{56mn} & = &  -\left[\frac{\cos \th \, \sin \th_6}
{\sqrt{1-\sin^2 \th \, \sin^2
    \th_6}} \right] F_{56mn} -   \left[\frac{\cos \th_6}
{1-\sin^2 \th \, \sin^2
    \th_6} \right] F_{6mn\,11}, \nonu \\
&& (m, n) =(7,8), (7,9), (7,10), (8,9), (8, 10), (9,10),
\nonu \\
\widetilde{F}_{5mn\,11} & = & \left[\frac{\sin \th}{\sqrt{1-\sin^2 \th \, \sin^2
    \th_6}} \right] F_{5mn\, 11}, 
\qquad  (m, n) =(6,7), (6, 8), (6,9), (7,8), (7,9),
\nonu \\
\widetilde{F}_{57mn} & = &  -\left[\frac{\cos \th \, \sin \th_6}
{\sqrt{1-\sin^2 \th \, \sin^2
    \th_6}} \right] F_{57mn} -   \left[\frac{\cos \th_6}{1-\sin^2 \th \, \sin^2
    \th_6} \right] F_{7mn\,11}, \nonu \\
&& (m, n) =(8,9), (8, 10), (9,10), 
\nonu \\
\widetilde{F}_{mnp\,11} & = & \left[\frac{\sin \th \, \cos \th_6}
{\sqrt{1-\sin^2 \th \, \sin^2
    \th_6}} \right] F_{5mnp} -   \left[\frac{\sin \th_6 \, \cos
 \th \, \sin \th}
{1-\sin^2 \th \, \sin^2
    \th_6} \right] F_{mnp\,11}, \nonu \\
& & (m, n, p =6, \cdots, 9),
\nonu \\
\widetilde{F}_{mn\,10\,11} & = & \left[\frac{\sin \th \, \cos \th_6}
{\sqrt{1-\sin^2 \th \, \sin^2
    \th_6}} \right] F_{5mn\,10}  - \left[\frac{\sin \th \, \cos
 \th \, \sin  \th_6}
{1-\sin^2 \th \, \sin^2
    \th_6} \right] F_{mn\,10\,11}, \nonu \\
&& (m, n)=(6,7), (6, 8), (6, 9), (7,8), (7,9).
\label{4-form11}
\eea
The $\widetilde{F}_{123\,11}$ is new, compared to the $SU(2) \times
U(1) \times
U(1)_R$-invariant flow.
At the IR fixed point in 4-dimensions, the components $\widetilde{F}_{1235},
\widetilde{F}_{123\,11},
 \widetilde{F}_{4mnp}$ and
$\widetilde{F}_{45mn}$
vanish.

The 4-form field strengths with upper indices can be obtained from
those with lower indices (\ref{4-form11}) by multiplying the
11-dimensional inverse metric (\ref{11d2}) and they are given by as follows:
\bea
\widetilde{F}^{1234} & = & F^{1234}, \qquad
\widetilde{F}^{1235}   =  -\left[\frac{\cos \th \, \sin \th_6}
{\sqrt{1-\sin^2 \th \, \sin^2
    \th_6}} \right] F^{1235}, \nonu \\
\widetilde{F}^{123\,11}  & = &  -\left[\frac{\csc \th \, \cos \th_6}
{\sqrt{1-\sin^2 \th \, \sin^2
    \th_6}} \right] F^{1235}, 
\nonu \\
\widetilde{F}^{45mn} & = & 
 \left[\cos \th_6
 \right] F^{4mn\, 11}
-\left[\frac{\cos \th \, \sin \th_6}
{\sqrt{1-\sin^2 \th \, \sin^2
    \th_6}} \right] F^{45mn}, 
\nonu \\ && 
(m, n)=(6,8), (6, 9), (6,10), (7,8), (7,9), (7,10), (8,9), (8,10), (9,10),
\nonu \\
\widetilde{F}^{4mn\,10} & = & F^{4mn\, 10}, 
\qquad (m, n) =(6,8), (6, 9), (7,8), (7,9), (8,9),
\nonu \\
\widetilde{F}^{4mn\,11} & = &
 - \left[\sin
  \th_6 \, \cot \th  \right] F^{4mn\,11}
 -\left[\frac{\csc \th \, \cos \th_6}
{\sqrt{1-\sin^2 \th \, \sin^2
    \th_6}} \right] F^{45mn}, \nonu \\
&& (m, n) =(6,8), (6,9), (6,10), (7, 8), (7,9), (7,10), (8,9), (8,10), (9,10), 
\nonu \\
\widetilde{F}^{5mn\,10} & = & 
-   \cos \th_6\, F^{mn\,10\,11}
-\left[\frac{\cos \th \, \sin \th_6}
{\sqrt{1-\sin^2 \th \, \sin^2
    \th_6}} \right] F^{5mn\, 10}, \nonu \\
&& (m,n) =(6,8),(6,9), (7,8), (7,9), (8,9),
\nonu \\
\widetilde{F}^{5mn\,11} & = & \left[\csc \th \, 
\sqrt{1-\sin^2 \th \, \sin^2 \th_6} 
\right] F^{5mn\, 11}, \nonu \\
&&  (m, n) =(6, 8), (6,9), (6, 10),(7,8),
(7,9), (7,10), (8,10), (9,10),
\nonu \\
\widetilde{F}^{mn\,10\,11} & = & - \left[\cot \th \, 
\sin \th_6 \right] F^{mn\, 10\,11}  
 +  \left[\frac{\csc \th \,  \cos \th_6}
{\sqrt{1-\sin^2 \th \, \sin^2
    \th_6}}  \right] F^{5mn\,10}, 
\nonu \\ 
&& (m, n) =(6, 8), (6, 9), (7, 8), (7, 9), (8,9).
\label{4-form12}
\eea
There is a new 4-form  
$
\widetilde{F}^{123\,11}$, 
compared to the 4-form field strengths (\ref{4-form10}) for $SU(2)
\times U(1) \times
U(1)_R$-invariant flow.
At the IR fixed point in 4-dimensions, the components $\widetilde{F}^{1235},
\widetilde{F}^{123\,11}, \widetilde{F}^{4mnp}$ and
$\widetilde{F}^{45mn}$
vanish.
By substituting the Ricci tensor (\ref{Ricci6}) and 
4-form field strengths (\ref{4-form11}) and (\ref{4-form12}) into the
Einstein equation in (\ref{fieldequations}), one can check that 
the LHS coincides with the RHS exactly. In doing this, 
one uses the fact that the solution characterized by (\ref{Ricci5}),
(\ref{4-form9}) 
and (\ref{4-form10}) for $SU(2) \times U(1) \times U(1)_R$-invariant flow
satisfies the Einstein equation. Therefore one concludes that 
the Einstein
equation for $SU(2) \times U(1)$-invariant flow is satisfied. 

\subsection{The left hand side of Maxwell equation}

The nonzero components of left hand side of Maxwell
equations in terms of (\ref{4-form9}) are given as follows:
\bea
(123) & = & 
F_{479\,11} F_{568\,10} - F_{479\,10} F_{568\,11}
-  F_{478\,11} F_{569\,10}+ F_{478\,10} F_{569\,11} \nonu \\
& - & F_{469\,11} F_{578\,10} + 
F_{469\,10} F_{578\,11} + F_{468\,11} F_{579\,10}   
-  F_{468\,10} F_{579\,11} \nonu \\
& + & F_{4579} F_{68\,10\,11} - F_{4578}
F_{69\,10\,11}  
- F_{4569} F_{78\,10\,11} + F_{4568} F_{79\,10\,11}, \nonu \\
(468) & = & -
F_{1235} \, F_{79\,10\,11}, \qquad
(469)  =  
F_{1235} \, F_{78\,10\,11}, \qquad
(46\,10)   =  - 
F_{1235} \, F_{789\,11}, \nonu \\
(478)  & = &   
F_{1235} \, F_{69\,10\,11}, \qquad
(479)   =   -
F_{1235} \, F_{68\,10\,11}, \qquad
(47\,10)  =  
F_{1235}  F_{689\,11}, \nonu \\
(489) & = & 
F_{1235}   F_{67\,10\,11}, \qquad
(48\,10)  =  -
F_{1235}   F_{679\,11}, \qquad
(49\,10)  =  
F_{1235}  F_{678\,11}, \nonu \\
(568) & = & -\left[\cos \th_6\right] 
( F_{1235}  F_{479\,10} - F_{1234}  F_{579\,10}) 
 - \left[
\frac{
 \cos \th \, \sin \th_6}{\sqrt{1-\sin^2 \th \, \sin^2 \th_6}} \right] 
F_{1234} \, F_{79\,10\,11}, \nonu 
\\
(569) & = & \left[\cos \th_6 \right] 
( F_{1235}  F_{478\,10} - F_{1234}  F_{578\,10}) 
 + \left[
\frac{
 \cos \th \, \sin \th_6}{\sqrt{1-\sin^2 \th \, \sin^2 \th_6}} \right]
F_{1234} \, F_{78\,10\,11}, \nonu \\
(56\,10) & = & -\left[\cos \th_6\right] 
( F_{1235}  F_{4789} - F_{1234}  F_{5789}) 
 - \left[
\frac{
 \cos \th \, \sin \th_6}{\sqrt{1-\sin^2 \th \, \sin^2 \th_6}} \right]
F_{1234} \, F_{789\,11}, \nonu \\
(578) & = & \left[\cos \th_6\right] 
( F_{1235}  F_{469\,10} - F_{1234}  F_{569\,10}) 
 + \left[
\frac{
 \cos \th \, \sin \th_6}{\sqrt{1-\sin^2 \th \, \sin^2 \th_6}} \right]
F_{1234} \, F_{69\,10\,11}, \nonu \\
(579) & = & -\left[\cos \th_6\right] 
(F_{1235}  F_{468\,10} - F_{1234}  F_{568\,10}) 
 -  \left[\frac{
 \cos \th \, \sin \th_6}{\sqrt{1-\sin^2 \th \, \sin^2 \th_6}}\right] 
F_{1234}  \, F_{68\,10\,11},
\nonu \\
(57\,10) & = & 
 \left[\cos \th_6 \right] 
(F_{1235}\, F_{4689} -F_{1234}\, F_{5689})+   
\left[\frac{\cos \th \, \sin \th_6}{ \sqrt{1-\sin^2 \th \,
  \sin^2 \th_6}}
\right] F_{1234} \, F_{689\,11},  
\nonu \\
(589) & = & 
\left[\cos \th_6 \right] 
(F_{1235}\, F_{467\,10} -F_{1234}\, F_{567\,10})+
  \left[ \frac{\cos \th \, \sin \th_6}{ \sqrt{1-\sin^2 \th \,
  \sin^2 \th_6}}
\right] F_{1234} \, F_{67\,10\,11},  
\nonu \\
(58\,10) & = & 
 -     
\left[\cos \th_6 \right] (F_{1235}\, F_{4679} -F_{1234}\, F_{5679})
   -\left[\frac{\cos \th \, \sin \th_6}{ \sqrt{1-\sin^2 \th \,
  \sin^2 \th_6}}
\right] F_{1234} \, F_{679\,11},  
\nonu \\
(59\,10) & = & 
\left[\cos \th_6 \right] (F_{1235}\, F_{4678} -F_{1234}\, F_{5678})+
  \left[\frac{\cos \th \, \sin \th_6}{ \sqrt{1-\sin^2 \th \,
  \sin^2 \th_6}}
\right] F_{1234} \, F_{678\,11},  
\nonu \\
(68\,10) & = & 
F_{1235} F_{479\,11} 
- F_{1234} F_{579\,11},
\nonu \\
(68\,11) & = & \left[\cot \th  \sin
\th_6 \right]
(F_{1235}  F_{479\,10} - F_{1234}  F_{579\,10}) 
-  \left[\frac{
 \csc \th \, \cos \th_6}{\sqrt{1-\sin^2 \th \, \sin^2 \th_6}} 
\right] F_{1234} \,  F_{79\,10\,11},
\nonu \\
(69\,10) & = & 
-F_{1235} F_{478\,11} + 
F_{1234} F_{578\,11},
\nonu \\
(69\,11) & = & \left[\cot \th  \sin \th_6\right] 
(-F_{1235}  F_{478\,10} + F_{1234}  F_{578\,10}) 
 +  \left[\frac{
 \csc \th \, \cos \th_6}{\sqrt{1-\sin^2 \th \, \sin^2 \th_6}} 
\right] F_{1234}  \, F_{78\,10\,11},
\nonu \\
(6\,10\,11) & = & \left[\cot \th  \sin \th_6 \right] 
(F_{1235}  F_{4789} - F_{1234}  F_{5789}) 
 - \left[ \frac{
 \csc \th \, \cos \th_6}{\sqrt{1-\sin^2 \th \, \sin^2 \th_6}} 
\right] F_{1234}  \, F_{789\,11},
\nonu \\
(78\,10) & = & 
-F_{1235} F_{469\,11} + 
F_{1234} 
F_{569\,11},
\nonu \\
(78\,11) & = & \left[\cot \th  \sin \th_6 \right] 
(-F_{1235}  F_{469\,10} + F_{1234}  F_{569\,10}) 
 + \left[\frac{
 \csc \th \, \cos \th_6}{\sqrt{1-\sin^2 \th \,  \sin^2 \th_6}} 
\right] F_{1234}\,  F_{69\,10\,11},
\nonu \\
(79\,10) & = & 
F_{1235} F_{468\,11} - 
F_{1234} 
F_{568\,11},
\nonu \\
(79\,11) & = & \left[\cot \th  \sin \th_6\right] 
(F_{1235}  F_{468\,10} - F_{1234}  F_{568\,10}) 
 - \left[\frac{
 \csc \th \, \cos \th_6}{\sqrt{1-\sin^2 \th \,  \sin^2 \th_6}} 
\right] F_{1234}\,  F_{68\,10\,11},
\nonu \\
(7\,10\,11) & = & \left[\cot \th  \sin \th_6\right]
(-F_{1235}  F_{4689} + F_{1234}  F_{5689}) 
 +  \left[\frac{
 \csc \th  \, \cos \th_6}{\sqrt{1-\sin^2 \th \, \sin^2 \th_6}} 
\right] F_{1234} \, F_{689\,11},
\nonu \\
(89\,10) & = & 
-F_{1235} F_{467\,11} + 
F_{1234} 
F_{567\,11},
\nonu \\
(89\,11) & = & \left[\cot \th  \sin \th_6 \right] 
(-F_{1235}  F_{467\,10} + F_{1234}  F_{567\,10}) 
 + \left[ \frac{
 \csc \th \, \cos \th_6}{\sqrt{1-\sin^2 \th \, \sin^2 \th_6}} 
\right]
F_{1234}\,  F_{67\,10\,11},
\nonu \\
(8\,10\,11) & = & \left[\cot \th  \sin \th_6\right] 
(F_{1235}  F_{4679} - F_{1234}  F_{5679}) 
 -  \left[\frac{
 \csc \th \, \cos \th_6}{\sqrt{1-\sin^2 \th \, \sin^2 \th_6}} 
\right]
F_{1234}\,  F_{679\,11},
\label{max2} 
\\
(9\,10\,11) & = & \left[\cot \th  \sin \th_6 \right] 
(-F_{1235}  F_{4678} + F_{1234}  F_{5678}) 
 +  \left[\frac{
 \csc \th \, \cos \th_6}{\sqrt{1-\sin^2 \th \, \sin^2 \th_6}} 
\right] F_{1234} \,  F_{678\,11}.
\nonu
\eea
One can easily check that 
the RHS of Maxwell equation (\ref{fieldequations})
with (\ref{4-form11}) and (\ref{11d2}) is exactly coincident with the
above
LHS of Maxwell equation (\ref{max2}).



\begin{thebibliography}{99}

\bibitem{ABJM}
  O.~Aharony, O.~Bergman, D.~L.~Jafferis and J.~Maldacena,
  ``N=6 superconformal Chern-Simons-matter theories, M2-branes and their
  gravity duals,''
  JHEP {\bf 0810}, 091 (2008).
  [arXiv:0806.1218 [hep-th]].

\bibitem{KW}
  I.~R.~Klebanov and E.~Witten,
  ``Superconformal field theory on threebranes at a Calabi-Yau  singularity,''
  Nucl.\ Phys.\  B {\bf 536}, 199 (1998)
  [arXiv:hep-th/9807080].

\bibitem{dN}
  B.~de Wit and H.~Nicolai,
  ``N=8 Supergravity,''
  Nucl.\ Phys.\  B {\bf 208}, 323 (1982);
%
  B.~de Wit and H.~Nicolai,
  ``N=8 Supergravity With Local SO(8) X SU(8) Invariance,''
  Phys.\ Lett.\  B {\bf 108}, 285 (1982).

\bibitem{Maldacena}
  J.~M.~Maldacena,
  ``The large N limit of superconformal field theories and supergravity,''
  Adv.\ Theor.\ Math.\ Phys.\  {\bf 2}, 231 (1998)
  [Int.\ J.\ Theor.\ Phys.\  {\bf 38}, 1113 (1999)]
  [arXiv:hep-th/9711200];
%
  E.~Witten,
  ``Anti-de Sitter space and holography,''
  Adv.\ Theor.\ Math.\ Phys.\  {\bf 2}, 253 (1998)
  [arXiv:hep-th/9802150];
%
  S.~S.~Gubser, I.~R.~Klebanov and A.~M.~Polyakov,
  ``Gauge theory correlators from non-critical string theory,''
  Phys.\ Lett.\  B {\bf 428}, 105 (1998)
  [arXiv:hep-th/9802109].

\bibitem{AP}
  C.~Ahn and J.~Paeng,
  ``Three-dimensional SCFTs, supersymmetric domain wall and renormalization
  group flow,''
  Nucl.\ Phys.\  B {\bf 595}, 119 (2001)
  [arXiv:hep-th/0008065].

\bibitem{AW}
  C.~Ahn and K.~Woo,
  ``Supersymmetric domain wall and RG flow from 4-dimensional gauged N = 8
  supergravity,''
  Nucl.\ Phys.\  B {\bf 599}, 83 (2001).
  [arXiv:hep-th/0011121].

\bibitem{AR99}
  C.~Ahn and S.~J.~Rey,
  ``More CFTs and RG flows from deforming M2/M5-brane horizon,''
  Nucl.\ Phys.\  B {\bf 572}, 188 (2000).
  [arXiv:hep-th/9911199].

\bibitem{AI}
  C.~Ahn and T.~Itoh,
  ``An N = 1 supersymmetric G(2)-invariant flow in M-theory,''
  Nucl.\ Phys.\  B {\bf 627}, 45 (2002).
  [arXiv:hep-th/0112010].

\bibitem{CPW}
  R.~Corrado, K.~Pilch and N.~P.~Warner,
  ``An N = 2 supersymmetric membrane flow,''
  Nucl.\ Phys.\  B {\bf 629}, 74 (2002)
  [arXiv:hep-th/0107220].

\bibitem{Ahn0806n2}
  C.~Ahn,
  ``Holographic Supergravity Dual to Three Dimensional N=2 Gauge Theory,''
  JHEP {\bf 0808}, 083 (2008).
  [arXiv:0806.1420 [hep-th]].

\bibitem{BKKS}
  M.~Benna, I.~Klebanov, T.~Klose and M.~Smedback,
  ``Superconformal Chern-Simons Theories and $AdS_4/CFT_3$ Correspondence,''
  JHEP {\bf 0809}, 072 (2008).
  [arXiv:0806.1519 [hep-th]];
%
  I.~Klebanov, T.~Klose and A.~Murugan,
  ``$AdS_4/CFT_3$ -- Squashed, Stretched and Warped,''
  JHEP {\bf 0903}, 140 (2009).
  [arXiv:0809.3773 [hep-th]];
%
  I.~R.~Klebanov, S.~S.~Pufu and F.~D.~Rocha,
  ``The Squashed, Stretched, and Warped Gets Perturbed,''
  JHEP {\bf 0906}, 019 (2009)
  [arXiv:0904.1009 [hep-th]].

\bibitem{Ahn0806n1}
  C.~Ahn,
  ``Towards Holographic Gravity Dual of N=1 Superconformal Chern-Simons Gauge
  Theory,''
  JHEP {\bf 0807}, 101 (2008)
  [arXiv:0806.4807 [hep-th]].

\bibitem{Ahn0812}
  C.~Ahn,
  ``Comments on Holographic Gravity Dual of N=6 Superconformal Chern-Simons
  Gauge Theory,''
  JHEP {\bf 0903}, 107 (2009).
  [arXiv:0812.4363 [hep-th]].

\bibitem{BHPW}
  N.~Bobev, N.~Halmagyi, K.~Pilch and N.~P.~Warner,
  ``Holographic, N=1 Supersymmetric RG Flows on M2 Branes,''
  JHEP {\bf 0909}, 043 (2009)
  [arXiv:0901.2736 [hep-th]].

\bibitem{GW02}
  C.~N.~Gowdigere and N.~P.~Warner,
  ``Flowing with eight supersymmetries in M-theory and F-theory,''
  JHEP {\bf 0312}, 048 (2003)
  [arXiv:hep-th/0212190].

\bibitem{PW03}
  C.~N.~Pope and N.~P.~Warner,
  ``A dielectric flow solution with maximal supersymmetry,''
  JHEP {\bf 0404}, 011 (2004)
  [arXiv:hep-th/0304132].

\bibitem{AW09}
  C.~Ahn and K.~Woo,
  ``N=8 Gauged Supergravity Theory and N=6 Superconformal Chern-Simons Matter
  Theory,''
  arXiv:0901.4666 [hep-th];
%
  C.~Ahn and K.~Woo,
  ``Are There Any New Vacua of Gauged N=8 Supergravity in Four Dimensions?,''
  arXiv:0904.2105 [hep-th].

\bibitem{Ahn0905}
  C.~Ahn,
  ``The Gauge Dual of Gauged N=8 Supergravity Theory,''
  arXiv:0905.3943 [hep-th].

\bibitem{AW0907}
  C.~Ahn and K.~Woo,
  ``Perturbing Around A Warped Product Of $AdS_4$ and Seven-Ellipsoid,''
  JHEP {\bf 0908}, 065 (2009)
  [arXiv:0907.0969 [hep-th]].

\bibitem{AW0908}
  C.~Ahn and K.~Woo,
  ``The Gauge Dual of A Warped Product of $AdS_4$ and A Squashed and Stretched
  Seven-Manifold,''
  Class.\ Quant.\ Grav.\  {\bf 27}, 035009 (2010)
  [arXiv:0908.2546 [hep-th]].

\bibitem{FKR}
  S.~Franco, I.~R.~Klebanov and D.~Rodriguez-Gomez,
  ``M2-branes on Orbifolds of the Cone over $Q^{1,1,1}$,''
  JHEP {\bf 0908}, 033 (2009)
  [arXiv:0903.3231 [hep-th]].

\bibitem{Ahn0909}
  C.~Ahn,
  ``New N=2 Supersymmetric Membrane Flow In Eleven-Dimensional Supergravity,''
  JHEP {\bf 0911}, 022 (2009)
  [arXiv:0909.3745 [hep-th]].

\bibitem{Ahn0910}
  C.~Ahn,
  ``The Eleven-Dimensional Uplift of Four-Dimensional Supersymmetric RG Flow,''
  arXiv:0910.3533 [hep-th].

\bibitem{AW1001}
  C.~Ahn and K.~Woo,
  ``An N=1 SU(3)-Invariant Supersymmetric Membrane Flow In Eleven-Dimensional
  Supergravity,''
  arXiv:1001.3783 [hep-th].

\bibitem{dWNW}
  B.~de Wit, H.~Nicolai and N.~P.~Warner,
  ``The Embedding Of Gauged N=8 Supergravity Into D = 11 Supergravity,''
  Nucl.\ Phys.\  B {\bf 255}, 29 (1985).

\bibitem{dN87}
  B.~de Wit and H.~Nicolai,
  ``The Consistency of the S**7 Truncation in D=11 Supergravity,''
  Nucl.\ Phys.\  B {\bf 281}, 211 (1987).

\bibitem{Warner83}
  N.~P.~Warner,
  ``Some New Extrema Of The Scalar Potential Of Gauged N=8 Supergravity,''
  Phys.\ Lett.\  B {\bf 128}, 169 (1983);
  N.~P.~Warner,
  ``Some Properties Of The Scalar Potential In Gauged Supergravity Theories,''
  Nucl.\ Phys.\  B {\bf 231}, 250 (1984).

\bibitem{CJS}
  E.~Cremmer, B.~Julia and J.~Scherk,
  ``Supergravity theory in 11 dimensions,''
  Phys.\ Lett.\  B {\bf 76}, 409 (1978).

\bibitem{DNP}
  M.~J.~Duff, B.~E.~W.~Nilsson and C.~N.~Pope,
  ``Kaluza-Klein Supergravity,''
  Phys.\ Rept.\  {\bf 130}, 1 (1986).

\bibitem{AI02}
  C.~Ahn and T.~Itoh,
  ``The 11-dimensional metric for AdS/CFT RG flows with common SU(3)
  invariance,''
  Nucl.\ Phys.\  B {\bf 646}, 257 (2002)
  [arXiv:hep-th/0208137].

\bibitem{FR}
  P.~G.~O.~Freund and M.~A.~Rubin,
  ``Dynamics Of Dimensional Reduction,''
  Phys.\ Lett.\  B {\bf 97}, 233 (1980).

\bibitem{BL}
  J.~Bagger and N.~Lambert,
  ``Gauge Symmetry and Supersymmetry of Multiple M2-Branes,''
  Phys.\ Rev.\  D {\bf 77}, 065008 (2008)
  [arXiv:0711.0955 [hep-th]];
%
  J.~Bagger and N.~Lambert,
  ``Modeling multiple M2's,''
  Phys.\ Rev.\  D {\bf 75}, 045020 (2007)
  [arXiv:hep-th/0611108];
%
  J.~Bagger and N.~Lambert,
  ``Comments On Multiple M2-branes,''
  JHEP {\bf 0802}, 105 (2008)
  [arXiv:0712.3738 [hep-th]].

\bibitem{mass}
  J.~Gomis, A.~J.~Salim and F.~Passerini,
  ``Matrix Theory of Type IIB Plane Wave from Membranes,''
  JHEP {\bf 0808}, 002 (2008)
  [arXiv:0804.2186 [hep-th]];
%
  K.~Hosomichi, K.~M.~Lee and S.~Lee,
  ``Mass-Deformed Bagger-Lambert Theory and its BPS Objects,''
  Phys.\ Rev.\  D {\bf 78}, 066015 (2008)
  [arXiv:0804.2519 [hep-th]];
%
  J.~Gomis, D.~Rodriguez-Gomez, M.~Van Raamsdonk and H.~Verlinde,
  ``A Massive Study of M2-brane Proposals,''
  JHEP {\bf 0809}, 113 (2008)
  [arXiv:0807.1074 [hep-th]].

\bibitem{GMSW}
  J.~P.~Gauntlett, D.~Martelli, J.~Sparks and D.~Waldram,
  ``Sasaki-Einstein metrics on S(2) x S(3),''
  Adv.\ Theor.\ Math.\ Phys.\  {\bf 8}, 711 (2004)
  [arXiv:hep-th/0403002].



\end{thebibliography}
\end{document}